\documentclass[11pt,letterpaper]{article}
\usepackage[margin=1in]{geometry}
\usepackage[T1]{fontenc}
\usepackage[utf8]{inputenc}
\usepackage{amsmath}
\usepackage{newtxtext,newtxmath}
\usepackage{textcomp}
\usepackage{siunitx}
\usepackage{multirow,booktabs}
\usepackage{graphicx}
\usepackage[table]{xcolor}
\usepackage{array}
\usepackage{microtype}
\usepackage{setspace}
\usepackage[font={small,singlespacing},labelfont=bf,labelsep=period,justification=justified,singlelinecheck=off,skip=6pt]{caption}

\usepackage{cite}
\makeatletter

\renewcommand\citepunct{,\penalty\@m\ }
\renewcommand{\@biblabel}[1]{#1.}
\renewcommand\section{\@startsection{section}{1}{\z@}{-3.25ex\@plus -1ex \@minus -.2ex}{1.2ex \@plus .2ex}{\normalfont\large\bfseries}}
\renewcommand\subsection{\@startsection{subsection}{2}{\z@}{-2.5ex\@plus -1ex \@minus -.2ex}{0.8ex \@plus .2ex}{\normalfont\normalsize\bfseries}}
\makeatother

\usepackage[hidelinks]{hyperref}
\newcommand\smref[1]{\ref*{#1}}
\newcommand\mref[1]{\ref*{#1}}

\newif\ifsubmission
\submissiontrue

\begin{document}

\begin{flushleft}
{\Large\bfseries The typical red cell makes too little vacuole material for coalescence to build the large vacuoles of pocked cells\par}
\medskip
\textbf{Short title:} Vacuole supply limits coalescence in pocked cells\par
\medskip
Zhaojie Chai$^{1\dagger}$, Jianlu Zheng$^{2\dagger}$, He Li$^{3}$, Ming Dao$^{2}$, George Em Karniadakis$^{1}$
\par\medskip
{\small
$^{1}$Division of Applied Mathematics, Brown University, Providence, RI 02912, USA.\\
$^{2}$Department of Materials Science and Engineering, Massachusetts Institute of Technology, Cambridge, MA 02139, USA.\\
$^{3}$School of Chemical, Materials and Biomedical Engineering, University of Georgia, Athens, GA 30602, USA.\\
$^{\dagger}$These authors contributed equally to this work.\par}
\end{flushleft}

\section*{Abstract}
Red blood cells carrying intracellular vacuoles, called pocked cells, are a sensitive marker of splenic function: about $0.5\%$ of cells in healthy people, $20$ to $70\%$ after splenectomy. Vacuoles too small to see are common, and visible ones are thought to grow when small ones merge. We tested that idea with a particle model that resolves the cell's membrane, its contents and every vacuole, and with an inventory of the vacuole volume a cell can make. In tube, shear and vascular flows, a cell's vacuoles met within a second, and, allowed to fuse, ten merged into one to five. Meeting is not what limits growth; material is. At the rate measured after splenectomy, all the vacuoles an average cell makes in its $120$-day life fuse into one body $0.6~\mu$m across, the size healthy cells carry, but $16$ times smaller in volume than the $1.5$-$\mu$m bodies of patients without a spleen. Taking vacuole size as set at formation, a population model in which vacuoles are made mainly in young cells, and far faster in a small minority that alone makes material for a large one, accounts for the published counts, sizes and age profiles and, on first-hand measurements, favours a splenic clearance that rises several-fold as cells age; letting that minority's vacuoles fuse fits nearly as well. At a model splenic slit, the cell's spare surface decided passage, and a vacuole mattered only when it led. A slit experiment would test whether brief encounters fuse vacuoles.

\section*{Teaser}
Most red cells make too little vacuole material in a lifetime for fusion to build the large vacuoles seen without a spleen.

\section*{Introduction}

A mature human red blood cell (RBC) has no nucleus, no mitochondria and no other organelle: all are cleared during terminal erythropoiesis, leaving a disc filled with haemoglobin. That simplicity is what makes the cell so deformable. It passes capillaries of $\sim\!3~\mu$m and the inter-endothelial slits of the spleen, $0.2$--$1.2~\mu$m wide, without rupture~\cite{Mebius2005Structure,pivkin2016biomechanics,Li2018Mechanics,moreau2023physical,dao2021erythrocyte}. Yet some circulating RBCs are not empty. They carry vacuoles, and are called ``pocked'' cells. In healthy subjects they are about $0.5\%$ of RBCs, with an upper reference limit of $2\%$; after splenectomy, in congenital asplenia and in sickle cell disease they are $20$--$70\%$, which is why their count serves as a sensitive and specific index of splenic filtration~\cite{buchanan1987formation,buchanan1987pocked,dumas2023intra,sissoko2022erythrocytic}. How the vacuoles arise, and how they grow, is only partly understood.

Three observations frame the question. The first is about size. By differential interference contrast (DIC) microscopy, pocked cells from splenectomized subjects carry one to over twenty vacuoles of $0.2$--$1.5~\mu$m, whereas healthy subjects usually show a single, smaller one of $0.2$--$0.8~\mu$m, and no normal cell carries more than five~\cite{holroyde1970acquisition}. Below what DIC can see there are more: by transmission electron microscopy, vacuoles of $130\pm72$~nm are found in about $13\%$ of red cell sections in healthy subjects and four times as often after splenectomy~\cite{reinhart1988red}. The second is about timing. Vacuoles appear in cells that are already mature: normal RBCs transfused into a hyposplenic recipient acquire them~\cite{holroyde1970acquisition}, mature erythrocytes form them spontaneously during incubation in vitro~\cite{sills1988physiologic,colin1991spontaneous}, and after splenectomy the pocked fraction rises only slowly, passing $20\%$ at $20$--$40$ days and levelling off at $40$--$60\%$ at $60$--$100$ days~\cite{buchanan1987formation}.

The third is about cell age, and it is the one the coalescence picture was built to explain: the burden runs in \emph{opposite} directions on the two sides of splenectomy. With a spleen, the vacuole-bearing fraction \emph{falls} steeply with cell density, from $5.4\pm2.3\%$ in the lightest fraction to $0.4\pm0.5\%$ in the densest, on the same gradient that puts the reticulocytes at $10.5\%$, $1.2\%$ and $0.2\%$~\cite{sills1988formation}. After splenectomy it \emph{rises}: pocked cells are more frequent in the densest fraction~\cite{reinhart1988red}, and vacuoles are reported to develop early in cell life and to persist~\cite{dehaan1988vacuoles}. To reconcile the first observation with the third, Reinhart and Chien concluded that small vacuoles ``tend to cluster and fuse during cell aging, and that the spleen is capable of removing these structures when they reach a certain size''~\cite{reinhart1988red}. A recent review takes this up as one hypothesis that reconciles all the information currently available: smaller vacuoles created during erythropoiesis coalesce progressively, in vivo and in vitro, and so appear optically in the mature cell from elements already present~\cite{dumas2023intra}.

For two vacuoles to fuse they must first meet, so coalescence is a transport problem before it is a membrane problem. The cytosol must carry them together, and the cytosol moves only because the host cell deforms; they must then stay side by side long enough for their bilayers to merge. Shear can drive the fusion of lipid vesicles in viscous solution~\cite{kogan2014shear}. The flows an RBC meets in the microcirculation differ widely even at vanishing Reynolds number: curvature, bifurcations and local dilations impose shear gradients, diverging streamlines and regions where cells linger~\cite{popel2005microcirculation,kihm2021lingering,cebral2005characterization,xiang2011hemodynamics,cai2021artificial}, and a cell responds by deforming and rotating in ways that depend on shear rate, suspending viscosity and confinement~\cite{abkarian2007swinging,fedosov2014deformation,lei2013blood}. Whether these differences change how often, and for how long, the vacuoles inside one cell meet has not been measured.

Coalescence is also an arithmetic problem, and that side has not been checked. Fusion conserves volume. However fast vacuoles fuse and however favourable the flow, the largest body a cell can ever hold is therefore no larger than all the vacuole it has made in its life. Both terms of that limit can be measured. The smallest bodies seen, the candidate starting material, measure $130\pm72$~nm~\cite{reinhart1988red}. And the rate at which vacuoles appear when nothing removes them was measured directly in six patients splenectomized for trauma, bled within the day before surgery and at $48$, $96$ and $144$~h after, whose pocked counts ran $0.4 \rightarrow 1.7 \rightarrow 3.3 \rightarrow 4.1\%$~\cite{sills1988physiologic}. A lifetime at that rate yields a finite amount of membrane and of enclosed volume, which can simply be compared with the sizes observed.

We take the question from both ends. The paths of vacuoles inside a deforming RBC cannot be followed in vivo, so we begin with simulation. We use a dissipative particle dynamics (DPD) model~\cite{hoogerbrugge1992simulating,espanol1995statistical,groot1997dissipative} of the RBC~\cite{fedosov2010multiscale,Fedosov2011Multiscale,li2014erythrocyte} that resolves the host membrane, the cytosol and every vacuole membrane, as the parasite inside a malaria-infected red cell was resolved in earlier simulations~\cite{Fedosov2011Quantifying}, and in which two vacuoles of the same cell merge once their surfaces have stayed within $\delta_f$ for a time $t_f$. Coarse-grained particle models of this kind have been used for red cells and other spectrin--actin skeletons~\cite{tang2017openrbc,chai2023dynamics}, whose periodic layout in the axon leads to sodium nanodomains~\cite{chai2022periodic}, for reticulocytes and stomatocytes~\cite{chai2026reticulocyte,chai2026quantifying} and, in a lysosomal storage disease, for red cells whose rheology is altered by the disease itself~\cite{chai2025silico,chai2026gaucher}, and particle dynamics has also been used to model collective cell migration and crowd evacuation~\cite{zhang2020deep,zhang2021deep}. With this model we ask which flows bring vacuoles into contact, whether those contacts last or turn over, and how a population of ten vacuoles coarsens once contacts can end in fusion. We then do the arithmetic, and the two answers point the same way: the flow gives coalescence every opportunity it could ask for, and a cell nucleating at the measured rate never makes the volume to use them. Supposing instead that a vacuole's size is set at birth, we build a model of the age-structured red cell population and ask what source and what splenic sink account for the published record. A coalescence model given the same source fits it nearly as well, so what the record constrains is the source, not the route by which a vacuole reaches its size. A size set at birth fixes not only how large a vacuole load is but how it is divided, so we next ask at what size, and in what number, vacuoles hold their host cell at a splenic-like constriction~\cite{deplaine2011sensing,safeukui2018sensing}. Finally, we ask what squeezing a cell repeatedly through a slit, the experiment this account proposes, does to its vacuoles.

\section*{Results}

\subsection*{A red cell model with membrane-bound vacuoles that can fuse}

\begin{figure}[!tbp]
\centering
\includegraphics[width=\textwidth,height=0.56\textheight,keepaspectratio]{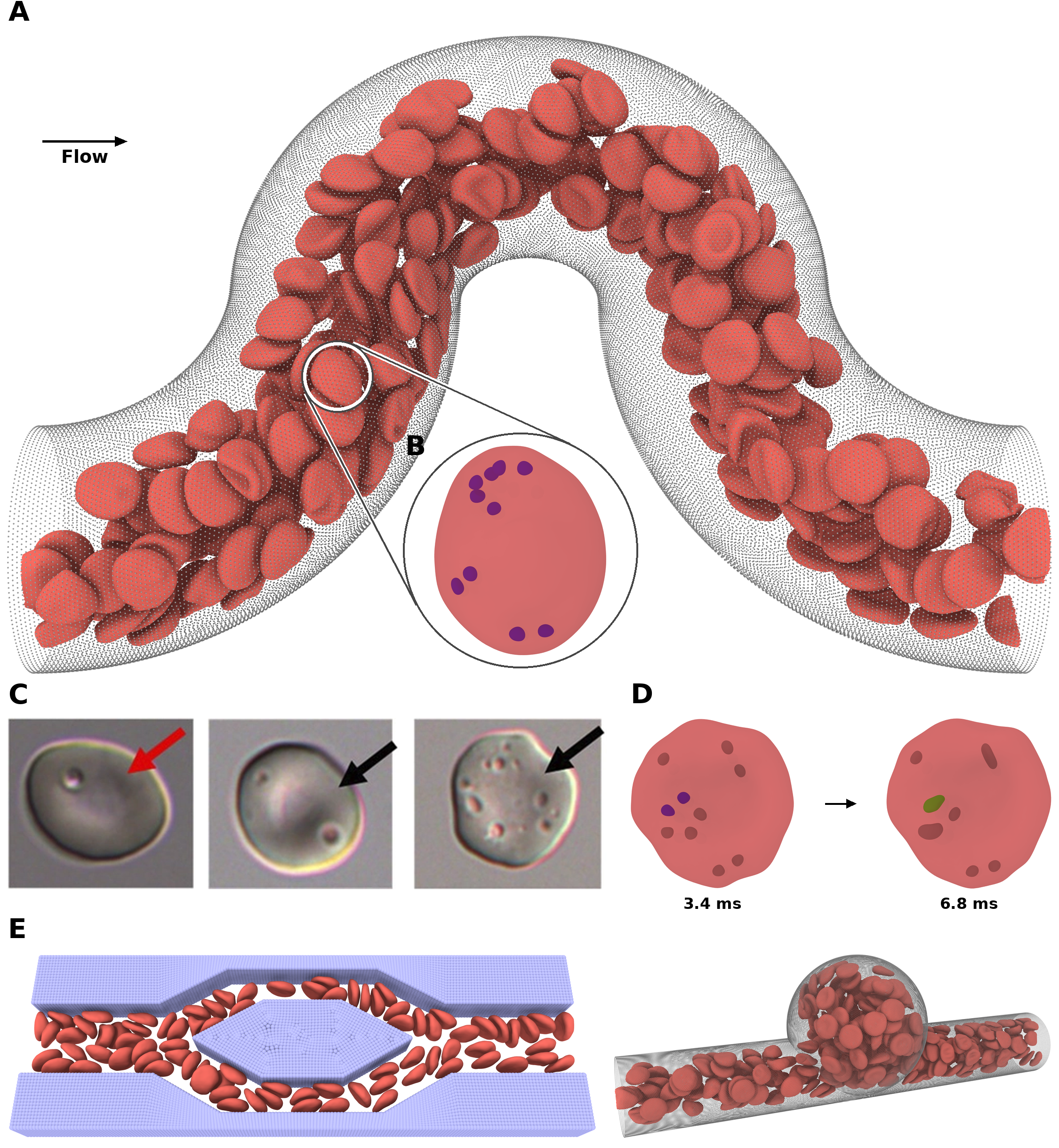}
\caption{\textbf{Pocked red cells and the model.}
\textbf{(A)} The curved vessel at step $150\,000$ ($785$~ms), seen obliquely from above with the flow from left to right: $233$ cells carrying $2301$ vacuoles, the membranes drawn opaque, which hides the vacuoles, and the wall drawn as its frozen particles.
\textbf{(B)} The circled cell of (A), enlarged, seen face-on and drawn translucent: a membrane of $500$ vertices enclosing ten vacuoles of $D = 0.6~\mu$m (purple).
\textbf{(C)} Pocked RBCs by differential interference contrast microscopy; arrows mark pocked cells. Cropped from Fig.~1D of Dumas et al.~\cite{dumas2023intra}, reproduced under CC BY 4.0.
\textbf{(D)} The fusion rule, from the cascade run at $\delta_f = 0.7~\mu$m and $t_f = 3.6$~ms: two vacuoles in contact at $3.4$~ms (purple) merge into one product (green) shown at $6.8$~ms, after it has relaxed; the other vacuoles of the cell are grey. The product takes the summed volume and the summed membrane area of its parents.
\textbf{(E)} The bifurcation (left) and the dilation (right), seen obliquely, with opaque membranes and the walls drawn as their frozen particles; the floor and the lid of the bifurcation are left out. (A) and (E) redraw the frames of Fig.~\ref{fig:survey}, C to E. The panels are not to a common scale.}
\label{fig:model}
\end{figure}

A pocked cell carries one or several vacuoles large enough to be seen by DIC microscopy (Fig.~\ref{fig:model}C). The model builds such a cell from two kinds of object. The RBC is a triangulated membrane of $500$ vertices enclosing DPD cytosol, with the shear modulus and bending rigidity of a healthy human RBC, $4.73~\mu$N/m and $2.4\times10^{-19}$~J, and with its area and volume held at $132.87~\mu$m$^2$ and $92.46~\mu$m$^3$. Each vacuole is a smaller closed vesicle of the same kind, softer than the host membrane ($2.6~\mu$N/m, $1.1\times10^{-19}$~J), whose area and volume are held near those of a sphere of prescribed diameter $D$ (Fig.~\ref{fig:model}B; Materials and Methods). The vacuoles are neutrally buoyant and are carried by the cytosol. They feel the host membrane and one another only through short-range repulsion, so nothing draws them together: any contact between them is made by the motion of the cell and its contents. Two vacuoles of the same cell merge into one once their surfaces have stayed within $\delta_f$ for a time $t_f$. The product takes both the summed volume and the summed membrane area of its parents, so it is not a sphere (Fig.~\ref{fig:model}D). All vacuole sizes below are diameters. Every duration is converted with the time scale of its own geometry, because the DPD time unit follows from the solvent density, which differs from one set-up to another (Materials and Methods; table~\smref{tab:runs}).

Throughout, two vacuoles of the same cell are said to be in contact when their centroids are closer than $d_c = 1.2~\mu$m. For vacuoles of $D = 0.6~\mu$m that leaves a gap of about $0.6~\mu$m between their surfaces, over a hundred times the thickness of a bilayer. What the model resolves is therefore proximity maintained by the flow, not the molecular apposition that fusion itself requires.

\subsection*{The type of flow sets whether vacuole contacts turn over}

\begin{figure}[!tbp]
\centering
\includegraphics[width=\textwidth,height=0.58\textheight,keepaspectratio]{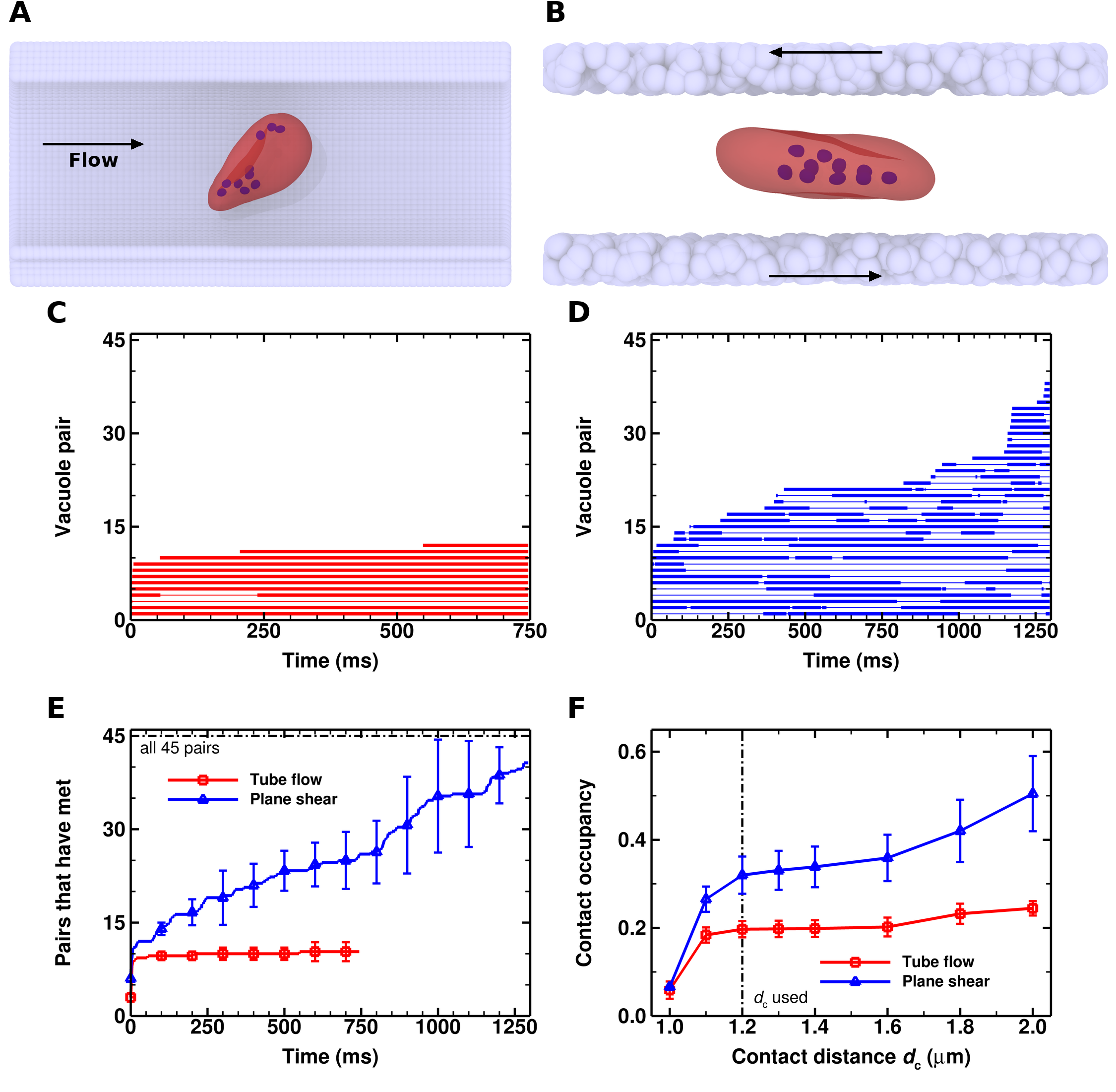}
\caption{\textbf{Vacuole contacts in tube flow and in plane shear at a matched shear rate.}
One cell with ten vacuoles of $D = 0.6~\mu$m, that is $45$ intracellular pairs, three seeds per flow; the seeds of a flow share their initial placement and differ in thermal noise. Contact is scored at $d_c = 1.2~\mu$m between vacuole centroids except in (F).
\textbf{(A)} The cell in the tube at the end of the run, seen from the side; the arrow gives the direction of flow.
\textbf{(B)} The cell in the plane gap between two walls moving in opposite directions (arrows). In (A) and (B) the membrane is translucent, so the vacuoles (purple) show through it.
\textbf{(C and D)} Contact raster for seed~1 of each flow, tube (C), $746$~ms, and plane shear (D), $1296$~ms: one row per pair, in the order of first contact, the pairs that never met left empty at the top. A bar marks a contact, and a thin line the time a pair spends apart after its first contact; contacts shorter than $3$~ms are drawn $3$~ms long, and separations shorter than $3$~ms are not drawn.
\textbf{(E)} Number of distinct pairs that have been in contact, against time (mean $\pm$ SD over three seeds); the dash-dotted line marks all $45$ pairs.
\textbf{(F)} Contact occupancy, the fraction of the time that two vacuoles of the same cell are in contact, taken over all pairs, against the threshold $d_c$; the dash-dotted line marks the value used elsewhere, $d_c = 1.2~\mu$m.}
\label{fig:contacts}
\end{figure}

We began with the simplest comparison: one cell carrying ten vacuoles of $D = 0.6~\mu$m, placed in two flows driven to the same shear rate. One is a tube of lumen $10.8~\mu$m; the other is a plane gap of $6~\mu$m between two walls moving in opposite directions at $\pm463~\mu$m/s (Fig.~\ref{fig:contacts}, A and B). Matching them takes some care, because the tube profile is blunted, $u/u_c = 0.70$ against $0.50$ for a parabola, and its nominal wall shear rate depends on the convention: $4u/R_t = 194\pm1~\mathrm{s}^{-1}$, $2u_c/R_t = 138\pm1~\mathrm{s}^{-1}$ and a directly measured near-wall gradient of $111\pm1~\mathrm{s}^{-1}$. The plane gap was matched to the second of these, $140.4\pm0.9~\mathrm{s}^{-1}$, to within $2\%$. Each flow was run with three seeds. The seeds of a flow share their initial placement of the ten vacuoles and differ only in thermal noise, so they are replicates of the noise and not of the configuration. That placement already puts three pairs in contact in the tube and six in plane shear at $t = 0$.

The two flows differ first in how many pairs ever meet. Ten vacuoles make $45$ pairs. In the tube, $12$, $9$ and $10$ of them made contact at any time, and $9$ had done so within the first $10$~ms. In plane shear, $38$, $40$ and $44$ pairs did, only $11$ of them within $10$~ms, so the rest were brought together later in the run (Fig.~\ref{fig:contacts}E). The flows differ more in how contacts end. In the tube, $8$, $9$ and $8$ pairs stayed in contact for more than $90\%$ of the record, and over $746$~ms only $2$, $1$ and $1$ contact episodes both began and ended within it. In plane shear, at most two pairs were that persistent, and $88$, $148$ and $118$ episodes ran their course, with median durations of $87.8$, $87.2$ and $106.0$~ms (Fig.~\ref{fig:contacts}, C and D). The plane-shear records are the longer, $1296$ to $1302$~ms against $745$ to $747$~ms, because the two set-ups differ in time scale; scored again over the tube's window, every comparison survives, the persistence one narrowing to $8$--$9$ pairs against $2$--$3$ (section~\smref{sec:contact}). The share of its time that a pair spends in contact separates the two flows far less: $0.197\pm0.019$ against $0.320\pm0.042$ (mean $\pm$ SD over three seeds, $d_c = 1.2~\mu$m). It rises with $d_c$, from $0.037$--$0.075$ against $0.065$--$0.068$ at $1.0~\mu$m to $0.23$--$0.26$ against $0.41$--$0.57$ at $2.0~\mu$m (Fig.~\ref{fig:contacts}F).

We do not report a rate at which new contacts form, because it cannot be defined cleanly. Vacuoles do not collide; they repel one another through a soft barrier. Entering contact is then only the crossing of the surface that $d_c$ defines, and a pair resting against the barrier jitters back and forth across any threshold set within its range: counted at $d_c = 1.0~\mu$m the crossings are more frequent in the tube, and at $1.2~\mu$m eightfold more frequent in plane shear (section~\smref{sec:contact}). The comparison above rests instead on how contacts end, scored at one threshold in both flows.

A cell in a tube therefore carries a few pairs of vacuoles that lock together and are never separated, whereas the same cell in plane shear keeps exchanging partners. Fusion requires apposition to be held, not merely made, so the two flows should differ in outcome even though their time-averaged contact differs by less than a factor of two.

\subsection*{Contacts, once made, persist in three vascular geometries}

\begin{figure}[!tbp]
\centering
\includegraphics[width=\textwidth,height=0.68\textheight,keepaspectratio]{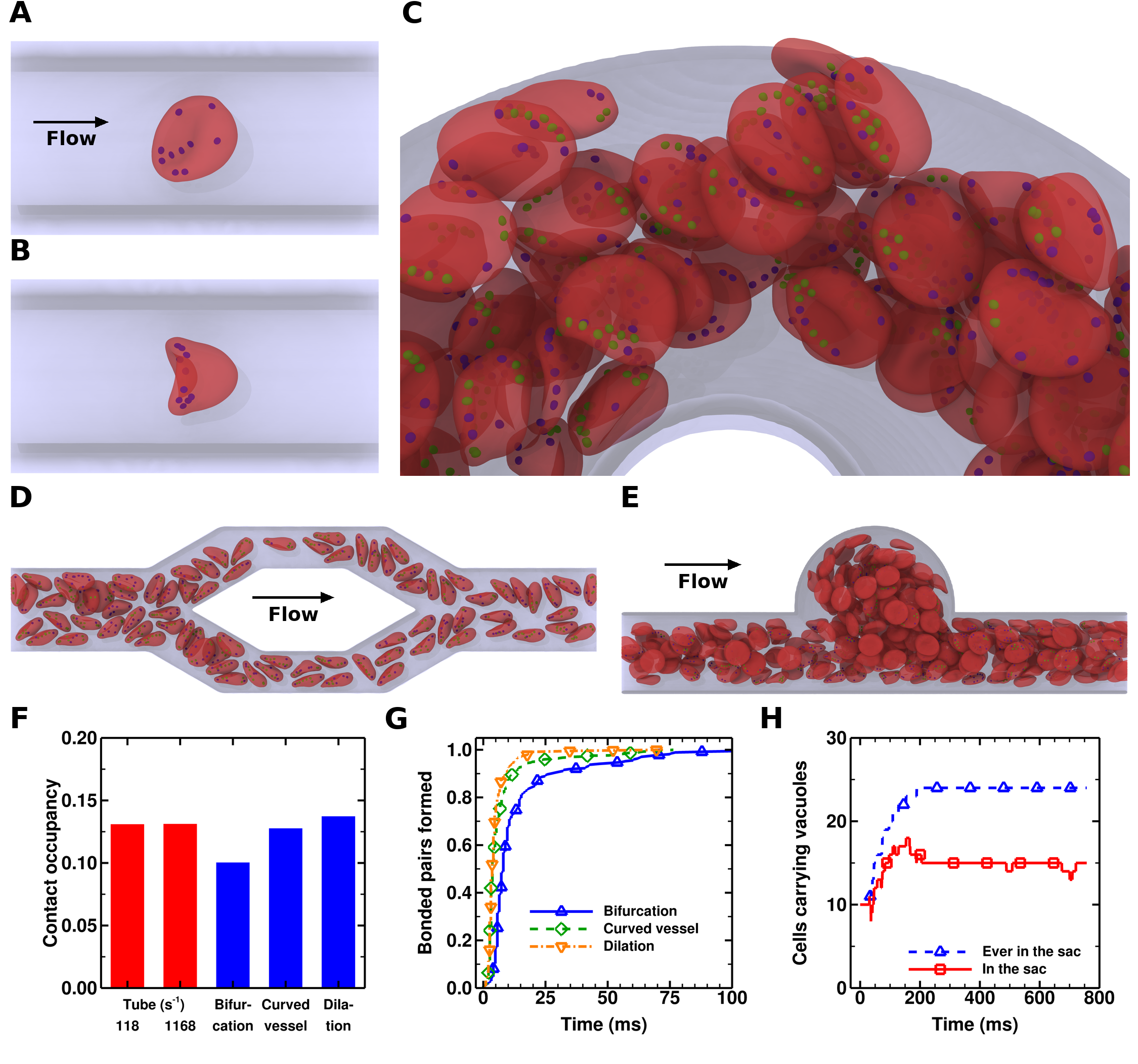}
\caption{\textbf{Vacuole contact across a microvascular survey.}
\textbf{(A and B)} One cell with ten vacuoles in a tube at $4u/R_t = 118~\mathrm{s}^{-1}$ (A) and $1168~\mathrm{s}^{-1}$ (B), seen from the side at step $150\,000$ ($562$ and $561$~ms), at a common magnification.
\textbf{(C)} The curved vessel at $150\,000$ steps ($785$~ms), close up on one bend and seen obliquely, the half of its wall nearer the viewer cut away.
\textbf{(D)} The bifurcation at $150\,000$ steps ($1206$~ms), its lumen split by a hexagonal obstacle.
\textbf{(E)} The aneurysm-like dilation (bulge-to-neck ratio $2.0$), viewed along the $z$ axis with the sac above the parent lumen, at step $150\,000$ ($758$~ms). All cells are drawn in the same red, including the $100$ that pre-fill the sac, so the cells that carry vacuoles are told apart by their vacuoles; at this instant $15$ of them are inside the sac. The panels are not to a common scale: the lumen is $10.8$, $20$ and $29.8~\mu$m across in the tube, the bifurcation and the curved vessel, and the sac $50.0~\mu$m. In (A) to (E) the membranes are translucent and the walls half transparent; in (C) to (E) the vacuoles of the bonded pairs of (G) are green and the others purple; arrows in (A), (D) and (E) give the direction of flow. The frames of (C) to (E) are redrawn in Fig.~\ref{fig:model}, A and E.
\textbf{(F)} Contact occupancy in each simulation at $d_c = 1.2~\mu$m: the two tubes in red, labelled by $4u/R_t$, and the three vessels in blue.
\textbf{(G)} Cumulative fraction of the bonded pairs (in contact in more than $90\%$ of the frames in which they are observed) that have made their first contact, against time since the pair was first observed: a start-up transient, since the record begins just after the vacuole--vacuole barrier replaces the softer repulsion used to prepare the simulations.
\textbf{(H)} Cells carrying vacuoles that are inside the sac of the dilation at that instant (red) and that have ever been inside it (blue), of the $90$ in the simulation.}
\label{fig:survey}
\end{figure}

We next asked whether that picture holds where the flow has structure along its length. One cell carrying ten vacuoles was driven through a tube at two drives, giving $4u/R_t$ of $118$ and $1168~\mathrm{s}^{-1}$ ($2u_c/R_t$ of $83$ and $844~\mathrm{s}^{-1}$; fig.~\smref{fig:tubeshear}). Populations of cells were then driven through three model vessels: a bifurcation ($80$ cells, $762$ vacuoles), a curved vessel ($233$ cells, $2301$ vacuoles) and an aneurysm-like dilation ($190$ cells, $90$ of them carrying $856$ vacuoles and the other $100$, free of vacuoles, pre-filling the sac) (Figs.~\ref{fig:model}, A and E, and~\ref{fig:survey}, A to E). A caution about these flows comes first. For plasma at these sizes and speeds the Reynolds number would be $1.6\times10^{-3}$, $1.3\times10^{-2}$ and $1.9\times10^{-2}$. The DPD solvent, which our unit mapping makes two to four thousand times denser than plasma, runs at about $6$, $29$ and $40$ (section~\smref{sec:reynolds}), and in the dilation this inertia shapes the flow in the sac, as we show below.

How much of the time vacuoles spend in contact barely depends on the environment. We measure it by the occupancy: the fraction of the time that two vacuoles of the same cell are in contact, taken over all pairs. It is $0.131$ and $0.131$ in the two tubes, although their drives differ tenfold, $0.100$ in the bifurcation, $0.128$ in the curved vessel and $0.137$ in the dilation (Fig.~\ref{fig:survey}F). It is carried by a minority of pairs that touch and then stay together: $272$ of $3400$ pairs in the bifurcation, $818$ of $10\,470$ in the curved vessel and $236$ of $3825$ in the dilation are in contact for more than $90\%$ of the frames in which they are observed, that is $8.0$, $7.8$ and $6.2\%$ of pairs.

These occupancies need two qualifications: the three records are not equally long, and none has settled. Restricted to the $758$~ms that the three share, they become $0.095$, $0.126$ and $0.137$, so the ordering survives, but each is still rising between the first and the last quarter of that window (section~\smref{sec:contact}).

The bonded pairs form almost at once, and we read this as a start-up effect rather than a result of the flow. The vessel simulations are prepared with a softer vacuole--vacuole repulsion, which is replaced by the production barrier just before the record begins, and $68$, $88$ and $91\%$ of the bonded pairs have formed within $10$~ms (Fig.~\ref{fig:survey}G; section~\smref{sec:contact}). What the record does establish is that contact, once made, is not undone by any of these geometries over the $0.76$--$1.2$~s they cover.

Where a cell is in the vessel matters little in two of the three. The occupancy varies by no more than $4\%$ between the five regions of the bifurcation and $5\%$ along the curved vessel. In the dilation it is $0.136$--$0.148$ in the three sections of the parent tube but $0.096$ and $0.092$ in the lower and the upper half of the sac: contact is lower inside the sac. No cell leaves the lumen of any of the three vessels (section~\smref{sec:checks}), but each vessel was run once, so the spread between geometries rests on a single realization of each.

The dilation takes up cells early and then keeps them. The number of vacuole-carrying cells inside the sac rises from $10$ at the start to $18$ at $154$~ms and then holds at $15.0\pm0.6$. No cell enters for the first time after $189$~ms, so only $24$ of the $90$ are ever inside it, and $15$ of them are still there when the record ends at $758$~ms (Fig.~\ref{fig:survey}H). We quote no residence time, because the sac is pre-filled with a second, vacuole-free population. Nor may the retention carry over to plasma. A single vortex fills the simulated sac. Cell-free Navier--Stokes flow through the same geometry has a similar vortex at the model's Reynolds number, but at plasma's it has no reverse flow anywhere in the sac (section~\smref{sec:reynolds}), so the retention may owe more to the model's inertia than to the geometry.

Retention does not raise contact, although at first sight it seems to. A cell's occupancy is $0.111$ while it is inside the sac against $0.103$ in the parent tube, and it is higher in the sac in $12$ of the $14$ cells that spend time in both, a cell being judged inside the sac by its height above the parent tube. But occupancy climbs throughout this record, and a cell's tube frames largely precede its sac frames, so that comparison measures elapsed time rather than position. The cells that never enter the sac are the control. Between the first $200$~ms and the remainder of the record they climb further than the entering cells, by $0.086$ against $0.021$, and end higher, at $0.217$ against $0.131$. In the simulated flow the sac holds cells, but it does not bring the vacuoles inside them together.

\subsection*{Ten vacuoles coarsen to four or five, then stop}

We then let contacts end in fusion. The same cell of ten vacuoles was run in tube flow at the nine combinations of $\delta_f \in \{0.5, 0.7, 0.9\}~\mu$m and $t_f \in \{1.2, 3.6, 11.9\}$~ms, with one combination repeated at a second thermostat seed. Every run started from the same placement of the ten vacuoles, in which $2$, $5$ and $8$ of the $45$ pairs already have their surfaces within $0.5$, $0.7$ and $0.9~\mu$m. The first fusions are therefore fixed by that starting configuration, and the runs are not independent samples of a cascade.

In all six runs that reached the end of their cascade, the ten vacuoles fell to four or five through five or six fusions (Fig.~\ref{fig:fusion}, A and B, and movie S1). The fusions come early. Across all ten runs, every fusion whose product contained three or fewer of the original vacuoles had been executed by $41.4$~ms, whereas the solvent reaches $95\%$ of its terminal mean speed only at $112$--$122$~ms (section~\smref{sec:fusion}): the cascade is over before the flow it runs in is established. The time of the first fusion followed the persistence requirement rather than the rate of encounter. It came at $2.2$, $1.6$ and $1.6$~ms for $t_f = 1.2$~ms, at $5.1$~ms for $3.6$~ms and at $17.7$~ms for $11.9$~ms, as expected when the pairs are already in contact at $t = 0$.

\begin{figure}[p]
\centering
\includegraphics[width=\textwidth,height=0.96\textheight,keepaspectratio]{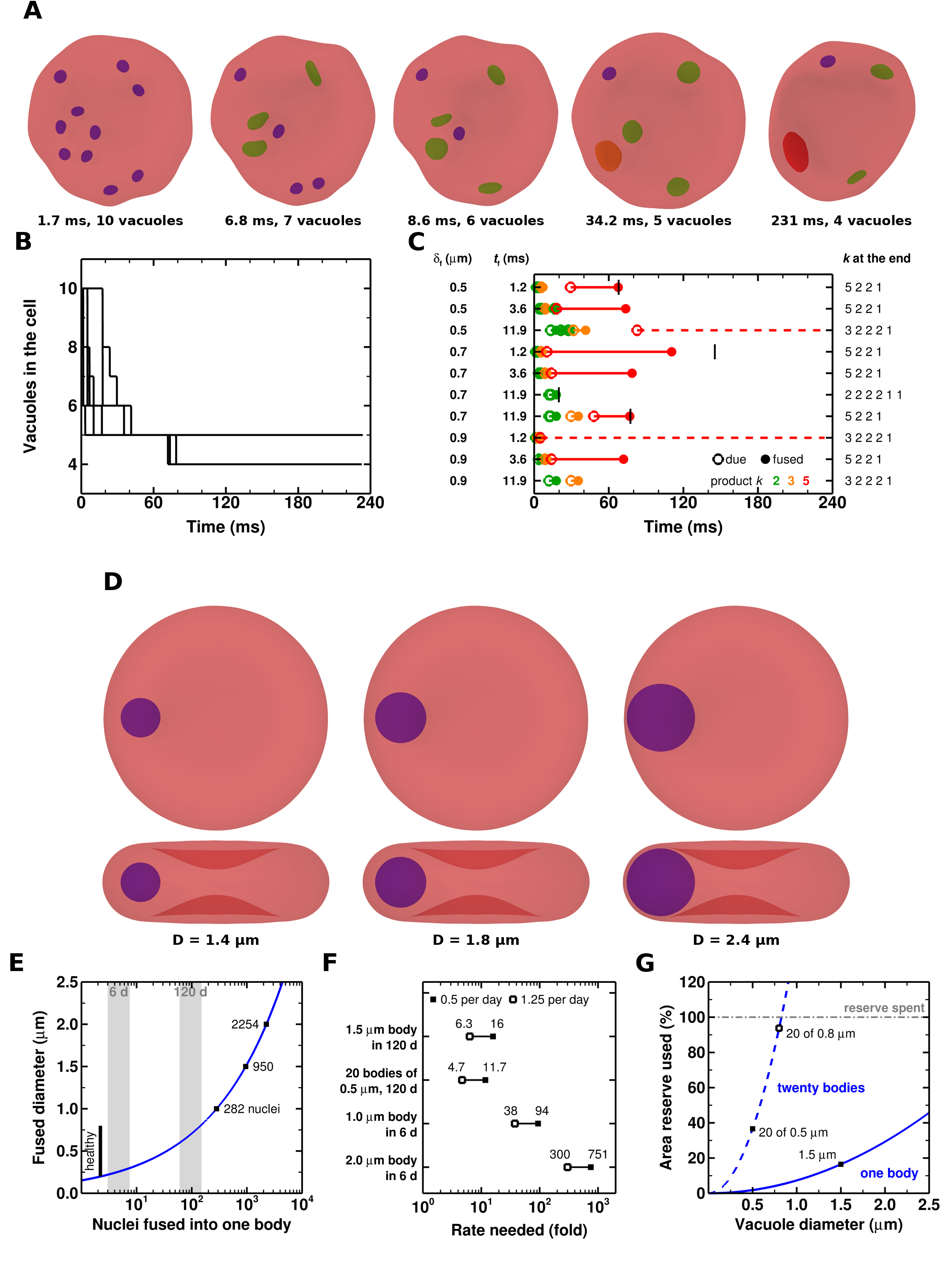}
\end{figure}
\begin{figure}[!tp]
\caption{\textbf{Fusion coarsens vacuoles, but at the measured rate a cell cannot nucleate the volume of a large one.}
\textbf{(A to C)} Ten vacuoles of $D_0 = 0.6~\mu$m in one cell in tube flow, at $\delta_f \in \{0.5, 0.7, 0.9\}~\mu$m and $t_f \in \{1.2, 3.6, 11.9\}$~ms, all from one initial placement; one combination was repeated at a second thermostat seed.
\textbf{(A)} One cascade, at $\delta_f = 0.7~\mu$m and $t_f = 3.6$~ms, at $1.7$, $6.8$, $8.6$, $34.2$ and $231$~ms, holding $10$, $7$, $6$, $5$ and $4$ vacuoles, all at one magnification. Colour gives the number of original vacuoles a body contains: purple $1$, green $2$, orange $3$, red $5$.
\textbf{(B)} Number of vacuoles against time, for the six runs that reached the end of their cascade; the four that stopped early on a vacuole-mesh bond are omitted.
\textbf{(C)} Every fusion of all ten runs, one row per run: open symbols where a pair had been in contact for $t_f$ and the fusion became due, filled symbols where it was executed, joined by a line; colour as in (A). A black bar marks where a truncated run stopped, and a red dashed line a five-parent fusion that became due and was never executed. The right-hand column gives the composition of the vacuoles left at the end.
\textbf{(D)} The model cell at rest carrying one vacuole of $D = 1.4$, $1.8$ and $2.4~\mu$m in its rim (purple), face-on above and edge-on below, at one camera distance: the starting states of the four-bead filter runs. A vacuole of $2.5~\mu$m no longer fits inside the membrane, and in the dimple even one of $2.4~\mu$m does not.
In (E) to (G), calculated curves are blue, observations black, and budgets and reference lines grey.
\textbf{(E)} The diameter of a single body made by fusing $N$ nuclei drawn from the measured size law, whose mean volume is $1.86\times10^{-3}~\mu$m$^3$~\cite{reinhart1988red}. Grey bands are the nuclei a cell makes in six days and in its $120$-day life at the two published rates, $0.5$~\cite{sills1988physiologic} and $1.25$ per cell per day~\cite{colin1991spontaneous}: a lifetime fuses to one body of $0.60$--$0.81~\mu$m, from the middle to the top of the healthy range (black bar, $0.2$--$0.8~\mu$m~\cite{holroyde1970acquisition}). Squares are the large bodies observed, with the nuclei each needs: $1.0~\mu$m after $144$~h in culture~\cite{sills1988physiologic}, $1.5~\mu$m in splenectomized patients~\cite{holroyde1970acquisition} and $2.0~\mu$m after $144$~h in culture~\cite{colin1991spontaneous}.
\textbf{(F)} The multiple of the measured nucleation rate that each observation needs, at $0.5$ (filled) and $1.25$ per day (open), over the time in which it was made: a lifetime for the single $1.5~\mu$m body and for the twenty bodies of $0.5~\mu$m of the most loaded cells~\cite{holroyde1970acquisition}, and $144$~h for the two culture observations.
\textbf{(G)} The membrane the vacuoles would take from the cell surface if they grew after birth, as a share of the cell's area reserve ($43~\mu$m$^2$, by which $140~\mu$m$^2$ exceeds the sphere of $90~\mu$m$^3$), for one body (solid) and twenty (dashed); at $100\%$ (dash-dotted) the cell would be a sphere. The largest body observed costs $16\%$, the twenty of the most loaded cells $37\%$, and twenty of $0.8~\mu$m, the top of the healthy range (open), $94\%$.}
\label{fig:fusion}
\end{figure}

What ends the cascade is not a shortage of contacts but the difficulty of placing the product. The simulation runs in segments, and a pair that has qualified is merged at the next break between them; the merger is put off whenever the product cannot be placed inside the host with a clearance of $0.3~\mu$m. In the six runs that executed it, the fusion that made a body of five original vacuoles was the last of the run and had waited $28.9$--$100.4$~ms after qualifying. In two more it never took place: the driver retried $26$ and $421$ times to the end of the record, and those runs ended with five vacuoles rather than four (Fig.~\ref{fig:fusion}C). Where the cascade stops is therefore set by where its products can be placed, and so by the shape the rule gives them: with near-spherical products the same ten vacuoles went on to one or two in the repeated-passage runs below. Four of the ten runs stopped early, when a bond of a vacuole mesh, never of the host membrane, exceeded its maximum extension (section~\smref{sec:fusion}). A truncated run is not an outcome, so Fig.~\ref{fig:fusion}B shows only the six that completed. The surviving set is not a full factorial and cannot be read as a test of sensitivity to $\delta_f$ or $t_f$.

The sizes the cascade produces are those that conservation of volume predicts. A body made of $k$ original vacuoles of $D_0 = 0.6~\mu$m has a volume-equivalent diameter $D_0 k^{1/3}$, that is $0.756$, $0.865$ and $1.026~\mu$m at $k = 2$, $3$ and $5$; the products measured $0.76$, $0.86$ and $1.04~\mu$m. Because the rule sums membrane area as well as volume, a product has more membrane than a sphere of its volume needs, and its reduced volume falls as $0.97/\sqrt{k}$. The products are flaccid bodies of the right volume rather than larger spheres.

\subsection*{At the measured rate a cell cannot nucleate the volume of a large vacuole}

The cascade above only rearranges what the cell already holds. Fusion joins vacuoles but makes no new material, so however often they meet, a cell's vacuoles can grow only as large as the material the cell has made. How large a vacuole a cell can reach is therefore a question of bookkeeping, not of flow, and the books can be balanced with numbers that are already published. Room is not the constraint: the model cell at rest holds a vacuole of $2.4~\mu$m in its rim, larger than any that has been reported, and only at $2.5~\mu$m does a vacuole no longer fit inside the membrane (Fig.~\ref{fig:fusion}D).

Two numbers are needed: how big a vacuole is when it forms, and how often one forms. For the first, we take the sizes of newly formed vacuoles, the nuclei from which any larger body would have to be built, to be lognormal with the reported mean and spread, $130\pm72$~nm, truncated to $40$--$330$~nm~\cite{reinhart1988red}. Because volume goes as the cube of the diameter, the larger ones count for more, and the mean volume is $1.86\times10^{-3}~\mu$m$^3$, about $1.6$ times that of a sphere of $130$~nm. For the second, there is only one direct measurement in people. Sills et al.\ followed six patients for six days after their spleens were removed for trauma, when nothing was left to clear the vacuoles that formed. Matching their curve with a coalescence model of the cell population gives $J_m \approx 0.5$ nuclei per cell per day~\cite{sills1988physiologic} (section~\smref{sec:pop}; table~\smref{tab:coalescence}).

Now the sum. A red cell lives about $120$ days, so at that rate it makes $60$ nuclei in its life, $0.111~\mu$m$^3$ in all. Even if every one of them fused into a single body, that body would be $0.597~\mu$m across (Fig.~\ref{fig:fusion}E). That is the size healthy adults with a spleen carry. Their vacuoles are ``usually single and small, $0.2$--$0.8~\mu$m'', multiple in under $0.01\%$ of cells, and no normal adult cell in that series carried more than five~\cite{holroyde1970acquisition}. The limit sits inside that range, and nothing was adjusted to put it there: a rate measured in Buffalo in $1988$ and a size range measured in Philadelphia in $1970$ agree with each other once volume is conserved.

For a cell nucleating at that rate, the same budget rules out everything larger. A body of $1.5~\mu$m, the largest reported in splenectomized patients, has a volume of $1.77~\mu$m$^3$. It needs about $950$ nuclei, or $7.9$ per cell per day, which is $16$ times the measured rate (Fig.~\ref{fig:fusion}, E and F). Twenty bodies of $0.5~\mu$m, the load of the most affected cells in the same series~\cite{holroyde1970acquisition}, need about $700$ nuclei, or $11.7$ times the rate. Cells do differ in rate, but at this mean a single broad distribution does not close the gap. Suppose the rate is gamma-distributed among cells with a coefficient of variation of $1.5$. Then a fraction $1.4\times10^{-4}$ of cells nucleates fast enough to meet the $1.5~\mu$m budget, and the largest body the coalescence model builds in $2000$ cells is $0.76~\mu$m. Widening the distribution to a coefficient of variation of $4$ raises that only to $1.08~\mu$m, and it drives the prevalence without a spleen down, from $29.4\%$ to $12.0\%$, against an observed $20$--$70\%$~\cite{holroyde1970acquisition,sills1988formation,buchanan1987pocked,buchanan1987formation} (table~\smref{tab:coalescence}). Raising the mean sixteen-fold instead breaks the six-day curve on which the rate rests. A separate minority of fast cells escapes this trade-off, and the population model below requires one; its budget is drawn there.

The contradiction is sharpest inside a single paper. Sills et al.\ incubated normal adult erythrocytes for $144$~h and saw intracellular vesicles of $0.05$--$1.0~\mu$m. In the same work, $^{31}$P NMR confirmed that the cells held normal 2,3-DPG, inorganic phosphate and ATP throughout, so these were not dying cells~\cite{sills1988physiologic}. To build a $1.0~\mu$m body by fusion takes $282$ nuclei. In $144$~h that is $47$ nucleations per day in that one cell, $94$ times the rate that the same paper's own pit counts imply for the population.

The rate is the one input that could be argued with, and the argument has been had in print. Colin and Schrier ran the same incubation protocol ``without significant modification'' and measured an adult acquisition at $144$~h of $10.5\%$, against $3.7$ and $4.6\%$ in the two Buffalo series: two to three times faster~\cite{colin1991spontaneous,sills1988formation,sills1988physiologic}. Their rate corresponds to $J_m = 1.25$ per cell per day. A lifetime is then $150$ nuclei and one fused body of $0.81~\mu$m; the $1.5~\mu$m body still needs $6.3$ times the rate, twenty bodies of $0.5~\mu$m need $4.7$ times, and the $1.0~\mu$m six-day vesicle $38$ times (Fig.~\ref{fig:fusion}F). Those multiples are for a cell at the mean rate, and at this rate the spread between cells does close the gap. With a mean of $1.0$ per day and a coefficient of variation of $2$, the coalescence model keeps both prevalences in range, at $2.4$ and $35.7\%$. It builds a body of $1.6~\mu$m among $2000$ cells and puts $28$ visible bodies in its most loaded cell, with a six-day acquisition of $8.2\%$, which is $2.2$ times the in vivo measurement and below Colin and Schrier's own (table~\smref{tab:coalescence}). At the faster rate, therefore, the budget excludes coalescence in the ordinary cell but not in a population whose cells differ this much. Nor does the \emph{agreement} at the healthy end survive the higher rate, where $0.81~\mu$m sits at the very top of the observed $0.2$--$0.8~\mu$m. Part of the gap between the laboratories may be where each microscope stopped: Colin and Schrier give their freshly drawn pits as $0.1$--$0.5~\mu$m~\cite{colin1991spontaneous}, Holroyde and Gardner $0.2$--$0.8~\mu$m~\cite{holroyde1970acquisition}. We therefore report both rates. The same paper also reports normal adult cells carrying bodies ``up to $2~\mu$m diameter'' after $144$~h~\cite{colin1991spontaneous}. That would take $2254$ nuclei, or $376$ per cell per day, $300$ times that laboratory's own rate. The laboratory that reports the fastest formation reports the largest bodies, and falls the furthest short.

Two further things could void this budget. The first is that vacuoles grow after birth, by osmotic swelling or by continued delivery of membrane. Both routes need a supply that a mature erythrocyte does not obviously have. Swelling requires solute added in proportion to volume, and a parcel of already-degraded cargo is diluted by its own growth. Delivery requires area, which grows as the two-thirds power of volume: taking one body from $130$~nm to $1.5~\mu$m multiplies its surface $133$-fold. A cell that synthesises no lipid and has emptied itself of organelles can draw that area only from its own plasma membrane. Twenty bodies of $0.5~\mu$m are $15.7~\mu$m$^2$ of membrane, about $11\%$ of the cell's $\sim\!140~\mu$m$^2$, and twenty of $0.8~\mu$m would be $29\%$. But a red cell of $140~\mu$m$^2$ and $90~\mu$m$^3$ holds only $43~\mu$m$^2$, some $31\%$, in excess of the sphere of equal volume, and that excess is the reserve that lets it cross a splenic slit. Twenty bodies of $0.8~\mu$m would spend $94\%$ of it (Fig.~\ref{fig:fusion}G) and leave the cell a surface-to-volume ratio of $1.11~\mu$m$^{-1}$, below the $1.22$--$1.33~\mu$m$^{-1}$ that simulated cells need to pass one~\cite{li2021How}. The second possibility is that vacuoles are born at the size at which they are seen, and what is known of their origin allows it. Colin and Schrier found that spontaneous vacuoles were not labelled by a fluid-phase albumin marker, at $0$ and at $144$~h, whereas vacuoles induced by primaquine in the same cells were. They concluded that the bodies are autophagic: degraded haemoglobin, membrane material and remnant organelles rather than trapped plasma~\cite{colin1991spontaneous}. The vacuole membrane is not labelled when the cell surface is coated with cationized ferritin~\cite{sills1988formation,sills1988physiologic}, and the first electron micrographs of these bodies show a mitochondrion and a vacuole discharging their contents into a larger preformed one~\cite{holroyde1970acquisition}. An invagination has a natural size, that of the neck that closes it; a parcel of the cell's own debris has the size of its cargo. The $130$~nm figure is the mean of the numerous \emph{small} bodies that dominate any section count, not a measurement of how large the large ones were when they formed. We take the second reading, and ask what follows from it.

\subsection*{A decaying source with a fast minority fits the record}

If a vacuole's size is set at birth, no transport model is needed to explain it, and everything that has been measured can be computed from a model of the red cell population, structured by age. Bodies are nucleated at a rate $J(a)$ per cell of age $a$, with diameters drawn from a fixed law, and each is removed by the spleen at a rate $R(D,a)$ that switches on above a gate diameter $D_g$. That the spleen removes the body and returns the cell is measured, not assumed. Vacuolated red cells transfused into a recipient with an intact spleen lose their vacuoles within $48$--$72$~h in humans~\cite{holroyde1970acquisition} and $3$--$6$~h in dogs~\cite{buchanan1987formation}, while the label on the cells themselves persists; in a splenectomized recipient they lose none~\cite{buchanan1987formation}. The model needs no simulation. Because births are Poisson and a size-dependent clearance only thins them, the number of bodies of each size in a cell of age $a$ stays Poisson, and every prevalence, ratio and maximum below is an integral (Materials and Methods).

\begin{figure}[!tbp]
\centering
\includegraphics[width=\textwidth,height=0.55\textheight,keepaspectratio]{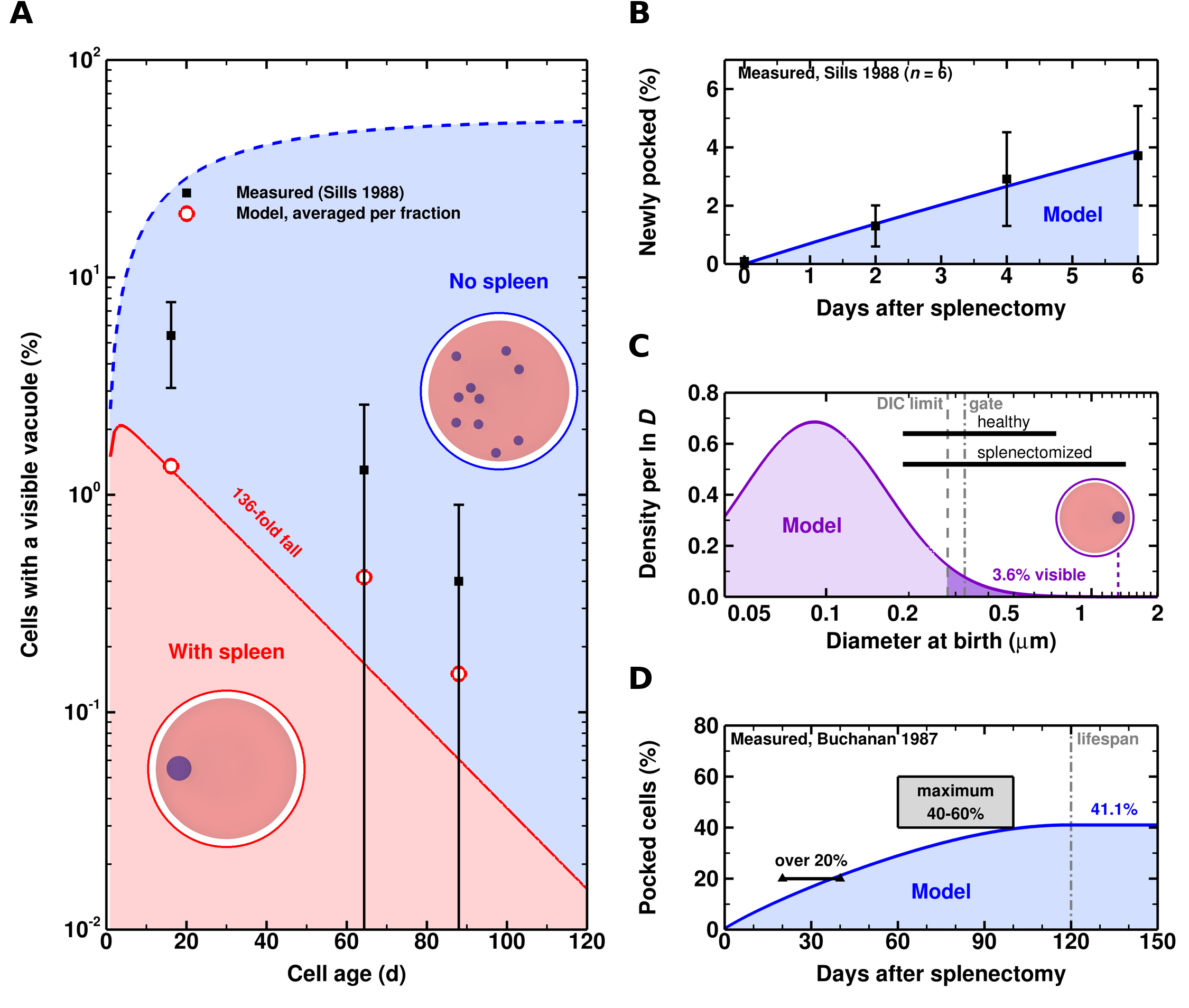}
\caption{\textbf{The population model against the published record.}
Model series are coloured, red with a spleen, blue without one or after its removal and purple for the birth-size law; measured series are black and reference lines grey.
The discs are simulated cells shown for illustration, not output of the population model: in (A), one with ten vacuoles of $D = 0.6~\mu$m (Fig.~\ref{fig:contacts}) and one with a single vacuole of $1.4~\mu$m (Fig.~\ref{fig:arrest}); in (C), the latter above its vacuole's diameter.
\textbf{(A)} The fraction of cells carrying a DIC-visible vacuole against cell age, with the spleen present (solid) and absent from birth (dashed). With a spleen the burden peaks at four days and falls $136$-fold over the lifespan; without one, vacuoles accumulate. Filled squares are the density fractions of Sills et al.~\cite{sills1988formation}, plotted at the mean age that each fraction's reticulocyte content implies, with SD bars (the two that reach zero run to the foot of the axis). They should be compared with the fraction-averaged model values (open circles), not with the per-age curve, because a gradient fraction averages the curve rather than sampling it. Those sit $2.7$ to $4.0$ times below the measurements while the unseparated prevalence is right, $0.48\%$ against $0.5\%$: the shape is reproduced and the level is not (section~\smref{sec:limits}).
\textbf{(B)} Acquisition of vacuoles in the six days after splenectomy, among cells carrying nothing at $t = 0$; only the $144$~h point was a scored target. The counts of Sills et al.~\cite{sills1988physiologic}, $0.4$ / $1.7$ / $3.3$ / $4.1\%$, are baseline-corrected to $0$ / $1.31$ / $2.91$ / $3.71\%$, with SD bars over six patients.
\textbf{(C)} The birth-size law, median $90$~nm and log-standard-deviation $0.65$. Only $3.6\%$ of births (shaded deeper) are above the DIC limit. Horizontal bars are the observed size ranges in healthy and splenectomized subjects~\cite{holroyde1970acquisition}; their heights carry no information. The dashed line is the DIC limit, $0.30~\mu$m, and the dash-dotted line the splenic gate, $0.35~\mu$m. The upper end of the splenectomized range is a scored target---the model's largest body is $1.57~\mu$m against $1.5$---but the shape of the law below it was not fitted to these ranges.
\textbf{(D)} The pocked fraction in the months after splenectomy, from the same parameters with nothing refitted; it reaches its plateau, the spleen-absent prevalence of $41.1\%$, at the red cell lifespan of $120$~days (dash-dotted line). In black, six children splenectomized for haematological disease, as their authors summarize them~\cite{buchanan1987formation}: over $20\%$ within $20$--$40$~days (bar) and a maximum of $40$--$60\%$ within $60$--$100$~days (box). Only the plateau level was a scored target.
Every measured value in this figure, with its source, its reported dispersion and the number of subjects behind it, is tabulated in tables~\smref{tab:measA} to~\smref{tab:measD}.}
\label{fig:popmodel}
\end{figure}

\begin{figure}[!tbp]
\centering
\includegraphics[width=\textwidth,height=0.52\textheight,keepaspectratio]{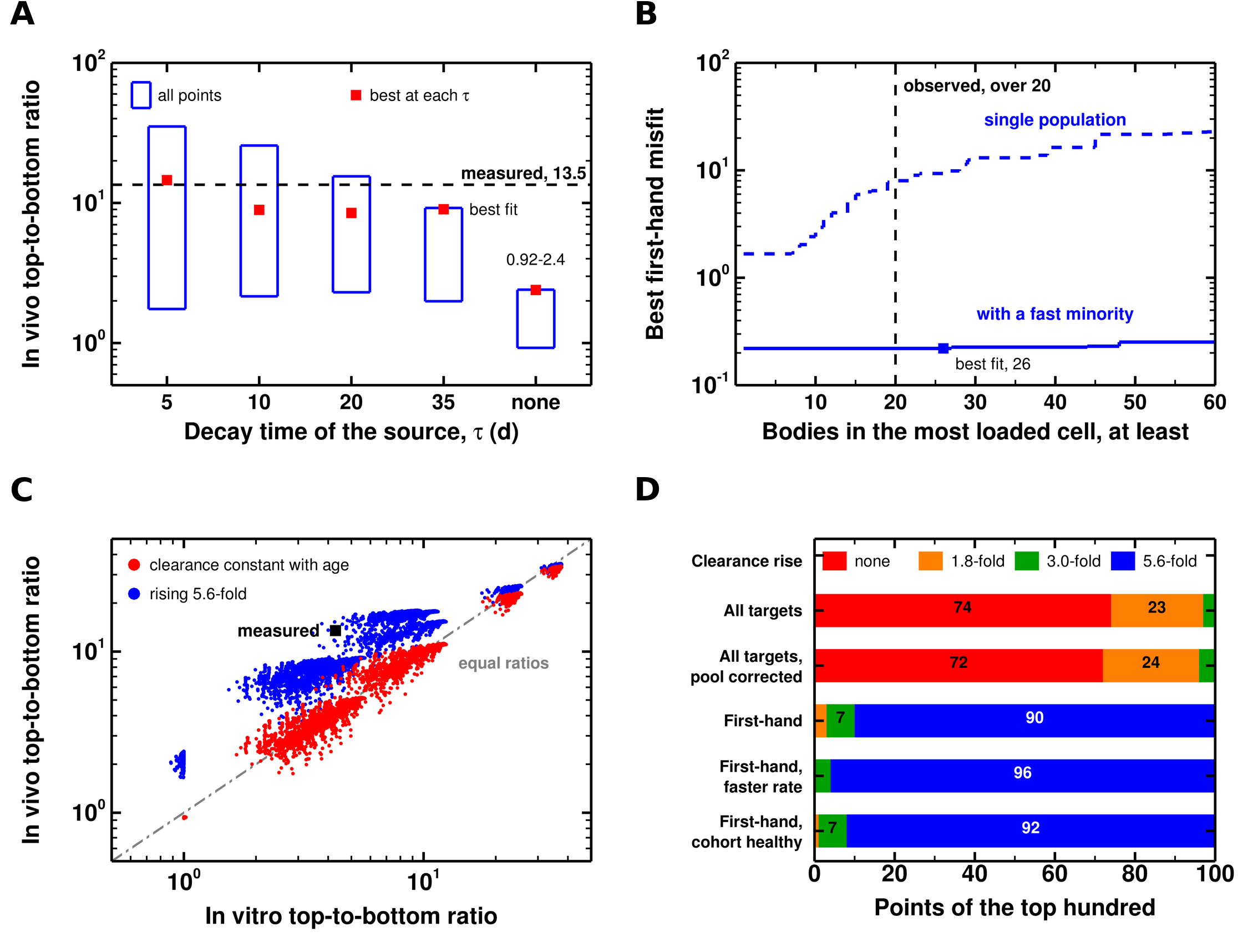}
\caption{\textbf{What the record forces on the population model, and where the record divides.}
Every point is one of the $32\,400$ parameter combinations of the scan (Materials and Methods). Measured values are black and reference lines grey. Where the rise of splenic clearance over the life of the cell is shown (C and D), colour gives it: red none, orange $1.8$-fold, green $3.0$-fold and blue $5.6$-fold.
\textbf{(A)} The source must switch off. The in vivo top-to-bottom density ratio against the decay time $\tau$ of the source: the range over every point with that $\tau$ (open blue bars) and the point with the best first-hand score at each (red squares), against the $13.5$ measured~\cite{sills1988formation}. A source that never decays gives $0.92$--$0.94$ at age-independent clearance and no more than $2.4$ anywhere in the scan.
\textbf{(B)} A minority must nucleate faster. The best first-hand score attainable when the most loaded cell must carry at least the number of bodies on the axis, with every cell nucleating at one rate (dashed) and with a fast minority (solid); lower is better. The best single-population point fits every other target as well as the fit we report, $0.224$ against $0.220$, but carries six bodies, and pays $1.45$ for missing the ``one to over $20$'' of the most affected cells~\cite{holroyde1970acquisition}; made to carry twenty, a single population scores $8.0$. The square is the fit we report, with $26$.
\textbf{(C)} The measured pair favours a clearance that rises with age. In vivo against in vitro top-to-bottom ratio for every point with clearance constant in age (red) and rising $5.6$-fold over the $120$-day life (blue). At constant clearance both ratios inherit their profile from the source and lie along equality (dash-dotted), the in vivo never more than $1.82$ times the in vitro; the measured pair, $13.5$ and $4.3$~\cite{sills1988formation}, is $3.14$-fold apart and falls among the points with rising clearance.
\textbf{(D)} How the hundred best points divide by the rise of clearance, under five rankings: every published target; the same with the per-cell estimate of Reinhart and Chien~\cite{reinhart1988red} corrected for section thickness; the first-hand targets alone; and the first-hand targets with the faster acquisition of a second laboratory~\cite{colin1991spontaneous}, or with the healthy prevalence narrowed to the largest cohort~\cite{buchanan1987formation}.}
\label{fig:forced}
\end{figure}

Two features of the model are forced by the data rather than chosen. The first is that the source must switch off as the cell ages, $J(a) = J_0(e^{-a/\tau} + \phi)$. The reason is that clearance is fast compared with ageing. If the source did not depend on age, a body above the gate would be swept long before the cell had measurably aged, the prevalence would be the same at every age, and the healthy density profile would come out flat: a top-to-bottom ratio of $0.92$--$0.94$, against the $13.5$ measured~\cite{sills1988formation}. A source concentrated in young cells gives both age profiles at once. With the spleen present, bodies are made in young cells and swept within a day or two, so prevalence follows $J(a)$ and falls with age. With the spleen absent they are never removed, so they accumulate, and prevalence rises with age (Fig.~\ref{fig:popmodel}A). No point with a non-decaying source reaches the top hundred under any of the scores below, and the control at $\tau = 10^6$~d never gives a ratio above $2.4$, even with the age-dependent clearance introduced below (Fig.~\ref{fig:forced}A). The timescale itself is not pinned down: the best point has $\tau = 35$~d, but $5$, $10$ and $20$~days all appear in the top twenty, trading against $J_0$. All are longer than the $\approx\!2$~d that a strictly reticulocytic source would give, so the capacity to make vacuoles outlives the reticulocyte stage.

The second is that a minority of cells must nucleate far faster than the rest. One distribution of rates cannot deliver both the prevalence without a spleen, which wants a narrow spread, and the ``one to over $20$ vacuoles'' of the most affected cells, which wants a wide one. The best single-population points carry only six or seven bodies in their most loaded cell, although the best of them fits every other target as well as the model we report; made to carry twenty, a single population fits the rest far worse (Fig.~\ref{fig:forced}B). We therefore give a fraction $f_{ab}$ of cells a nucleation rate $m_J$ times that of the rest. Heterogeneity of this kind is itself an observation: pitted cells are significantly more frequent among pathologically shaped red cells than among discocytes, and the number of pits per cell shows great individual variation~\cite{demeter1989preferential}.

We scanned $32\,400$ parameter combinations, with each density fraction entering as a tilt in age matched to its reticulocyte content (Materials and Methods). The best point has $J_0 = 0.60$/day, $\tau = 35$~d, $\phi = 0$, a birth-size median of $90$~nm with log-standard-deviation $0.65$ (Fig.~\ref{fig:popmodel}C), $f_{ab} = 0.01$ and $m_J = 25$ (table~\smref{tab:popmodel}). Of the six-day curve after splenectomy only the $144$~h endpoint is a scored target, yet the model passes through the two intermediate points, at $1.38$ and $2.66\%$ against a measured $1.31$ and $2.91$, well inside the spread of the six patients (Fig.~\ref{fig:popmodel}B). The most loaded cell carries $26$ bodies and the largest is $1.57~\mu$m, against ``one to over $20$'' and $1.5~\mu$m. The vacuole membrane of a minority cell at $120$ days is $24\%$ of the host's area, so the load is physically containable.

The volume budget of such a cell is not that of the rest. A minority cell nucleates $508$ bodies in its life, $1.45~\mu$m$^3$ in all, which gathered into one would measure $1.40~\mu$m; a cell of the majority nucleates $20$ bodies, $0.058~\mu$m$^3$ in all, or $0.48~\mu$m. At the faster rate discussed below the minority cell nucleates $2.89~\mu$m$^3$, more than the $1.77~\mu$m$^3$ of a $1.5~\mu$m body, and across the hundred points that best fit the first-hand targets (defined below) its lifetime volume is never below $1.18~\mu$m$^3$. The material for a large vacuole therefore exists in these cells and in no others, and in them the budget cannot tell bodies born large from bodies assembled by fusion.

Nor can the record. We gave a coalescence model the same source, the same minority and the age-dependent clearance introduced next, put a free fusion efficiency in place of the size law, and scanned it over the same grid. Scored on the first-hand targets, where a lower score is a better fit, it reaches $0.349$, inside the hundred best points of the scan above ($0.220$ to $0.373$). Like the born-large model, it fails without a minority ($1.90$) or with a constant source ($5.56$) (section~\smref{sec:pop}). Its best points ask the fast cells for $3.4$ to $5.2~\mu$m$^3$ of vacuole and put the prevalence without a spleen at $75$ to $77\%$, above the observed range. What the record requires is therefore the source and the minority, by either route to size.

The last feature is one we did not put in, and it is a property of the spleen. The same donors' fractions were measured twice, once in vivo and once after $144$~h of incubation, where by construction there is no spleen~\cite{sills1988formation}. In vivo the prevalence is a steady state, $J(a)/R$; in vitro the acquisition is $J(a)$. If the clearance rate did not depend on cell age, both would inherit their profile from $J(a)$, and the two top-to-bottom ratios would have to be \emph{equal}. They are not: $13.5$ in vivo against $4.3$ in vitro, $3.14$-fold apart. At constant $R$ no point of the scan puts them more than $1.82$-fold apart, and none more than $2.49$-fold or $3.12$-fold with a clearance that rises $1.8$-fold or three-fold over the life of the cell (Fig.~\ref{fig:forced}C). Matching the measured pair therefore takes a splenic clearance that rises at least three-fold with age. That is what an organ filtering on deformability should do, given that red cells stiffen as they age, but here it is derived from two tables of one paper rather than assumed. The optical limit sets the size of that bound but not its existence: at a DIC limit of $0.40~\mu$m rather than $0.30$, a $1.8$-fold rise falls $1\%$ short of the measured pair, at $0.20~\mu$m a three-fold rise still falls $13\%$ short, and at both, constant clearance falls well short (section~\smref{sec:pop}). The direction of the rise is firmer than its size, because both ratios divide by the densest fraction, measured at $0.4\pm0.5\%$ in vivo and $2.4\pm2.0\%$ in vitro.

That requirement is also where the model parts company with the literature, and the split is clean. We scored all $32\,400$ points twice: once against every published target, and once against the first-hand targets, which leave out the four targets that describe the pool of vacuoles below the optical limit. Of the first ranking's top hundred, $74$ have age-independent clearance and none a $5.6$-fold rise; of the second ranking's top hundred, $90$ have the $5.6$-fold rise and none age-independent clearance (Fig.~\ref{fig:forced}D). The four targets come from a single paper. Reinhart and Chien report about $20$ small vacuoles per erythrocyte, a four-fold rise after splenectomy, and a frequency similar in density-separated cells both before and after~\cite{reinhart1988red}. The rise our model reproduces, at $3.2$-fold, without being fitted to it. The estimate of $20$ is a conversion from vacuoles per cell section to vacuoles per cell that divides by the section thickness, whereas a section samples the thickness plus the body diameter. With their own $t = 80$~nm and $D = 130$~nm the factor is $2.6$, and the estimate becomes $7.7$--$9.6$ per cell according to how $D$ is read; our prediction of $5.6$, and the $11.2$ that follows from the faster acquisition rate discussed below, bracket it. What is left is the flatness with age. Scored again with the corrected estimate in place of $20$, the top hundred still holds $72$ points at age-independent clearance against $74$ before, so it is the flatness alone that divides the two rankings. We report the sub-visible pool as a prediction rather than a fitted quantity.

Two of the choices behind that split are ours rather than the data's, so we scored the same points again with each of them changed in turn. Replacing the six-day target with the faster acquisition of Colin and Schrier moves exactly one parameter, $J_0$, from $0.60$ to $1.20$ per day, and leaves every other at the same grid value; $96$ of the top hundred then have the $5.6$-fold rise and none age-independent clearance, against $90$ before. Narrowing the healthy target from the $0.5$--$2.6\%$ range to the $0.52\pm0.04\%$ (mean $\pm$ SE) of the largest cohort gives $92$ and none, and a best point that scores within $1\%$ of the one we report. So whichever of the two rates is right, and wherever the healthy target is placed within the published spread, the model that fits the first-hand record is the same. The faster rate changes the predicted sub-visible pool, from $5.6$ to $11.2$ bodies per cell. The all-targets ranking, by contrast, has a physical cost: its best point requires the vacuole membrane of a minority cell to reach $84\%$ of the host's area by day $120$, against $24\%$ for the first-hand ranking.

Three further predictions follow with no more fitting. The first is the whole shape of the rise after splenectomy. A cell of age $a$ at time $t$ after surgery has been without a spleen for only $\min(a,t)$, so the count cannot saturate before $120$ days. But because the source is concentrated in young cells, most of the rise happens well before that: the modelled count crosses $20\%$ at $37$ days, reaches $96\%$ of its plateau by $100$ days and its plateau at $120$ (Fig.~\ref{fig:popmodel}D). The published time courses have that shape. Serial counts in six children splenectomized for haematological disease rose ``to over a value of $20\%$ within $20$--$40$ days'' and reached their maximum of $40$--$60\%$ ``within $60$--$100$ days''~\cite{buchanan1987formation}, where the modelled count reaches $40\%$ at $103$ days, just after that window. Another series rose continuously between $10$ and $130$ days~\cite{dehaan1988vacuoles}. Thirty-one adults plateaued sooner, at nine to twelve weeks, where our curve stands at $73$--$89\%$ of its final value, within the $\pm20\%$ by which duplicate counts of this assay differ~\cite{buchanan1987formation}. Only the plateau level was a scored target; its approach contains no parameter beyond the red cell lifespan and the decay time of the source. The second concerns neonatal blood. Being rich in reticulocytes, it must show high counts at every density and a flatter profile than adult blood, and the measured neonatal fractions are $19.3$, $8.2$ and $9.9\%$: flat and non-monotonic where the adult profile falls steeply~\cite{sills1988formation}. The third is that functional hyposplenism should sit between the two branches, at a partly open gate, as the reported $8.5\%$ does~\cite{sills1988formation}.

\subsection*{A slit holds cells by surface reserve, not by vacuole}

Two earlier steps rest on the same filter. The population model favours a splenic clearance that rises with the age of the cell (Fig.~\ref{fig:forced}C), and the budget of Fig.~\ref{fig:fusion}G rests on the claim that a cell short of surface cannot cross a splenic slit. What is known of that filter points to the cell's surface. Simulated red cells cross an interendothelial slit only above a surface-to-volume ratio of $1.22$--$1.33~\mu$m$^{-1}$~\cite{li2021How}. In microfluidic channels a moderate fall in that ratio markedly widens the narrowest channel a red cell can traverse, while drastic stiffening has no measurable effect~\cite{namvar2021surface}. And the perfused human spleen retains cells increasingly with the surface they have lost, trapping all of those whose ratio is reduced by more than $27\%$~\cite{safeukui2012quantitative}. What a vacuole adds to this is not known: it enlarges the contents of the cell without adding to its surface, and it could itself be the obstacle. We therefore pushed the model cell at a wall pierced by one aperture of about the section of an interendothelial slit. The cell carried one vacuole of $D = 0.8$ or $1.4~\mu$m in its trailing or its leading rim; its surface-to-volume ratio was set to $1.41$, $1.20$ or $1.09~\mu$m$^{-1}$ by raising its volume at fixed area; and it was driven at two strengths, $f = 8$ and $4$, with three seeds for each combination (Fig.~\ref{fig:arrest}, A to E; Materials and Methods). The model has no fission, so a cell that does not cross is held.

\begin{figure}[p]
\centering
\includegraphics[width=\textwidth,height=0.96\textheight,keepaspectratio]{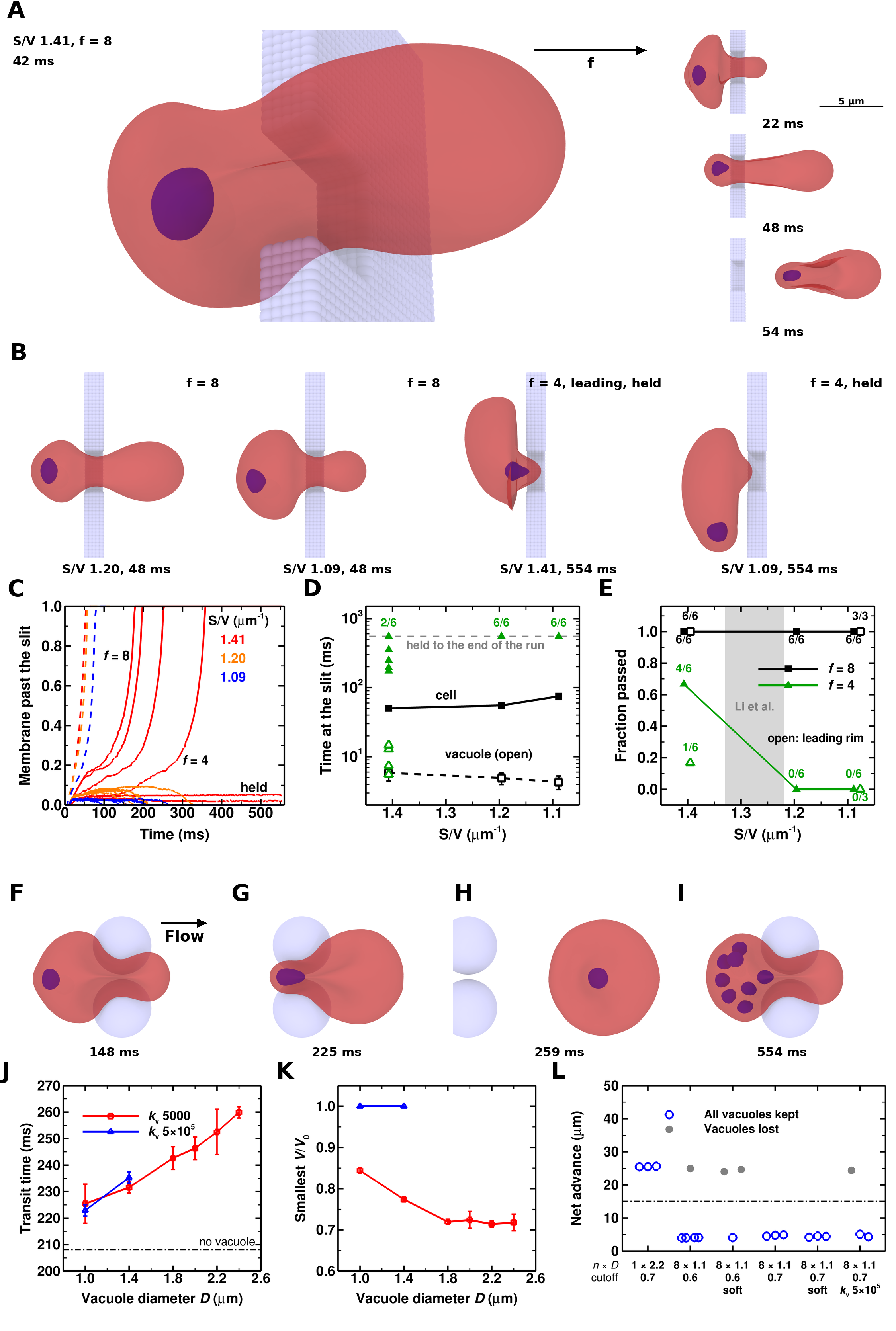}
\end{figure}
\begin{figure}[!tp]
\caption{\textbf{Arrest at a slit and at a microsphere filter.}
\textbf{(A to E)} A cell carrying one vacuole (purple) is pushed at a slit, a wall pierced by one aperture, by a force $f$ on its membrane and cytosol (arrow), seen along the long side of the aperture with the wall cut at mid-height, so that the aperture appears as a recess; times are from the start of the drive, and the bar is $5~\mu$m in the mid-plane of the aperture.
\textbf{(A)} A cell with a surface-to-volume ratio of $1.41~\mu$m$^{-1}$ and a $1.4~\mu$m vacuole in its trailing rim at $f = 8$: half through, seen obliquely (left), and its front through the aperture, the vacuole entering, and past (right).
\textbf{(B)} Cells at $1.20$ and $1.09~\mu$m$^{-1}$ with the same vacuole at $f = 8$, at $48$~ms, still in the aperture; and at $f = 4$ at the end of the run, held: at $1.41~\mu$m$^{-1}$ with the vacuole in the leading rim, at the mouth of the aperture, and at $1.09~\mu$m$^{-1}$ with it trailing.
\textbf{(C)} Fraction of the membrane past the downstream face of the wall against time, vacuole trailing, coloured by surface-to-volume ratio: every run at $f = 4$ (solid) and the mean of the six runs at $f = 8$ (dashed).
\textbf{(D)} Time at the slit against surface-to-volume ratio, vacuole trailing, of the cell (filled), from its front reaching the wall to its rear leaving it, and of the vacuole (open): mean $\pm$ SD of the six runs at $f = 8$ (black; $D = 0.8$ and $1.4~\mu$m pooled) and every run at $f = 4$ (green). Cells that had not left when the run ended are drawn at its end (dashed), with their number.
\textbf{(E)} Fraction of runs in which the cell crossed, at $f = 8$ (squares) and $f = 4$ (triangles): filled, the vacuole trailing ($D = 0.8$ and $1.4~\mu$m, three seeds each); open, leading ($D = 0.8$ and $1.4~\mu$m at $1.41~\mu$m$^{-1}$, $1.4~\mu$m at $1.09~\mu$m$^{-1}$), drawn $0.012~\mu$m$^{-1}$ to the right. The grey band is the $1.22$--$1.33~\mu$m$^{-1}$ below which the simulated cells of Li et al.\ did not cross~\cite{li2021How}.
\textbf{(F to H)} A cell at a four-bead filter with one vacuole of $D = 1.4~\mu$m and the volume constraint stiffened a hundredfold, at $148$, $225$ and $259$~ms: approaching the pore at the centre of four beads, in it, and past it.
\textbf{(I)} A cell carrying eight vacuoles of $D = 1.1~\mu$m at $554$~ms, still held at the layer. In (F) to (I) the vacuoles are purple and the arrow gives the direction of flow.
\textbf{(J)} Cell transit through the bead layer against the diameter of a vacuole in the trailing rim (mean $\pm$ SD over three seeds): red squares, the default volume constraint; blue triangles, the hundredfold stiffer constraint, shown at the two diameters where every seed met the containment criterion. The dash-dotted line is the cell without a vacuole, $208.2$~ms.
\textbf{(K)} Smallest vacuole volume over the run, relative to its value at the first frame, against diameter; same runs and symbols as (J).
\textbf{(L)} Net advance of the cell over the run, by group: one vacuole of $D = 2.2~\mu$m and eight of $D = 1.1~\mu$m under each setting of the vacuole--vacuole pair, the membrane--vacuole cutoff and the volume constraint. Open symbols are runs that kept every vacuole inside the host; filled grey symbols are runs that lost one. The dash-dotted line at $15~\mu$m is the threshold below which a cell is scored as held.}
\label{fig:arrest}
\end{figure}

At the stronger drive, $f = 8$, every cell crossed, and the less surface reserve it had, the longer it spent at the slit: $50.2\pm1.4$~ms at $1.41~\mu$m$^{-1}$, $55.4$~ms, identical in all six runs, at $1.20$, and $75.1\pm1.5$~ms at $1.09$ (Fig.~\ref{fig:arrest}, A, B, and D). At the weaker drive, $f = 4$, the ratio decided whether the cell crossed at all (Fig.~\ref{fig:arrest}, C and E). Four of the six cells at $1.41~\mu$m$^{-1}$ crossed, but only after a wait: half the membrane was through the aperture $150$--$329$~ms after the cell reached it, and the rest followed in $24$--$26$~ms. The other two were still at the wall when their runs ended, $549$~ms after they reached it. None of the twelve cells at $1.20$ and $1.09~\mu$m$^{-1}$ crossed, although, holding more cytosol, they were pushed harder, with $81$--$87$ against $75$~pN. Each put $3$--$9\%$ of its membrane through the aperture within the first $150$~ms and then withdrew it. In this model the boundary therefore lies between $1.41$ and $1.20~\mu$m$^{-1}$, where Li et al.\ place theirs~\cite{li2021How}.

A vacuole in the trailing rim changed none of this. It crossed with the rear of the cell: in $19$ of the $22$ passages it left the slit in the same frame as the rear, and in the other three one frame, $1.85$~ms, before it. It spent $4$--$7$~ms in the aperture at $f = 8$ and $6$--$15$~ms at $f = 4$ (Fig.~\ref{fig:arrest}, A and D). A vacuole in the leading rim is another matter, because it reaches the wall first. At $f = 8$ all nine cells so loaded crossed. A $0.8~\mu$m vacuole there cost nothing, but a $1.4~\mu$m one took $31$--$39$~ms to clear the aperture and lengthened the crossing of a cell at $1.41~\mu$m$^{-1}$ to $57$--$63$~ms, against $50$--$52$~ms with the same vacuole trailing. At $1.09~\mu$m$^{-1}$, where the cell itself is the obstacle, it added nothing measurable. At $f = 4$ and $1.41~\mu$m$^{-1}$, one of the six cells with a leading vacuole crossed, against four of six with a trailing one, and all three with a $1.4~\mu$m vacuole were held with it at the mouth of the aperture (Fig.~\ref{fig:arrest}B; section~\smref{sec:slit}). With three runs to a group, the difference is a tendency rather than a result (Materials and Methods).

The slit, then, selects the cell and not the vacuole. A vacuole the cell can host tips the outcome only when it arrives first, and only for a cell already at the margin. The rise of clearance with age that the population model requires would then be a property of the host, not of its load. Suppose the spleen removes vacuoles from the cells it holds, which is the premise of pitting and which this model does not attempt. A vacuole then meets removal at the rate at which its host is held, and that rate is set by the host's deformability, of which the surface reserve is only one part. Rabbit red cells of known age lost $10.5\%$ of their surface area but also $8.4\%$ of their volume over $50$ days, which left the ratio of the two almost unchanged, while their deformability fell through the loss of area and through dehydration~\cite{waugh1992rheologic}. The cost set out in Fig.~\ref{fig:fusion}G is also a real one: a cell that had built twenty $0.8~\mu$m vacuoles from its own surface would stand near $1.11~\mu$m$^{-1}$, where the model cell at $1.09$ never crossed at $f = 4$ and took half as long again at $f = 8$. One limit should be kept in mind. Because only the cell is driven, a held cell is pressed against the wall by the force on its own contents (section~\smref{sec:slit}); what the model supports is the comparison between surface reserves and vacuole positions at a common drive, not the force at which a real cell is held.

\subsection*{Divided vacuole volume holds the model cell at a microsphere filter}

Vacuoles in hyposplenic subjects reach $1.5~\mu$m~\cite{dumas2023intra}, comparable to the constrictions an RBC must pass. We therefore drove single cells through a layer of four beads of diameter $4.02~\mu$m on a square lattice of pitch $4.75~\mu$m, the in vitro analogue of splenic filtration by microsphere layers~\cite{deplaine2011sensing} (Fig.~\ref{fig:arrest}, F to I). The cell passes through the pore at the centre of the four beads. Its throat measures $2.72~\mu$m between bead particle centres, but the frozen beads exclude solvent from a shell around them, so the effective throat is $1.75~\mu$m (section~\smref{sec:geometry}). The drive is the smallest that carries a cell without a vacuole through the layer, and such a cell crossed in $208.2\pm2.1$~ms.

With one vacuole in the trailing rim, no size that a cell can host held it. Transit rose monotonically with $D$, from $225.4\pm7.4$~ms at $D = 1.0~\mu$m to $259.9\pm2.1$~ms at $D = 2.4~\mu$m, a delay of a quarter over the empty cell, and every valid run passed (Fig.~\ref{fig:arrest}J). But the passage was bought in ways a real vacuole cannot pay for. At the default volume constraint the vacuole deflated as it crossed, to $0.69$--$0.85$ of its initial volume (Fig.~\ref{fig:arrest}K). That is a water efflux not available to a membrane-bounded body in milliseconds: compressing it by a quarter takes an osmotic pressure four orders of magnitude above the $\sim\!30$~Pa that the constraint supplies (section~\smref{sec:filter}). With the volume held by a hundredfold stiffer constraint, the cell still passed at $D = 1.0$ and $1.4~\mu$m ($222.9\pm2.1$ and $235.3\pm2.1$~ms), but it did so by stretching the vacuole membrane by $5$--$15\%$ in area, up to several times the few percent at which bilayers rupture~\cite{rawicz2000effect,olbrich2000water}. At $D \geq 1.8~\mu$m with the volume held, all twelve runs failed the containment criterion at the pore, where the barrier that keeps the vacuole inside the host gave way. That is a limit of the model, and we report no passage outcome there. The model cell thus lacks the property a real RBC would use, excess membrane area. What the sweep establishes is the absence of arrest by deflation or stretch, not that a $2.4~\mu$m vacuole is harmless.

Where the vacuole rides matters here as it did at the slit. Every run above placed it in the trailing rim; driving it there with the fluid it displaces rather than with the nominal share changed no outcome, and changed the transit by $3$--$5\%$ (section~\smref{sec:filter}). In the leading rim, a $1.4~\mu$m vacuole held its cell at the layer for the whole run in two of three seeds, and the third cell crossed only after $510$~ms. A $2.2~\mu$m one delayed every crossing, to $384$--$414$~ms. We have not resolved why the smaller body is the more obstructive. That a single vacuole does not hold its cell at the filter is therefore a statement about a vacuole that trails.

How the same volume is divided changes the outcome. Eight vacuoles of $D = 1.1~\mu$m have together the volume of one of $D = 2.2~\mu$m. They held their host at the bead layer in all $13$ runs that kept every vacuole inside the cell: over the run such a cell advanced $4.36\pm0.37~\mu$m, against $25.5\pm0.1~\mu$m for the cell carrying one $2.2~\mu$m vacuole, which passed in all three of its runs (Fig.~\ref{fig:arrest}, I and L), and in all three again, if slowly, when the vacuole led. The arrest persisted with a plain soft vacuole--vacuole pair interaction, with the membrane--vacuole cutoff widened from $0.6$ to $0.7~\mu$m, and with the volume constraint stiffened a hundredfold. Four runs lost a vacuole through the host membrane during the crossing; each of those cells then passed, and they are excluded. One caveat is structural rather than statistical. The repulsive shells around eight vacuoles exclude $43$--$55~\mu$m$^3$ of the $92.5~\mu$m$^3$ cell, against $22~\mu$m$^3$ for a single $2.2~\mu$m vacuole, so part of the effect is the excluded volume the model gives to a divided load.

\subsection*{Squeezing changes which vacuoles meet, not how many fuse}

\begin{figure}[!tbp]
\centering
\includegraphics[width=\textwidth,height=0.56\textheight,keepaspectratio]{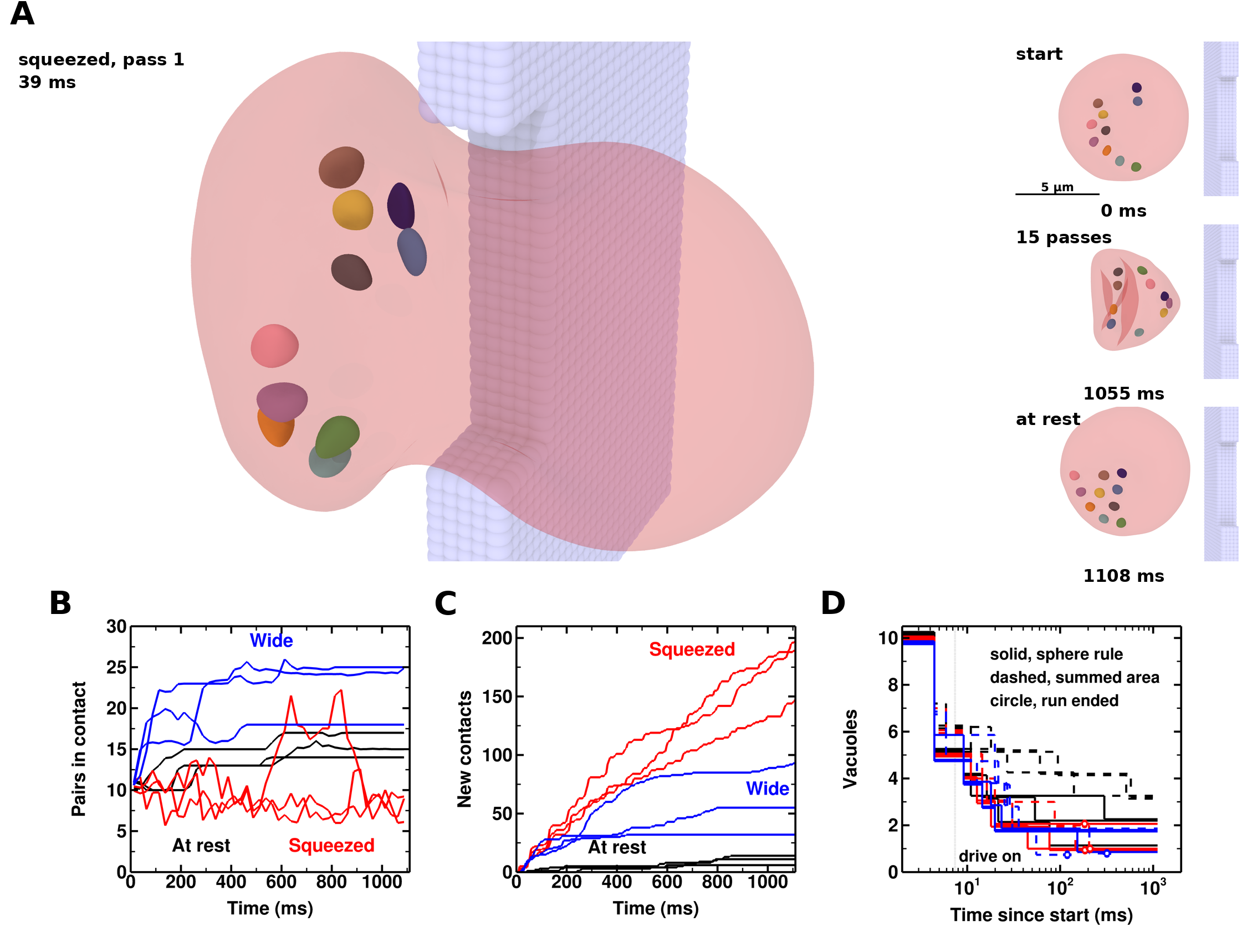}
\caption{\textbf{Repeated passage through the slit.}
\textbf{(A)} A cell like that of Fig.~\ref{fig:fusion}A, its ten vacuoles of $D = 0.6~\mu$m each in its own colour, in runs without fusion: squeezed, halfway through the slit on its first pass ($39$~ms), seen obliquely with the near half of the wall cut away (left); and face-on (right, top to bottom) at the start of the drive, common to the squeezed and resting arms, squeezed and back at its starting place after $15$ passes ($1055$~ms), and at rest at the end of its run ($1108$~ms), the three cut to one frame so that the slit lines up. In the face-on views the wall is cut at the cell's mid-plane, so the aperture shows as the gap at its centre. Scale bar, $5~\mu$m, for the face-on views.
\textbf{(B)} Pairs of vacuoles in contact, of $45$, their surfaces within $\delta_f = 0.7~\mu$m, against time since the drive started, in $25$~ms means: at rest (black), squeezed through the slit (red) and driven through the wide opening (blue), three runs each, without fusion.
\textbf{(C)} New contacts that lasted $t_f = 3.6$~ms, accumulated from the start of the drive; same runs.
\textbf{(D)} Vacuoles per cell in the runs with fusion, $\delta_f = 0.7~\mu$m and $t_f = 3.6$~ms, against time since the start of the run, which includes $7.4$~ms without drive (dotted line): solid, each product a near-sphere of the summed volume; dashed, each product given the summed area of its parents. Colours as in (B). Runs are offset vertically by up to $0.26$ so that runs with the same count stay visible; an open circle marks a run that ended on a vacuole-mesh bond.}
\label{fig:passage}
\end{figure}

The experiment this account leaves open is to squeeze pocked cells repeatedly through a narrow slit and count their vacuoles (Discussion). We first asked the model what squeezing does to the vacuoles of one cell. The cell of Fig.~\ref{fig:fusion}A, with its ten vacuoles of $D = 0.6~\mu$m, was turned edge-on to the slit of Fig.~\ref{fig:arrest}A and driven at $f = 8$ for $1108$~ms. Because the box is periodic, the cell returned to the wall once per lap and crossed the aperture $15$ or $16$ times, about every $70$~ms (Fig.~\ref{fig:passage}A and movie S2). Two controls ran for the same time. In one, the same cell was driven through an opening of $9.9\times11.2~\mu$m cut in the same wall, which it crossed $63$--$68$ times without being squeezed; in the other, the cell was left at rest. Each of the three arms was run three times without fusion, to follow contacts, and six times with the rule of Fig.~\ref{fig:fusion}, $\delta_f = 0.7~\mu$m and $t_f = 3.6$~ms: three with each product taking the summed area of its parents, and three with the area of a near-sphere of the summed volume (Materials and Methods). Contact is scored here as the fusion rule scores it: the surfaces of two vacuoles within $\delta_f$.

Squeezing moved the vacuoles past one another rather than together (Fig.~\ref{fig:passage}, B and C). At rest, $13.0$--$15.5$ of the $45$ pairs were in contact on average, and they were the same pairs throughout: in $1108$~ms only $6$--$14$ new contacts lasted $t_f$, among $5$--$9$ pairs. A squeezed cell had fewer pairs in contact at any moment, $8.4$--$11.5$, but it changed them at every pass. In it, $147$--$197$ new contacts lasted $t_f$, $9$ to $13$ per pass, among $24$--$43$ of the $45$ pairs, ten to thirty times as many as at rest. Through the wide opening the vacuoles instead crowded together, with $17.7$--$23.1$ pairs in contact, and only $32$--$93$ new contacts formed, fewer than two per lap. Passage pressed the vacuoles together only marginally harder. The fifth percentile of the separation of pairs in contact was $0.441$--$0.442~\mu$m in the squeezed cells against $0.453$--$0.455~\mu$m at rest, and both are at the floor the vacuole--vacuole barrier sets, so the model cannot say how hard a real passage would press two vacuole membranes together. What it does say is that passage changes which vacuoles meet.

With the fusion rule switched on, the count fell alike in all three arms (Fig.~\ref{fig:passage}D). Half of the fall came before any drive: the relaxed cell has $11$ of its $45$ pairs within $0.5~\mu$m, and their fusions had left five or six vacuoles when the drive started, $7.4$~ms into the run. With near-spherical products every run then went on to one or two vacuoles, the cells at rest as well as the squeezed ones, and in seven of the nine the last fusion had come by $69$~ms, before a squeezed cell reached the slit for the second time. With summed areas the cells at rest stopped at three vacuoles, each still holding qualified pairs whose products could not be placed, as in Fig.~\ref{fig:fusion}C. Every driven cell reached one or two, and five of the six then failed on a vacuole-mesh bond (section~\smref{sec:passage}).

Under a rule that fuses any contact held for $t_f$, squeezing therefore adds nothing: the vacuoles of a cell at rest are already in contact, and the count is set by how they were placed. Such a rule cannot describe real cells, which keep several vacuoles for weeks. What squeezing does supply is encounters, about ten new ones per pass in a cell of ten vacuoles. An experiment that squeezes cells therefore asks whether an encounter can fuse two vacuoles that a standing contact does not. It must compare the squeezed cells with the same cells kept at rest and with cells pushed through a channel too wide to squeeze them, since flow alone rearranges vacuoles, and differently.

\section*{Discussion}

The coalescence hypothesis makes two claims, one mechanical and one arithmetic, and we have tested both. The mechanical claim holds. The vacuoles of one cell come into contact within a second in a tube and in plane shear, and contacts once made persist in all three vascular geometries; the flow decides whether contacts turn over, not how much contact there is. Once contacts can end in fusion, ten vacuoles coarsen to four or five, or to one or two when the products are made near-spherical. The arithmetic claim does not hold for the ordinary cell. The volume a cell nucleates over $120$ days at the rate measured in patients makes one body of $0.6~\mu$m: what healthy cells carry, and $16$-fold short in volume of what splenectomized patients carry. If size is set at birth instead, a source that switches off as the cell ages and runs fastest in a minority of cells accounts for the published record and, scored against its first-hand part, favours a splenic clearance that rises several-fold over the cell's life. Letting the vacuoles of that minority fuse instead fits the first-hand record nearly as well, so the record requires the source and the minority and leaves open how the fast cells come by their large bodies. At a slit and at a filter, passage was decided by the cell's surface reserve and by how a vacuole load was placed and divided. Squeezed repeatedly, a cell's vacuoles changed partners at every pass without being pressed closer.

\textit{Transport is not the bottleneck.} Pocked cells accumulate over the $60$--$100$ days after splenectomy~\cite{buchanan1987formation}, and normal cells transfused into a splenectomized recipient acquire vacuoles over $15$ to $21$ days~\cite{holroyde1970acquisition}. Our cells make contacts within a second and, in a tube, hold them for the whole record. At least six orders of magnitude separate the two. The natural inference is that the slow step is the merging of two apposed membranes. But that does not rescue coalescence in a cell nucleating at the measured rate: there the sizes do not appear even when every contact is granted and allowed to succeed, because the material was never nucleated. A slow membrane step would explain why large vacuoles are rare; in such a cell it cannot explain how they exist at all. The question stays open only where the material exists, in the fast minority that the published counts require, whose lifetime output of $1.45~\mu$m$^3$ is within a fifth of the volume of a $1.5~\mu$m body. There the budget is silent and the published record does not decide, and the slit experiment below is the test.

\textit{The arithmetic is not ours.} Volume-conserving fusion places products on the ladder $D_0k^{1/3}$, and our measured products sit on it; a visible vacuole of $1.5~\mu$m has the volume of about $125$ vacuoles of $0.3~\mu$m. What gives the comparison its force is that the supply side has been measured: six trauma splenectomies give the rate at which vacuoles appear with the spleen removed~\cite{sills1988physiologic}, and a lifetime at that rate is $60$ nuclei. The sharpest comparison is internal to that same paper. Its $144$-h incubations produced $1.0~\mu$m vesicles that fusion would need $47$ nucleations per cell per day to build, against the $0.5$ that its own pit counts imply for the population. The argument is about the cell at the mean rate, and it shifts the question to the cells that are not at the mean: large vacuoles need a cell that makes six to sixteen times the vacuole volume of the ordinary one, according to which of the two rates is right, whichever way that volume is assembled.

\textit{Born young, and possibly born large.} Both experiments that asked what these vacuoles contain found them free of the medium the cell sat in, while drug-induced vacuoles in the same cells took it up: a fluid-phase marker stayed out of them~\cite{colin1991spontaneous}, and an extracellular stain was absent from $97$--$100\%$ of them~\cite{reinhart1988red}. They are parcels of the cell's own degraded contents, not invaginations of the mature cell's surface. Plasma membrane and fusion may nonetheless go into making them. In the last stage of reticulocyte maturation, large glycophorin A-containing vesicles form at the cytosolic face of the plasma membrane, are internalized, and fuse with autophagosomes before the autophagosomal contents are expelled~\cite{griffiths2012maturing}. Mature reticulocytes extrude these vesicles intact, and identical ones, about $1.4~\mu$m across, are found on the red cells of splenectomized patients, in whom cells bearing them are more frequent~\cite{mankelow2015autophagic}. A vesicle that the maturing cell fails to expel would be a large body present from birth. What the budget argues against is only the later assembly of large bodies by the fusion of small ones in a cell nucleating at the measured rate. One exploratory observation bears on this. In one sample of sickle cell blood, compact unstained dots of the kind that cells labelled pocked carry were found in reticulocytes about as often as in mature cells (ratio $1.5$, $95\%$ CI $0$--$4.4$), against the $0.03$ that the population model gives for bodies made after release; but only two reticulocytes carried a dot, and both were called equivocal (section~\smref{sec:dots}). What then has to be explained is when vacuoles are made, and the density fractions answer it. Only a source concentrated in young cells gives both the healthy profile, which falls thirteen-fold from the lightest to the densest fraction~\cite{sills1988formation}, and the profile without a spleen, which rises~\cite{reinhart1988red,dehaan1988vacuoles}. Such a source also removes a parameter rather than adding one: a constant source would need a sink for small bodies to stop the sub-visible pool growing throughout the cell's life, whereas a decaying source saturates that pool by itself.

\textit{A prediction about the spleen, from two tables of one paper.} If splenic clearance did not depend on cell age, the top-to-bottom density ratios measured in vivo and after $144$~h of incubation in the same donors would be equal~\cite{sills1988formation}. They differ threefold. The only way we have found to reconcile them is a clearance that rises at least three-fold over the cell's life, and most of the points that fit the first-hand record best sit at $5.6$-fold, the largest rise of the main scan, or between ten- and fifty-five-fold when the scan is extended (section~\smref{sec:pop}). The direction of the rise is better supported than its size, which moves with the optical limit (Results). A rise is what an organ filtering on deformability should show, given that red cells stiffen as they age~\cite{clark1988senescence,mohandas2008red}, but we did not assume it. The slit says where such a rise would come from. It selected the cell by its surface reserve and not by its vacuole (Fig.~\ref{fig:arrest}, A to E); and since aged cells lose volume along with surface~\cite{waugh1992rheologic}, the likelier lever is their dehydration, which raises the viscosity of their contents and which our model does not vary. Measuring a density-fractionated cohort both ways, in more than the original ten donors, would test it.

\textit{The slit experiment still has a use, with a different meaning.} Squeezing pocked cells repeatedly through a narrow slit suggests itself as a test of coalescence: forcing vacuoles into apposition should reduce their number if merging were limiting. The model says that squeezing does not force apposition. It presses vacuoles barely closer than rest does, but gives them new neighbours at every pass, ten to thirty times as many new contacts as a cell at rest makes (Fig.~\ref{fig:passage}). The experiment therefore asks whether an encounter can fuse two vacuoles where a standing contact does not, and in doing so it tests the born-large picture. Suppose vacuole number falls with repeated passage, faster than in the same cells kept at rest, while the total vacuole volume per cell does not. Then fusion is available and merely slow, and in the fast minority, where the material exists, it could assemble the large bodies. If neither changes, fresh encounters do not fuse vacuoles either, as expected if the cells carry what they were given at birth. The resting control is needed because cells go on acquiring vacuoles in vitro~\cite{sills1988formation}, and a channel too wide to squeeze the cells separates the constriction from the flow. A first look at the resting control exists. In time-lapse images of unstained blood, no merge was seen in $29$ cells carrying two or more compact dots over six hours at rest, a one-sided $95\%$ bound of $0.10$ per cell, although focus changes left the call uncertain in $15$ of them and the dots have yet to be identified as vacuoles by DIC (section~\smref{sec:dots}). Microfluidic chips with slits modelled on the spleen's exist~\cite{picot2015biomimetic,gambhire2017high,qiang2023microfluidic}, and vacuoles are counted cell by cell under differential interference contrast~\cite{dumas2023intra}, so the populations can be compared without sorting anything. A fall in total volume would indicate expulsion, which our model does not represent.

\textit{What a constriction selects.} If vacuole loads were assembled by coalescence, the way a given volume is divided would be a passing state on the way to a single large body. If the load is drawn at birth from a size law, its division is a stable property of the cell, and a constriction can select on it directly. At the four-bead filter, dividing the same vacuole volume among eight bodies changed passage into arrest. Two mechanisms are available in the model, and we have not separated them. Eight bodies cannot all be accommodated in the part of the cell that enters the pore first, which the leading-rim runs bear on; and the repulsive shells the model gives them exclude twice the volume that one large vacuole does, a modelling artefact in magnitude if not in sign. No experiment can yet settle this, since pocked cells cannot be sorted by load nor vacuoles isolated~\cite{dumas2023intra}, and a bed of microspheres~\cite{deplaine2011sensing,safeukui2018sensing} retains a population, not a chosen load. We therefore report the filter result as a property of the model, and make no claim about pitting or the expulsion of vacuoles, which the model does not represent.

\textit{Where we disagree with the record.} The population model contradicts one observation in the primary literature outright. Reinhart and Chien found the frequency of small vacuoles similar in density-separated fractions~\cite{reinhart1988red}, and our small pool is not flat. Their four-fold rise after splenectomy we do reproduce, at $3.2$-fold, and our predictions bracket their per-cell estimate once it is corrected for section thickness. Two things the same paper says bear on the disagreement. Reticulocytes, which in their words usually have vacuoles, were excluded from every count, so the flatness is measured on a population without the youngest cells, those in which our source is concentrated. And the authors set out both possibilities themselves, a pool of remnants carried through the cell's life or formation balanced by removal, and wrote that no data were available to decide between them. Scored both ways, the scan gives two rankings that want different models, and we report both. Neither's best hundred contains a point from the other's preferred end, and the one combination common to both, at a $1.8$-fold rise of clearance and ranked $37$th and $67$th, buys its place by shortening the source decay sevenfold. Sectioned electron microscopy of density-separated cells from a modern cohort would decide it.

A second disagreement lies within the record itself, and we take a side in it. Sills et al.\ found vacuole acquisition five times higher in cord than in adult cells incubated together, and read it as increased formation. Colin and Schrier, running the same protocol, found the two only $1.6$-fold apart and concluded that the neonatal excess must be reduced clearance by an immature macrophage system~\cite{sills1988formation,colin1991spontaneous}. Ours is an account of the source, so it agrees with the first. The two laboratories agree on the conditional acquisition of cord cells at $144$~h to within two percentage points, $17.2$ against $19.2\%$, and differ $2.8$-fold on adults, $10.5$ against $3.7\%$. The dispute is therefore about the adult denominator, and the neonatal rate is the best-replicated acquisition figure in this literature. Against ourselves we note that reticulocytes fall fifty-eight-fold across the neonatal gradient while vesicles fall only two-fold~\cite{sills1988formation}, which is uncomfortable for a pure source account in either laboratory's hands.

\textit{Limitations.} The nucleation rate behind the volume budget and the fitted source is one laboratory's six-day curve, and a second laboratory measured adult acquisition two to three times faster. For a cell at the mean rate, the budget's case against coalescence survives that, but its agreement with the healthy size range does not; and with cells differing widely in rate, the faster one lets a coalescence model reach the observed sizes (Results). The budget does not reach the fast minority at either rate, and given the same source and minority a coalescence model fits the first-hand record almost as well as the born-large one ($0.349$ against $0.220$), so the origin of the large bodies in those cells is not settled here. The population model is a mean-field description of a Poisson birth-and-clearance process. It cannot say why a source should decay, only that it must. The age dependence of its clearance is a single exponential, chosen for want of any measurement of its shape, which coupling to a model of macrophage-mediated clearance~\cite{chai2026multiscale} could supply. Its best point, one of a $32\,400$-point grid with no confidence region, marks the region the record admits rather than an estimate. On the simulation side, the fusion criterion is phenomenological, and the vacuoles we simulate, $0.6~\mu$m and larger, sit at the resolution limit of the DPD fluid, so the sub-$300$~nm vacuoles from which the coalescence hypothesis starts are not resolved. The cascade and the repeated-passage runs each share one initial placement, and several ended on a vacuole-mesh bond. The vascular simulations were run once each, their contact occupancy had not settled, and their solvent is two to four thousand times denser than plasma. At the slit only the cell is driven, the cytosol has the viscosity of plasma, and no run covers more than $1.3$~s of the weeks over which vacuoles accumulate. Independent placements are the first thing a follow-up should add, and automated tuning of the numerical schemes~\cite{toscano2026graftathena} may make such campaigns cheaper. Section~\smref{sec:limits} sets out each limitation in full.

Taken together, these results change the question about pocked red cells rather than answering the one that was asked. The mechanical premise of the coalescence hypothesis, that the vacuoles of a cell meet, is satisfied with room to spare. The sizes seen in patients still do not follow, because a red cell nucleating at the measured rate does not make enough vacuole volume in its whole life to build them; only a fast minority does. The question is therefore less how small vacuoles become large ones than which cells make so much vacuole material, whether it arrives large, and why that capacity belongs to young cells and to a minority of them. That reframing can be tested in three independent ways, none of which requires sorting or isolating vacuoles: the saturation of the rise after splenectomy at about the red cell lifespan, the dissociation between reticulocyte content and vesicle burden in neonates, and the age dependence of splenic clearance implied by measuring one cohort's density fractions both in the circulation and in a flask.

\section*{Materials and Methods}

\subsection*{Study design}

The study asked whether the vacuoles of pocked red cells could be built by the fusion of smaller ones. It combined three parts: DPD simulations of single red cells and of cell populations carrying vacuoles, in which we measured how often the vacuoles of a cell meet in tube, shear and vascular flows, how a population of vacuoles coarsens when contacts can end in fusion, and how a vacuole load affects passage through a slit and a microsphere filter; a volume budget built on published nucleation rates and vacuole sizes; and a population model of vacuole birth and splenic clearance scored against published measurements. Two exploratory observations on human blood are reported in the Supplementary Materials (section~\smref{sec:dots}).

\subsection*{Dissipative particle dynamics}

Simulations used dissipative particle dynamics~\cite{hoogerbrugge1992simulating,espanol1995statistical,groot1997dissipative} in an extended LAMMPS solver~\cite{fedosov2010multiscale,thompson2022lammps}. The force between beads $i$ and $j$ is the sum of conservative, dissipative and random terms, with the dissipative and random amplitudes related by the fluctuation--dissipation theorem, $\sigma^2 = 2\gamma k_{\mathrm{B}}T$~\cite{groot1997dissipative,Fedosov2011Multiscale}. Unless stated otherwise, pairs interact with $a = 30$, $\gamma = 15$, $\sigma = 2.3774$, weight exponent $0.25$ and cutoff $r_c = 1$, which fixes $k_{\mathrm{B}}T = \sigma^2/2\gamma = 0.1884$ in reduced units; the time step is $0.001$. Plasma and cytosol are the same DPD fluid, and walls and beads are frozen DPD particles. One length unit is $1~\mu$m and $k_{\mathrm{B}}T$ corresponds to $4.14\times10^{-21}$~J.

The time unit follows from matching the viscosity of the DPD fluid to that of plasma, $1.2$~mPa\,s. We measured the DPD viscosity in body-force-driven flow between frozen walls at six densities and fitted $\eta^{\mathrm{DPD}} = 0.840\,\rho^{1.958}$ for $\rho \geq 3$, so that $\tau = 54.6~\mathrm{ms}/\eta^{\mathrm{DPD}}$ and the physical duration of a step depends on the solvent density of each geometry (section~\smref{sec:units}; table~\smref{tab:runs}). Where the density is measured, and whether the law or an interpolation of the calibration converts it, changes $\tau$ by up to $13\%$, so we treat $\tau$ as uncertain by about $10\%$; this affects durations but not the dimensionless comparisons between simulations.

\subsection*{Red cell and vacuole model}

The RBC membrane is a network of $500$ vertices with worm-like-chain bonds, a bending energy on dihedral angles and global area and volume constraints, $A_0 = 132.87~\mu$m$^2$ and $V_0 = 92.46~\mu$m$^3$, following~\cite{fedosov2010multiscale,li2014erythrocyte}. Its shear modulus and bending rigidity are those of a healthy human RBC, $4.73~\mu$N/m and $k_c = 2.4\times10^{-19}$~J, which are $215$ and $12.6$ in reduced units with $k_c = \sqrt{3}\,k_b/2$. Each vacuole is a closed triangulated vesicle with the same kinds of potential and softer constants, $2.6~\mu$N/m and $1.1\times10^{-19}$~J: $34$ vertices at $D = 0.6$ and $1.1~\mu$m, and $154$ at the single vacuoles of $1.0$--$2.4~\mu$m used at the filter, with area and volume targets near those of a sphere of diameter $D$. Membrane and vacuole vertices repel through a barrier of $a = 600$, $\gamma = 0.4$, $\sigma = 0.3882$ with a cutoff of $0.6$ or $0.7~\mu$m, which leaves a gap of $0.54$ or $0.64~\mu$m and keeps every vacuole vertex inside its host. Vacuoles repel one another through a barrier of the same form with a cutoff of $0.5~\mu$m, whose closest approach is $0.43~\mu$m between the vertices of two vacuoles. At rest the cell accepts a single vacuole up to $D = 2.45~\mu$m, and only in its rim.

Flow is driven by a uniform body force on solvent particles and membrane vertices, except in plane shear, where the walls are moved, and at the slit, where only the cell is driven. Each vacuole vertex receives a share of the force its displaced fluid would receive, $f_v = \rho V_0 f / N$, so that vacuoles are nominally neutrally buoyant (section~\smref{sec:drive}); because the solvent stands off the vacuole's vertices, this share under-drives them, which matters little inside a cell, where the cytosol carries the vacuole along, and changed no outcome at the filter (section~\smref{sec:filter}). In the vascular simulations the vacuole--vacuole pair is soft during equilibration and is replaced by the barrier at the last stage, immediately before the production run, whose clock is reset to zero.

\subsection*{Fusion}

Runs with fusion proceed in segments. Between segments, two vacuoles of the same cell fuse once the smallest distance between their vertices has stayed below $\delta_f$ for a continuous time $t_f$. We swept $\delta_f$ over $0.5$, $0.7$ and $0.9~\mu$m and $t_f$ over $1.2$, $3.6$ and $11.9$~ms on the measured time scale of that simulation, which the driver set as nominal $1$, $3$ and $10$~ms (section~\smref{sec:fusion}); every time is quoted on the measured scale. The product takes the summed volume and the summed membrane area of its parents, is remeshed with a vertex count proportional to its area, and is placed at the volume-weighted centroid of the pair, shifted towards the centroid of the host until it clears the host membrane by half the membrane--vacuole cutoff; the cytosol particles it displaces are reinserted in the same cell. A fusion that cannot be placed is deferred and retried at the next segment boundary (section~\smref{sec:fusion}).

\subsection*{Contact analysis}

Vacuole positions were written every $20$ steps in the single-cell simulations and every $100$ steps in the vascular ones, and pairs were formed only between vacuoles of the same cell. A pair is in contact while the distance between the two centroids is below $d_c$. For each threshold we report the occupancy, the fraction of the observed frames, taken over all pairs, in which the pair is in contact, and the number and duration of episodes that both begin and end within the record, counting the others as censored. Figs.~\ref{fig:contacts} and~\ref{fig:survey} use $d_c = 1.2~\mu$m throughout, so that environments are compared at a common threshold; the sweep from $1.0$ to $2.0~\mu$m is in section~\smref{sec:contact}. In the runs of Fig.~\ref{fig:passage}, a pair is in contact while the smallest distance between the vertices of its two vacuoles is below $\delta_f = 0.7~\mu$m, and a new contact is one that begins after the drive has started and lasts at least $t_f = 3.6$~ms (section~\smref{sec:passage}).

\subsection*{Geometries and driving}

\textit{Encounter tube and plane shear.} The tube has a lumen of $10.8~\mu$m in a box of $30.4\times16\times16~\mu$m$^3$ and is driven at $f = 0.45$, a pressure gradient of $0.043$~Pa/$\mu$m. The plane-shear box is $20\times13\times16~\mu$m$^3$, with wall slabs moving at $\pm463~\mu$m/s that shear a gap of $6~\mu$m (section~\smref{sec:geometry}).

\textit{Survey tubes and vessels.} The two survey tubes use the same lumen at $f = 0.28$ and $2.5$, that is $0.026$ and $0.23$~Pa/$\mu$m. The bifurcation is a channel $10~\mu$m high whose $20~\mu$m lumen is divided by a hexagonal obstacle into two branches $10~\mu$m wide; the curved vessel is a lumen of $29.8~\mu$m on an $\Omega$-shaped centreline; and the dilation is a lumen of $25.0~\mu$m opening into a spherical sac of $50.0~\mu$m, a bulge-to-neck ratio of $2.0$ (section~\smref{sec:geometry}). All three are driven at $f = 0.5$, $0.031$--$0.040$~Pa/$\mu$m at their lumen densities; at $f = 2$ the drive pushed membrane vertices through the wall. Reynolds and capillary numbers are defined in section~\smref{sec:reynolds}.

\textit{Slit.} A wall of frozen particles $0.88~\mu$m thick closes a periodic box of $30\times12.5\times12.5~\mu$m$^3$ and is pierced by one rectangular aperture of $2.85\times5.48~\mu$m between particle centres; because the solvent stands off the wall particles, the section it fills within the aperture is equivalent to $1.74\times3.14~\mu$m, against $1.2\times4.0~\mu$m for the slit of Li et al.~\cite{li2021How}. The cell starts edge-on to the aperture, with its vacuole $2.6~\mu$m behind or ahead of its centre, and its volume was raised at fixed area to give surface-to-volume ratios of $1.41$, $1.20$ and $1.09~\mu$m$^{-1}$. The body force acts on membrane vertices and cytosol only, because driving the plasma as well makes the fluid cavitate behind a plugged aperture; the membrane--solvent repulsion is raised to $a = 100$ to keep the cytosol inside the cell, and each vacuole carries the force on the fluid it displaces. The drive is $f = 8$ or $4$, a total force on the cell of $150$--$174$ or $75$--$87$~pN, and each run proceeds for $100\,000$ or $150\,000$ steps, $370$ or $554$~ms. The construction is given in full in section~\smref{sec:slit}.

\textit{Microsphere filter.} A periodic box of $20\times9.5\times9.5~\mu$m$^3$ holds four beads, each $3232$ frozen particles within $2.008~\mu$m of its centre, on a square lattice of pitch $4.75~\mu$m. The cell starts centred on a pore with a single vacuole in its trailing rim and is driven at $f = 3$, a pressure gradient of $0.29$~Pa/$\mu$m; a control set drives the vacuole with the fluid it displaces and places it in either rim (section~\smref{sec:filter}). After a $2000$-step ramp of the membrane--vacuole barrier each run proceeds for $150\,000$ steps (section~\smref{sec:geometry}).

\textit{Repeated passage.} The cell of the fusion cascade, its ten vacuoles placed as there, was turned edge-on to the aperture of the slit set-up, relaxed, and driven as at the slit at $f = 8$; the box is periodic, so the cell returns to the wall once per lap. A second set-up pierces the same wall with an aperture of $9.87\times11.18~\mu$m between particle centres, wider than the cell, and in the third arm the cell is not driven. The runs without fusion last $300\,000$ steps; those with fusion take half the time step over the same $1108$~ms of drive, with each product given either the summed area of its parents or the area of a sphere of volume $V_0/0.97$, a near-sphere (section~\smref{sec:passage}).

\subsection*{Analysis of passage through the slit and the filter}

At the slit, a cell reaches the wall when its front comes within $0.8~\mu$m of the wall's upstream plane and leaves it when its rear is $0.8~\mu$m past the downstream plane, the vacuole likewise by its front and rearmost vertices, and a cell that has not left by the end of the run is held (section~\smref{sec:slit}). At the filter, a run is valid only if every vacuole stayed hosted and at least $99\%$ inside the membrane, tested by its generalized winding number, in every frame (section~\smref{sec:checks}); transit runs from the frame at which the cell front reaches the bead layer to that at which its rear leaves it, and a cell that advanced less than $15~\mu$m is scored as held (section~\smref{sec:filter}).

\subsection*{Population model}

Red cells are followed from $0$ to $120$ days, and because production is steady the population age distribution is uniform over that interval. A cell of age $a$ nucleates bodies at rate $J(a) = J_0(e^{-a/\tau} + \phi)$ with diameters drawn from a lognormal law of median $d_{50}$ and log-standard-deviation $\sigma$, represented on a $400$-point log-spaced grid. A body of diameter $D$ in a cell of age $a$ is removed by the spleen at rate $R(D,a) = R_{\mathrm{pass}}\,g(D)\,e^{a/\tau_R}$, where $g$ rises from $0$ to $1$ across a gate at $D_g = 0.35~\mu$m and $R_{\mathrm{pass}} = 2$/day is the number of splenic passes a cell makes per day. $R_{\mathrm{pass}}$ and $D_g$ are fixed, not fitted. Since $R \gg 1/\tau_R$, $R$ is evaluated at the cell's current age.

Births are Poisson and size-dependent clearance thins them independently, which preserves the Poisson property, so the expected number of surviving bodies per unit diameter in a cell of age $a$ is
\begin{equation}
\Lambda(D,a) = f(D)\int_0^a J(s)\,e^{-R(D,a)(a-s)}\,\mathrm{d}s ,
\end{equation}
with $f$ the birth-size density. The inner integral is closed form,
$J_0\left[e^{-a/\tau} - e^{-Ra}\right]/(R - 1/\tau)$ for the decaying part of the source, tending to $J_0\,a\,e^{-Ra}$ as $R \to 1/\tau$; the floor adds $J_0\phi\,(1 - e^{-Ra})/R$ (section~\smref{sec:pop}). Every observable is then an integral rather than a simulation: a cell of age $a$ is pocked with probability $1 - \exp\!\left[-\int_{0.3~\mu\mathrm{m}}^{\infty}\Lambda(D,a)\,\mathrm{d}D\right]$, taking $0.3~\mu$m as the DIC limit, and the reported prevalence is that probability averaged over age. The largest body and the most loaded cell are the corresponding extremes over $10^4$ cells, the size of a clinical series at $500$ consecutive cells per subject in twenty subjects.

A fraction $f_{ab}$ of cells carries $J_0$ multiplied by $m_J$ and birth sizes multiplied by $m_d$; the remainder is the majority population. The spleen-absent branch sets $R = 0$ for the time since splenectomy, so a cell of age $a$ at time $t$ after surgery has been cleared for $\max(0, a-t)$ of its life. In vitro acquisition is the same integral with $R = 0$ throughout and $a$ running over the donor's own age distribution.

Density fractions are not modelled as thirds of the lifespan, because they were not measured that way: the gradient samples are its uppermost and lowermost two to three layers plus several from the middle, which is why an equal-thirds average of the published fractions gives $2.4\%$ where the same donors' unseparated blood gives $0.5\%$. Each fraction is instead a weight $w(a) \propto e^{ba}$ over the lifespan, with $b$ solved so that the fraction's reticulocyte content, $(e^{bT_r}-1)/(e^{bT_L}-1)$ with $T_r = 1.8$~d and $T_L = 120$~d, matches the measured $10.5$, $1.2$ and $0.2\%$ against an unseparated $1.5\%$~\cite{sills1988formation}. That gives $b = -1/16$, $+1/275$ and $+1/37$ per day for the light, middle and dense fractions; the middle fraction is essentially the whole population.

The scan covered $32\,400$ combinations of $J_0$, $\tau$, $\phi$, $d_{50}$, $\sigma$, $f_{ab}$, $m_J$, $m_d$ and $\tau_R$, each scored twice: once against every published target, and once with the four targets that describe the pool below the optical limit set aside, that pool being an output of this model rather than one of its claims. Each score is a sum of squared log deviations, with a target that is a range contributing nothing inside it. Three further rankings each change one term: the six-day acquisition of $3.71\%$ gives way to the $10.47\%$ a second laboratory obtained on an explicitly unmodified copy of the protocol~\cite{colin1991spontaneous}; the healthy prevalence range $0.5$--$2.6\%$ to $0.44$--$0.60\%$, the mean $\pm\,2$ standard errors of the largest cohort~\cite{buchanan1987formation}; and the estimate of about $20$ sub-visible vacuoles per cell to the $7.7$--$9.6$ that follows from correcting it for section thickness. The grid, the targets and how each is scored, the best point of every ranking, and the scan extended to a fifty-five-fold rise of clearance and repeated at DIC limits of $0.20$ and $0.40~\mu$m are given in section~\smref{sec:pop}.

\subsection*{Statistics}

Each condition of the encounter, slit, filter and repeated-passage series was run with three or more seeds, and the cascade once per combination with one combination repeated. Values are given as mean $\pm$ SD over seeds, and per-seed values are listed where the seeds are few. No inferential test is reported for these: with three seeds per condition the comparisons rest on non-overlapping ranges, and a test would add no information. The one exception is a comparison of counts at the slit, where there are no ranges to compare: one crossing in six runs with the vacuole leading, against four in six with it trailing, has a one-sided Fisher's exact $p = 0.12$, and we report it as a tendency. The exploratory observations on human blood (section~\smref{sec:dots}) are given with $95\%$ intervals: a stratified bootstrap for the ratio of weighted proportions, a one-sided binomial bound where no event was seen, and Cohen's $\kappa$ for agreement between readers.

\nocite{Fedosov2011Predicting}
\begingroup
\small
\bibliography{ref_main}

\begin{thebibliography}{10}

\bibitem{Mebius2005Structure}
R.~E. Mebius, G.~Kraal, Structure and function of the spleen. \textit{Nat. Rev.
  Immunol.} \textbf{5}, 606--616 (2005).

\bibitem{pivkin2016biomechanics}
I.~V. Pivkin, Z.~Peng, G.~E. Karniadakis, P.~A. Buffet, M.~Dao, S.~Suresh,
  Biomechanics of red blood cells in human spleen and consequences for
  physiology and disease. \textit{Proc. Natl. Acad. Sci. U.S.A.} \textbf{113},
  7804--7809 (2016).

\bibitem{Li2018Mechanics}
H.~Li, L.~Lu, X.~Li, P.~A. Buffet, M.~Dao, G.~E. Karniadakis, S.~Suresh,
  Mechanics of diseased red blood cells in human spleen and consequences for
  hereditary blood disorders. \textit{Proc. Natl. Acad. Sci. U.S.A.}
  \textbf{115}, 9574--9579 (2018).

\bibitem{moreau2023physical}
A.~Moreau, F.~Yaya, H.~Lu, A.~Surendranath, A.~Charrier, B.~Dehapiot,
  E.~Helfer, A.~Viallat, Z.~Peng, Physical mechanisms of red blood cell splenic
  filtration. \textit{Proc. Natl. Acad. Sci. U.S.A.} \textbf{120}, e2300095120
  (2023).

\bibitem{dao2021erythrocyte}
M.~Dao, I.~MacDonald, R.~Asaro, Erythrocyte flow through the interendothelial
  slits of the splenic venous sinus. \textit{Biomech. Model. Mechanobiol.}
  \textbf{20}, 2227--2245 (2021).

\bibitem{buchanan1987formation}
G.~R. Buchanan, C.~A. Holtkamp, J.~A. Horton, Formation and disappearance of
  pocked erythrocytes: studies in human subjects and laboratory animals.
  \textit{Am. J. Hematol.} \textbf{25}, 243--251 (1987).

\bibitem{buchanan1987pocked}
G.~R. Buchanan, C.~A. Holtkamp, Pocked erythrocyte counts in patients with
  hereditary spherocytosis before and after splenectomy. \textit{Am. J.
  Hematol.} \textbf{25}, 253--257 (1987).

\bibitem{dumas2023intra}
L.~Dumas, C.~Roussel, P.~Buffet, Intra-erythrocytic vacuoles in asplenic
  patients: elusive genesis and original clearance of unique organelles.
  \textit{Front. Physiol.} \textbf{14}, 1324463 (2023).

\bibitem{sissoko2022erythrocytic}
A.~Sissoko, A.~Fricot-Monsinjon, C.~Roussel, S.~Manceau, L.~Dumas, C.~Capito,
  S.~Allali, N.~Yekkache, M.~Dussiot, Y.~Nguyen, A.~{Lefort Des Ylouses},
  B.~Aussilhou, M.~Tichit, D.~Hardy, B.~Ma{\^\i}tre, A.~Eckly, M.~{De
  Montalembert}, M.~Cavazzana, L.~Joseph, P.~Buffet, Erythrocytic vacuoles that
  accumulate a fluorescent dye predict spleen size and function in sickle cell
  disease. \textit{Am. J. Hematol.} \textbf{97}, E385--E388 (2022).

\bibitem{holroyde1970acquisition}
C.~P. Holroyde, F.~H. Gardner, Acquisition of autophagic vacuoles by human
  erythrocytes. Physiological role of the spleen. \textit{Blood} \textbf{36},
  566--575, PMID: 5473517 (1970).

\bibitem{reinhart1988red}
W.~H. Reinhart, S.~Chien, Red cell vacuoles: their size and distribution under
  normal conditions and after splenectomy. \textit{Am. J. Hematol.}
  \textbf{27}, 265--271 (1988).

\bibitem{sills1988physiologic}
R.~H. Sills, J.~H. Tamburlin, N.~J. Barrios, P.~L. Yeagle, C.~A. Glomski,
  Physiologic formation of intracellular vesicles in mature erythrocytes.
  \textit{Am. J. Hematol.} \textbf{28}, 219--226 (1988).

\bibitem{colin1991spontaneous}
F.~C. Colin, S.~L. Schrier, Spontaneous endocytosis in human neonatal and adult
  red blood cells: comparison to drug-induced endocytosis and to
  receptor-mediated endocytosis. \textit{Am. J. Hematol.} \textbf{37}, 34--40
  (1991).

\bibitem{sills1988formation}
R.~H. Sills, J.~H. Tamburlin, N.~J. Barrios, C.~A. Glomski, P.~L. Yeagle,
  Formation of intracellular vesicles in neonatal and adult erythrocytes:
  evidence against the concept of neonatal hyposplenism. \textit{Pediatr. Res.}
  \textbf{24}, 703--708 (1988).

\bibitem{dehaan1988vacuoles}
L.~D. De~Haan, J.~M. Werre, A.~M. Ruben, A.~H. Huls, J.~de~Gier, G.~E. Staal,
  Vacuoles in red cells from splenectomized subjects originate during cell
  life: association with glycosylated haemoglobin? \textit{Eur. J. Haematol.}
  \textbf{41}, 482--488 (1988).

\bibitem{kogan2014shear}
M.~Kogan, B.~Feng, B.~Nord{\'e}n, S.~Rocha, T.~Beke-Somfai, Shear-Induced
  Membrane Fusion in Viscous Solutions. \textit{Langmuir} \textbf{30},
  4875--4878 (2014).

\bibitem{popel2005microcirculation}
A.~S. Popel, P.~C. Johnson, Microcirculation and hemorheology. \textit{Annu.
  Rev. Fluid Mech.} \textbf{37}, 43--69 (2005).

\bibitem{kihm2021lingering}
A.~Kihm, S.~Quint, M.~W. Laschke, M.~D. Menger, T.~John, L.~Kaestner,
  C.~Wagner, Lingering dynamics in microvascular blood flow. \textit{Biophys.
  J.} \textbf{120}, 432--439 (2021).

\bibitem{cebral2005characterization}
J.~R. Cebral, M.~A. Castro, J.~E. Burgess, R.~S. Pergolizzi, M.~J. Sheridan,
  C.~M. Putman, Characterization of cerebral aneurysms for assessing risk of
  rupture by using patient-specific computational hemodynamics models.
  \textit{AJNR Am. J. Neuroradiol.} \textbf{26}, 2550--2559 (2005).

\bibitem{xiang2011hemodynamics}
J.~Xiang, S.~K. Natarajan, M.~Tremmel, D.~Ma, J.~Mocco, L.~N. Hopkins, A.~H.
  Siddiqui, E.~I. Levy, H.~Meng, Hemodynamic-morphologic discriminants for
  intracranial aneurysm rupture. \textit{Stroke} \textbf{42}, 144--152 (2011).

\bibitem{cai2021artificial}
S.~Cai, H.~Li, F.~Zheng, F.~Kong, M.~Dao, G.~E. Karniadakis, S.~Suresh,
  Artificial intelligence velocimetry and microaneurysm-on-a-chip for
  three-dimensional analysis of blood flow in physiology and disease.
  \textit{Proc. Natl. Acad. Sci. U.S.A.} \textbf{118}, e2100697118 (2021).

\bibitem{abkarian2007swinging}
M.~Abkarian, M.~Faivre, A.~Viallat, Swinging of red blood cells under shear
  flow. \textit{Phys. Rev. Lett.} \textbf{98}, 188302 (2007).

\bibitem{fedosov2014deformation}
D.~A. Fedosov, M.~Peltom{\"a}ki, G.~Gompper, Deformation and dynamics of red
  blood cells in flow through cylindrical microchannels. \textit{Soft Matter}
  \textbf{10}, 4258--4267 (2014).

\bibitem{lei2013blood}
H.~Lei, D.~A. Fedosov, B.~Caswell, G.~E. Karniadakis, Blood flow in small
  tubes: quantifying the transition to the non-continuum regime. \textit{J.
  Fluid Mech.} \textbf{722}, 214--239 (2013).

\bibitem{hoogerbrugge1992simulating}
P.~J. Hoogerbrugge, J.~M. V.~A. Koelman, Simulating microscopic hydrodynamic
  phenomena with dissipative particle dynamics. \textit{Europhys. Lett.}
  \textbf{19}, 155--160 (1992).

\bibitem{espanol1995statistical}
P.~Espa{\~n}ol, P.~Warren, Statistical mechanics of dissipative particle
  dynamics. \textit{Europhys. Lett.} \textbf{30}, 191--196 (1995).

\bibitem{groot1997dissipative}
R.~D. Groot, P.~B. Warren, Dissipative particle dynamics: Bridging the gap
  between atomistic and mesoscopic simulation. \textit{J. Chem. Phys.}
  \textbf{107}, 4423--4435 (1997).

\bibitem{fedosov2010multiscale}
D.~Fedosov, B.~Caswell, G.~Karniadakis, A multiscale red blood cell model with
  accurate mechanics, rheology, and dynamics. \textit{Biophys. J.} \textbf{98},
  2215--2225 (2010).

\bibitem{Fedosov2011Multiscale}
D.~A. Fedosov, H.~Lei, B.~Caswell, S.~Suresh, G.~E. Karniadakis, Multiscale
  Modeling of Red Blood Cell Mechanics and Blood Flow in Malaria. \textit{PLOS
  Comput. Biol.} \textbf{7}, e1002270 (2011).

\bibitem{li2014erythrocyte}
H.~Li, G.~Lykotrafitis, Erythrocyte membrane model with explicit description of
  the lipid bilayer and the spectrin network. \textit{Biophys. J.}
  \textbf{107}, 642--653 (2014).

\bibitem{Fedosov2011Quantifying}
D.~A. Fedosov, B.~Caswell, S.~Suresh, G.~E. Karniadakis, Quantifying the
  biophysical characteristics of \textit{Plasmodium-falciparum}-parasitized red
  blood cells in microcirculation. \textit{Proc. Natl. Acad. Sci. U.S.A.}
  \textbf{108}, 35--39 (2011).

\bibitem{tang2017openrbc}
Y.-H. Tang, L.~Lu, H.~Li, C.~Evangelinos, L.~Grinberg, V.~Sachdeva, G.~E.
  Karniadakis, OpenRBC: A fast simulator of red blood cells at protein
  resolution. \textit{Biophys. J.} \textbf{112}, 2030--2037 (2017).

\bibitem{chai2023dynamics}
Z.~Chai, S.~Gu, G.~Lykotrafitis, Dynamics of the axon plasma membrane skeleton.
  \textit{Soft Matter} \textbf{19}, 2514--2528 (2023).

\bibitem{chai2022periodic}
Z.~Chai, A.~V. Tzingounis, G.~Lykotrafitis, The periodic axon membrane skeleton
  leads to Na nanodomains but does not impact action potentials.
  \textit{Biophys. J.} \textbf{121}, 3334--3344 (2022).

\bibitem{chai2026reticulocyte}
Z.~Chai, J.~Zheng, H.~Li, M.~Dao, G.~E. Karniadakis, Quantifying reticulocyte
  biomechanics in health and disease. \textit{arXiv preprint arXiv:2607.21810}
  (2026).

\bibitem{chai2026quantifying}
Z.~Chai, J.~Zheng, H.~Li, M.~Dao, G.~E. Karniadakis, Quantifying the
  biophysical properties of stomatocytes in health and disease.
  \textit{Biophys. J.}, doi:10.1016/j.bpj.2026.10.007 (2026).

\bibitem{chai2025silico}
Z.~Chai, G.~Li, P.~A. Ndour, P.~Connes, P.~A. Buffet, M.~Franco, G.~E.
  Karniadakis, In silico biophysics and rheology of blood and red blood cells
  in Gaucher Disease. \textit{PLOS Comput. Biol.} \textbf{21}, e1012705 (2025).

\bibitem{chai2026gaucher}
Z.~Chai, M.~de~Person, P.~A. Buffet, M.~Franco, G.~E. Karniadakis, Quantifying
  the biophysical properties of red blood cells in {G}aucher disease.
  \textit{Biophys. J.}, in press (2026).

\bibitem{zhang2020deep}
Y.~Zhang, Z.~Chai, Y.~Sun, G.~Lykotrafitis, A deep reinforcement learning model
  based on deterministic policy gradient for collective neural crest cell
  migration. \textit{arXiv preprint arXiv:2007.03190} (2020).

\bibitem{zhang2021deep}
Y.~Zhang, Z.~Chai, G.~Lykotrafitis, Deep reinforcement learning with a particle
  dynamics environment applied to emergency evacuation of a room with
  obstacles. \textit{Physica A} \textbf{571}, 125845 (2021).

\bibitem{deplaine2011sensing}
G.~Deplaine, I.~Safeukui, F.~Jeddi, F.~Lacoste, V.~Brousse, S.~Perrot,
  S.~Biligui, M.~Guillotte, C.~Guitton, S.~Dokmak, B.~Aussilhou, A.~Sauvanet,
  D.~{Cazals Hatem}, F.~Paye, M.~Thellier, D.~Mazier, G.~Milon, N.~Mohandas,
  O.~Mercereau-Puijalon, P.~H. David, P.~A. Buffet, The sensing of poorly
  deformable red blood cells by the human spleen can be mimicked in vitro.
  \textit{Blood} \textbf{117}, e88--e95 (2011).

\bibitem{safeukui2018sensing}
I.~Safeukui, P.~A. Buffet, G.~Deplaine, S.~Perrot, V.~Brousse, A.~Sauvanet,
  B.~Aussilhou, S.~Dokmak, A.~Couvelard, D.~Cazals-Hatem,
  O.~Mercereau-Puijalon, G.~Milon, P.~H. David, N.~Mohandas, Sensing of red
  blood cells with decreased membrane deformability by the human spleen.
  \textit{Blood Adv.} \textbf{2}, 2581--2587 (2018).

\bibitem{li2021How}
H.~Li, Z.~L. Liu, L.~Lu, P.~Buffet, G.~E. Karniadakis, How the spleen reshapes
  and retains young and old red blood cells: A computational investigation.
  \textit{PLOS Comput. Biol.} \textbf{17}, e1009516 (2021).

\bibitem{demeter1989preferential}
J.~Demeter, Z.~Magyar, F.~Varga, M.~Szathm{\'a}ri, D.~Lehoczky, J.~Ellegaard,
  Preferential localisation of red cell vacuoles to pathologically shaped human
  red blood cells. \textit{Eur. J. Haematol.} \textbf{43}, 41--44 (1989).

\bibitem{namvar2021surface}
A.~Namvar, A.~J. Blanch, M.~W. Dixon, O.~M.~S. Carmo, B.~Liu, S.~Tiash,
  O.~Looker, D.~Andrew, L.-J. Chan, W.-H. Tham, P.~V.~S. Lee, V.~Rajagopal,
  L.~Tilley, Surface area-to-volume ratio, not cellular viscoelasticity, is the
  major determinant of red blood cell traversal through small channels.
  \textit{Cell. Microbiol.} \textbf{23}, e13270 (2021).

\bibitem{safeukui2012quantitative}
I.~Safeukui, P.~A. Buffet, G.~Deplaine, S.~Perrot, V.~Brousse, A.~Ndour,
  M.~Nguyen, O.~Mercereau-Puijalon, P.~H. David, G.~Milon, N.~Mohandas,
  Quantitative assessment of sensing and sequestration of spherocytic
  erythrocytes by the human spleen. \textit{Blood} \textbf{120}, 424--430
  (2012).

\bibitem{waugh1992rheologic}
R.~E. Waugh, M.~Narla, C.~W. Jackson, T.~J. Mueller, T.~Suzuki, G.~L. Dale,
  Rheologic properties of senescent erythrocytes: loss of surface area and
  volume with red blood cell age. \textit{Blood} \textbf{79}, 1351--1358
  (1992).

\bibitem{rawicz2000effect}
W.~Rawicz, K.~C. Olbrich, T.~McIntosh, D.~Needham, E.~Evans, Effect of chain
  length and unsaturation on elasticity of lipid bilayers. \textit{Biophys. J.}
  \textbf{79}, 328--339 (2000).

\bibitem{olbrich2000water}
K.~Olbrich, W.~Rawicz, D.~Needham, E.~Evans, Water permeability and mechanical
  strength of polyunsaturated lipid bilayers. \textit{Biophys. J.} \textbf{79},
  321--327 (2000).

\bibitem{griffiths2012maturing}
R.~E. Griffiths, S.~Kupzig, N.~Cogan, T.~J. Mankelow, V.~M.~S. Betin,
  K.~Trakarnsanga, E.~J. Massey, J.~D. Lane, S.~F. Parsons, D.~J. Anstee,
  Maturing reticulocytes internalize plasma membrane in glycophorin
  {A}-containing vesicles that fuse with autophagosomes before exocytosis.
  \textit{Blood} \textbf{119}, 6296--6306 (2012).

\bibitem{mankelow2015autophagic}
T.~J. Mankelow, R.~E. Griffiths, S.~Trompeter, J.~F. Flatt, N.~M. Cogan, E.~J.
  Massey, D.~J. Anstee, Autophagic vesicles on mature human reticulocytes
  explain phosphatidylserine-positive red cells in sickle cell disease.
  \textit{Blood} \textbf{126}, 1831--1834 (2015).

\bibitem{clark1988senescence}
M.~R. Clark, Senescence of red blood cells: progress and problems.
  \textit{Physiol. Rev.} \textbf{68}, 503--554 (1988).

\bibitem{mohandas2008red}
N.~Mohandas, P.~G. Gallagher, Red cell membrane: past, present, and future.
  \textit{Blood} \textbf{112}, 3939--3948 (2008).

\bibitem{picot2015biomimetic}
J.~Picot, P.~A. Ndour, S.~D. Lefevre, W.~El~Nemer, H.~Tawfik, J.~Galimand,
  L.~Da~Costa, J.-A. Ribeil, M.~de~Montalembert, V.~Brousse, B.~Le~Pioufle,
  P.~Buffet, C.~Le~Van~Kim, O.~Fran{\c{c}}ais, A biomimetic microfluidic chip
  to study the circulation and mechanical retention of red blood cells in the
  spleen. \textit{Am. J. Hematol.} \textbf{90}, 339--345 (2015).

\bibitem{gambhire2017high}
P.~Gambhire, S.~Atwell, C.~Iss, F.~Bedu, I.~Ozerov, C.~Badens, E.~Helfer,
  A.~Viallat, A.~Charrier, High aspect ratio sub-micrometer channels using wet
  etching: application to the dynamics of red blood cell transiting through
  biomimetic splenic slits. \textit{Small} \textbf{13}, 1700967 (2017).

\bibitem{qiang2023microfluidic}
Y.~Qiang, A.~Sissoko, Z.~L. Liu, T.~Dong, F.~Zheng, F.~Kong, J.~M. Higgins,
  G.~E. Karniadakis, P.~A. Buffet, S.~Suresh, M.~Dao, Microfluidic study of
  retention and elimination of abnormal red blood cells by human spleen with
  implications for sickle cell disease. \textit{Proc. Natl. Acad. Sci. U.S.A.}
  \textbf{120}, e2217607120 (2023).

\bibitem{chai2026multiscale}
Z.~Chai, N.~Ahmadi~Daryakenari, G.~E. Karniadakis, A multiscale
  signaling--biophysical framework reveals mechanisms of macrophage-mediated
  {RBC} clearance in sickle cell and {G}aucher disease. \textit{PNAS Nexus},
  pgag348 (2026).

\bibitem{toscano2026graftathena}
J.~D. Toscano, Z.~Chai, G.~E. Karniadakis, {GRAFT-ATHENA}: Self-Improving
  Agentic Teams for Autonomous Discovery and Evolutionary Numerical Algorithms.
  \textit{arXiv preprint arXiv:2605.11117} (2026).

\bibitem{thompson2022lammps}
A.~P. Thompson, H.~M. Aktulga, R.~Berger, D.~S. Bolintineanu, W.~M. Brown,
  P.~S. Crozier, P.~J. in~'t Veld, A.~Kohlmeyer, S.~G. Moore, T.~D. Nguyen,
  R.~Shan, M.~J. Stevens, J.~Tranchida, C.~Trott, S.~J. Plimpton, LAMMPS -- a
  flexible simulation tool for particle-based materials modeling at the atomic,
  meso, and continuum scales. \textit{Comput. Phys. Commun.} \textbf{271},
  108171 (2022).

\bibitem{Fedosov2011Predicting}
D.~A. Fedosov, W.~Pan, B.~Caswell, G.~Gompper, G.~E. Karniadakis, Predicting
  human blood viscosity in silico. \textit{Proc. Natl. Acad. Sci. U.S.A.}
  \textbf{108}, 11772--11777 (2011).

\end{thebibliography}
\endgroup

\section*{Acknowledgments}
We thank Pierre A. Buffet for discussions of pocked red cells and splenic filtration. Simulations were carried out at the Center for Computation and Visualization of Brown University.

\paragraph*{Funding:}
Research reported in this publication was supported by the National Heart, Lung, and Blood Institute (NHLBI) and the National Institute of General Medical Sciences (NIGMS) of the National Institutes of Health under award numbers R01HL154150 (G.E.K.) and R01GM163243 (H.L.). Award R01HL154150 totals \$1,269,793 to date and award R01GM163243 totals \$394,808 to date, each with 100\% from federal sources. The content is solely the responsibility of the authors and does not necessarily represent the official views of the National Institutes of Health.

\paragraph*{Author contributions:}
Conceptualization: Z.C., G.E.K.\\
Methodology: Z.C.\\
Software: Z.C.\\
Formal analysis: Z.C.\\
Investigation: Z.C., J.Z.\\
Data curation: Z.C., J.Z.\\
Visualization: Z.C.\\
Resources: M.D.\\
Supervision: M.D., G.E.K.\\
Funding acquisition: H.L., M.D., G.E.K.\\
Writing---original draft: Z.C.\\
Writing---review \& editing: Z.C., J.Z., H.L., M.D., G.E.K.

\paragraph*{Competing interests:}
The authors declare they have no competing interests.

\paragraph*{Data, code, and materials availability:}
All data needed to evaluate the conclusions in the paper are present in the paper and/or the Supplementary Materials. The simulations used an in-house extension of LAMMPS~\cite{thompson2022lammps}; every model equation and parameter is given in Materials and Methods and the Supplementary Materials. This study did not generate any new materials.

\section*{Supplementary Materials}
\noindent The Supplementary Materials follow on the next pages: Supplementary Text, Figs.~\smref{fig:tubeshear} and~\smref{fig:dotstrips}, Tables~\smref{tab:runs} to~\smref{tab:dots} and captions for Movies S1 and S2. Movies S1 and S2 are available as ancillary files of this article.

\clearpage
\setcounter{section}{0}\setcounter{figure}{0}\setcounter{table}{0}\setcounter{equation}{0}
\renewcommand{\thesection}{S\arabic{section}}
\renewcommand{\thefigure}{S\arabic{figure}}
\renewcommand{\thetable}{S\arabic{table}}
\renewcommand{\theequation}{S\arabic{equation}}
\renewcommand{\theHsection}{S\arabic{section}}
\renewcommand{\theHfigure}{S\arabic{figure}}
\renewcommand{\theHtable}{S\arabic{table}}
\renewcommand{\theHequation}{S\arabic{equation}}
\singlespacing
\setlength{\parindent}{0pt}\setlength{\parskip}{6pt}
\begin{center}{\Large\bfseries Supplementary Materials}\end{center}

\section*{Supplementary Text}

This document supports the main text in four parts. The first sets out the simulation method: the dissipative particle dynamics (DPD) formulation and pair coefficients (section~\ref{sec:dpd}); the mapping to physical units, with the viscosity calibration and the time scale of each geometry (section~\ref{sec:units}); the red blood cell (RBC) and vacuole models (section~\ref{sec:models}); and how the flows are driven and how vacuoles are kept neutrally buoyant (section~\ref{sec:drive}). The second describes how the runs were analysed and checked: the fusion algorithm and a table of every cascade run (section~\ref{sec:fusion}); the contact analysis and its threshold sweep (section~\ref{sec:contact}); and the containment, temperature and quality checks applied to every run (section~\ref{sec:checks}). The third goes through the simulations themselves: the geometries (section~\ref{sec:geometry}); the slit runs (section~\ref{sec:slit}); the filter runs (section~\ref{sec:filter}); the runs of repeated passage, with and without fusion (section~\ref{sec:passage}); and the Reynolds and capillary numbers, with a Navier--Stokes check of the flow in the dilation's sac (section~\ref{sec:reynolds}). The fourth covers the population model and the data: the numerical implementation, scan and scoring of the model, with the best point of every ranking and the scan's sensitivity to the range of the clearance rise and to the DIC limit (section~\ref{sec:pop}); the published measurements plotted in Fig.~\mref{fig:popmodel}, with their sources (section~\ref{sec:fig5data}); the unstained dots seen in reticulocytes and in cells at rest (section~\ref{sec:dots}); and the limitations of the model, set out in full (section~\ref{sec:limits}). Table~\ref{tab:runs} lists every simulation reported in this work.

\begin{table}[htbp]
\small
\centering
\renewcommand\arraystretch{1.2}
\caption{\textbf{Simulations reported in this work.}}
\label{tab:runs}
\resizebox{\textwidth}{!}{
\begin{tabular}{llllll}
\toprule
Simulation & Cells; vacuoles & Drive & $\tau$ (ms per $10^3$ steps) & Steps (duration) & Runs \\
\midrule
Encounter tube & $1$; $10\times0.6~\mu$m & $f = 0.45$ & $3.73$ & $200\,000$ ($745$--$747$~ms) & $3$ \\
Plane shear & $1$; $10\times0.6~\mu$m & walls $\pm463~\mu$m/s & $6.50$ & $200\,000$ ($1296$--$1302$~ms) & $3$ \\
Survey tubes & $1$; $10\times0.6~\mu$m & $f = 0.28$ / $2.5$ & $3.75$ / $3.74$ & $150\,000$ ($562$ / $561$~ms) & $1$ each \\
Bifurcation & $80$; $762$ & $f = 0.5$ & $8.04$ & $150\,000$ ($1206$~ms) & $1$ \\
Curved vessel & $233$; $2301$ & $f = 0.5$ & $5.23$ & $150\,000$ ($785$~ms) & $1$ \\
Dilation & $90$ with $856$; $200$ without & $f = 0.5$ & $5.05$ & $150\,000$ ($758$~ms) & $1$ \\
Fusion cascade & $1$; $10\times0.6~\mu$m & $f = 1.0$ & $3.89$ & $\sim\!60\,000$ ($231$--$234$~ms) & $10$ \\
Slit & $1$; $1\times0.8$ or $1.4~\mu$m & $f = 8$ / $4$, cell only & $3.70$ & $102\,000$ / $152\,000$ ($377$ / $562$~ms) & $54$ \\
Filter, one vacuole & $1$; none or $1\times1.0$--$2.4~\mu$m & $f = 3$ & $3.70$ & $152\,000$ ($562$~ms) & $43$ \\
Filter, eight vacuoles & $1$; $8\times1.1~\mu$m & $f = 3$ & $3.70$ & $152\,000$ ($562$~ms) & $17$ \\
Filter, displaced-fluid drive & $1$; $1\times1.4$ or $2.2~\mu$m, trailing or leading & $f = 3$ & $3.70$ & $152\,000$ ($562$~ms) & $12$ \\
Repeated passage & $1$; $10\times0.6~\mu$m & $f = 8$, cell only, or none & $3.70$ & $302\,000$ ($1116$~ms) & $9$ \\
Repeated passage, fusion & $1$; $10\times0.6~\mu$m & $f = 8$, cell only, or none & $3.70$ & $604\,000$ at $\Delta t = 0.0005$ ($1116$~ms) & $18$ \\
\bottomrule
\end{tabular}}
\\[4pt]
\parbox{0.97\textwidth}{\footnotesize Durations are the number of steps times the $\tau$ of that geometry, which follows from its solvent density (main text, Materials and Methods; section~\ref{sec:units}) and carries an uncertainty of about $10\%$. The slit series comprises $36$ runs with the vacuole trailing and $18$ with it leading. The filter series comprises $22$ runs at the default volume constraint and $21$ at the stiffened ones; the eight-vacuole series comprises $17$ runs across four settings of the vacuole--vacuole pair, the membrane--vacuole cutoff and the volume constraint. The cascade runs proceed in segments and end when the cascade completes or a vacuole mesh fails. The repeated-passage runs comprise three arms, squeezed through the slit, driven through a wide opening and at rest, with three seeds each without fusion and six with it, three under each area rule. The runs with fusion take half the time step, so their duration is half their steps times $\tau$. Five of the six driven runs with summed areas ended early on a vacuole-mesh bond (section~\ref{sec:fusion} and table~\ref{tab:passfusion}).}
\end{table}

\section{Dissipative particle dynamics}\label{sec:dpd}

The force on particle $i$ is the sum, over neighbours $j$ within the cutoff $r_c$, of conservative, dissipative and random pair forces~\cite{hoogerbrugge1992simulating,espanol1995statistical,groot1997dissipative},
\begin{align}
\mathbf{F}_{ij}^{\mathrm{C}} &= a_{ij}\left(1-r_{ij}/r_c\right)\mathbf{e}_{ij}, \\
\mathbf{F}_{ij}^{\mathrm{D}} &= -\gamma_{ij}\,\omega_{\mathrm{D}}(r_{ij})\,(\mathbf{e}_{ij}\cdot\mathbf{v}_{ij})\,\mathbf{e}_{ij}, \\
\mathbf{F}_{ij}^{\mathrm{R}} &= \sigma_{ij}\,\omega_{\mathrm{R}}(r_{ij})\,\xi_{ij}\,\Delta t^{-1/2}\,\mathbf{e}_{ij},
\end{align}
where $r_{ij}=\|\mathbf{r}_i-\mathbf{r}_j\|$, $\mathbf{e}_{ij}=(\mathbf{r}_i-\mathbf{r}_j)/r_{ij}$, $\mathbf{v}_{ij}=\mathbf{v}_i-\mathbf{v}_j$, and $\xi_{ij}$ is a Gaussian random number of zero mean and unit variance drawn for each pair at each step. The dissipative and random weights are
\begin{equation}
\omega_{\mathrm{D}}(r) = \left[\omega_{\mathrm{R}}(r)\right]^2 = \left(1-r/r_c\right)^{2k}, \qquad k = 0.25,
\end{equation}
while the conservative weight stays linear. The fluctuation--dissipation theorem ties the two amplitudes to the temperature of the bath, $\sigma_{ij}^2 = 2\gamma_{ij}k_{\mathrm{B}}T$~\cite{groot1997dissipative}. The input specifies $a$, $\gamma$ and $\sigma$ for every pair (table~\ref{tab:pairs}); every pair satisfies this relation with $k_{\mathrm{B}}T = 0.1884$, which is therefore the bath temperature in reduced units. All particles have unit mass, the time step is $\Delta t = 0.001$, and the equations of motion are integrated with a modified velocity-Verlet scheme~\cite{groot1997dissipative} in an extended LAMMPS solver~\cite{fedosov2010multiscale,thompson2022lammps}. The solvent starts at half the bath temperature.

Plasma and cytosol are the same DPD fluid. Walls and beads are DPD particles that are not integrated. They interact with the solvent, the membranes and the vacuoles through the pair forces of table~\ref{tab:pairs}; pairs of wall particles are excluded from the neighbour list, and no bounce-back condition is applied. The walls of the plane-shear cell are the exception: their particles move at a prescribed velocity (section~\ref{sec:geometry}).

\begin{table}[htbp]
\small
\centering
\renewcommand\arraystretch{1.25}
\caption{\textbf{DPD pair coefficients in reduced units.} Every pair uses the weight exponent $k = 0.25$ and satisfies $\sigma^2 = 2\gamma k_{\mathrm{B}}T$ with $k_{\mathrm{B}}T = 0.1884$.}
\label{tab:pairs}
\begin{tabular}{p{6.6cm}ccccp{2.8cm}}
\toprule
Pair & $a$ & $\gamma$ & $\sigma$ & $r_c$ & Used in \\
\midrule
All pairs not listed below (solvent, membrane, wall and bead particles) & 30 & 15 & 2.3774 & 1.0 & all \\
Membrane vertex--vacuole vertex (barrier) & 600 & 0.4 & 0.3882 & 0.6 or 0.7 & all \\
Vertices of different vacuoles (barrier) & 600 & 0.4 & 0.3882 & 0.5 & several vacuoles per cell \\
Vertices of the same vacuole & \multicolumn{4}{c}{excluded from the neighbour list} & all \\
Vacuole vertex--wall or bead particle & 600 & 20 & 2.7452 & 1.0 & filter \\
Membrane vertex--wall particle & 60 & 20 & 2.7452 & 1.0 & survey tubes, vessels \\
Wall--wall & \multicolumn{4}{c}{excluded from the neighbour list} & all \\
\bottomrule
\end{tabular}
\\[4pt]
\parbox{0.97\textwidth}{\footnotesize The cutoff of the membrane--vacuole barrier is the smallest that keeps every vacuole vertex inside its host at the strongest drive used. At $0.5~\mu$m a vacuole escapes while the cell is squeezed in the pore, because the membrane triangles there are stretched until their centroids lie further than $0.5~\mu$m from all three vertices; $0.6$ and $0.7~\mu$m both contain it. The cutoff also sets the gap the barrier holds open around a vacuole---$0.45$, $0.54$ and $0.64~\mu$m at cutoffs of $0.5$, $0.6$ and $0.7~\mu$m---and that gap adds to the size that has to pass a constriction, so the smaller is preferred where containment allows. A cutoff of $0.6~\mu$m is used in the encounter tube, the plane-shear cell, the survey tubes, the vessel simulations, the cascade and part of the filter series, and $0.7~\mu$m in the rest of the filter series. Vacuoles repel one another through a barrier of the same form, so that a vacuole is no softer against another vacuole than against its host membrane; the closest approach observed between two vacuole vertices is then $0.43~\mu$m, against $0.76~\mu$m under the ordinary solvent repulsion. One filter series replaces that barrier by the plain soft pair ($a = 30$, $\gamma = 15$, $\sigma = 2.3774$, $r_c = 1$) as a control. In the vessel simulations the vacuole--vacuole pair is the soft one while the system is prepared and is switched to the barrier at the last equilibration stage, immediately before production (section~\ref{sec:contact}). In the membrane--wall row, the wall particles of the survey tubes and the vessel simulations carry a stronger repulsion than the generic pair; the encounter tube, the plane-shear cell and the filter use the generic $a = 30$.}
\end{table}

\section{Units, the viscosity calibration and the time scale of each geometry}\label{sec:units}

One reduced length unit is $L_0 = 1~\mu$m. Identifying the bath temperature with $k_{\mathrm{B}}T = 4.14\times10^{-21}$~J at $37^\circ$C gives the energy unit $E_0 = 4.14\times10^{-21}~\mathrm{J}/0.1884 = 2.197\times10^{-20}$~J. Moduli per unit length map with $E_0/L_0^2 = 0.02197~\mu$N/m, bending energies with $E_0$, pressures with $E_0/L_0^3 = 0.02197$~Pa and pressure gradients with $E_0/L_0^4 = 0.02197$~Pa/$\mu$m. The time unit follows from matching the viscosity of the DPD fluid to that of plasma, $\eta_p = 1.2$~mPa\,s~\cite{Fedosov2011Predicting}:
\begin{equation}
\tau = \frac{\eta_p L_0^3}{\eta^{\mathrm{DPD}} E_0} = \frac{54.6~\mathrm{ms}}{\eta^{\mathrm{DPD}}},
\end{equation}
where $\eta^{\mathrm{DPD}}$ is the viscosity of the DPD fluid in reduced units and one reduced time unit is $10^3$ steps. Durations, velocities and shear rates are converted with the $\tau$ of their own geometry.

\textit{Viscosity of the DPD fluid.} $\eta^{\mathrm{DPD}}$ depends on the solvent density $\rho$, which differs between geometries, so we measured it by plane Poiseuille flow at six densities. Particles on an fcc lattice filled a periodic box of $15a\times24a\times15a$, where $a$ is the lattice constant. Particles with $y < 2a$ or $y > 22a$ were frozen, forming a wall of the same lattice across the periodic boundary and leaving a channel $20a$ wide. A body force $g = 0.05$ per particle drove the fluid along $x$ for $10^6$ steps, and the velocity profile was averaged in bins of $0.25$ from step $3\times10^5$ on. Momentum balance, $\eta^{\mathrm{DPD}}\,\mathrm{d}^2u/\mathrm{d}y^2 = -\rho g$, gives the viscosity from the curvature of a quadratic fit to the profile, independently of slip at the wall (table~\ref{tab:visc}). For $\rho \geq 3$ we use
\begin{equation}
\eta^{\mathrm{DPD}} = 0.840\,\rho^{1.958}.
\end{equation}
This law reproduces the measured viscosities within $4\%$ when $3$--$4~\mu$m next to each wall, where the frozen particles structure the fluid, are left out of the fit. Fits reaching closer to the wall give viscosities up to $6\%$ lower: margins of $2$, $3$ and $4~\mu$m give $\tau = 3.760$, $3.692$ and $3.630$~ms at $\rho = 4.33$, a spread of $\pm1.8\%$ about the central value, and a margin of $5~\mu$m gives $3.564$~ms, $3.5\%$ below it. The last three running averages, at $8$, $9$ and $10\times10^5$ steps, agree within $0.1\%$.

\begin{table}[htbp]
\small
\centering
\renewcommand\arraystretch{1.25}
\caption{\textbf{Viscosity of the DPD fluid from plane Poiseuille flow.} The bulk density is the particle density more than $1.5~\mu$m from the walls. The two viscosity columns differ in the margin, in $\mu$m, left out of the quadratic fit next to each wall.}
\label{tab:visc}
\begin{tabular}{cccccc}
\toprule
Lattice density & Bulk density & $\eta^{\mathrm{DPD}}$, margin $0.5$ & $\eta^{\mathrm{DPD}}$, margin $4$ & $0.840\,\rho^{1.958}$ & $\tau$ from the law (ms) \\
\midrule
1.8 & 1.805 & 3.25 & 3.86 & -- & -- \\
2.2 & 2.191 & 5.17 & 6.00 & -- & -- \\
3.0 & 2.991 & 6.78 & 7.13 & 7.22 & 7.56 \\
4.0 & 3.991 & 12.29 & 13.19 & 12.68 & 4.31 \\
6.0 & 5.992 & 26.42 & 28.29 & 28.05 & 1.95 \\
7.6 & 7.591 & 42.44 & 44.47 & 44.56 & 1.23 \\
\bottomrule
\end{tabular}
\end{table}

\textit{Time scale of each geometry.} We measured the solvent density of every geometry and converted it to a viscosity, and hence to $\tau$, in one of two ways (table~\ref{tab:tau}). The filter, the encounter simulations and the cascade use the law; the survey tubes and the vessels use a linear interpolation of the six measured viscosities of table~\ref{tab:visc} in $\log\eta$ against $\log\rho$. Only the bifurcation lies below the range of the fit, where the law fails (at the two lowest calibration densities it underestimates $\eta$ by about a third); applying the law there would lengthen $\tau$ by $8\%$. Everywhere else the two conversions agree closely. Applying the interpolation to the simulations that use the law would change $\tau$ by at most $4\%$ (encounter tube $3.73 \to 3.68$~ms, plane shear $6.50 \to 6.76$~ms). Applying the law to the survey tubes and the other vessels would lengthen it by $1$--$3\%$.

\begin{table}[htbp]
\small
\centering
\renewcommand\arraystretch{1.25}
\caption{\textbf{Solvent density and time scale of each geometry, with the method used to convert $\rho$ to $\eta^{\mathrm{DPD}}$.}}
\label{tab:tau}
\begin{tabular}{lccp{6.4cm}}
\toprule
Geometry & $\rho$ & $\tau$ (ms per $10^3$ steps) & Method and region \\
\midrule
Microsphere filter & 4.325 & 3.695 & law; box mean over the solvent-accessible volume. The drive builds a strong axial gradient, from $\rho \approx 8$ against the upstream face of the bead layer to $\approx 0.5$ within it and $\approx 3.3$ downstream, so no single region is representative \\
Encounter tube & $\approx 4.30$ & $3.733$, $3.727$, $3.734$ & law; lumen core, $r < 4~\mu$m about the axis, clear of the solvent seeded in the box corners; per seed \\
Plane shear & $\approx 3.24$ & $6.480$, $6.508$, $6.500$ & law; solvent in the gap, away from the cell; per seed \\
Fusion cascade & 4.21 & 3.89 & law; lumen core as in the encounter tube \\
Survey tubes & 4.239, 4.246 & 3.750, 3.738 & interpolation; lumen core, $r < 4~\mu$m, steps $110\,000$--$150\,000$ \\
Bifurcation & 2.79 & 8.04 & interpolation; solvent inside the lumen, in the two unambiguous single-channel spans ($x < 35$ and $x > 100~\mu$m), after the temperature plateaus at step $12\,000$ \\
Curved vessel & 3.60 & 5.23 & interpolation; solvent inside the lumen, within $15.0~\mu$m of the vessel's centreline, the radius of the innermost wall particles; quoted at step $150\,000$. Solvent crosses this frozen wall during the run: the lumen holds $98.4\%$ of it at step $0$ and $87.8\%$ at step $150\,000$, and a median over all of it, $3.43$, would lengthen $\tau$ by $11\%$. The lumen's own density falls as solvent leaves, from $3.86$ at step $0$, where $\tau$ would be $4.5$~ms \\
Dilation, lumen & 3.66 & 5.05 & interpolation; solvent inside the parent lumen, $r < 12.5~\mu$m about its axis; quoted at step $120\,000$, and $4.93$ at step $150\,000$ \\
\quad\quad sac & 3.71 & 4.90 & interpolation; solvent at $r \geq 12.5~\mu$m and $x = 55$--$106~\mu$m, at step $120\,000$; the mean axial speed here is $3\%$ of the lumen's, $30$ against $933~\mu$m/s \\
\bottomrule
\end{tabular}
\\[4pt]
\parbox{0.97\textwidth}{\footnotesize Where the density is measured matters as much as how it is converted. Measuring $\rho$ only in solvent more than $1.5~\mu$m from any membrane vertex, rather than over the whole flow region, raises it and shortens $\tau$. In the first seed of the encounter tube it goes from $4.2307$ to $4.4907 \pm 0.0179$, that is $\tau$ from $3.858$ to $3.433$~ms ($-11\%$). In the first seed of the plane-shear simulation it goes from $3.1963$ to $3.4767 \pm 0.0144$, that is $\tau$ from $6.680$ to $5.666$~ms ($-15\%$). The values in the table sit between these two conventions, and the second would shorten them by $8$ and $13\%$; we treat $\tau$ as uncertain by about $10\%$ throughout. This affects every physical duration but no dimensionless comparison, and it is common to all simulations, so it does not change any comparison between them.}
\end{table}

\section{Red blood cell and vacuole models}\label{sec:models}

\textit{Cell membrane.} Each RBC membrane is a triangulated network of $N_v = 500$ vertices~\cite{fedosov2010multiscale,li2014erythrocyte} with energy
\begin{equation}
U = U_{\mathrm{WLC}} + U_{\mathrm{POW}} + U_{\mathrm{bend}} + U_{\mathrm{area}} + U_{\mathrm{vol}}.
\end{equation}
Every edge is a worm-like-chain spring combined with a power-law repulsion,
\begin{equation}
U_{\mathrm{WLC}} = \frac{k_{\mathrm{B}}T_s\, l_{\max}}{4\lambda}\,\frac{3x^2-2x^3}{1-x}, \quad x = \frac{l}{l_{\max}}, \qquad U_{\mathrm{POW}} = \frac{k_p}{(m-1)\,l^{m-1}},
\end{equation}
with $l_{\max} = 2.2\,l_0$, where $l_0$ is the length of the edge in the initial mesh, $k_{\mathrm{B}}T_s = 0.093$ and $m = 2$. For each spring, $k_p$ is first chosen so that the spring exerts no force at $l_0$. The linear shear modulus of the network is then~\cite{fedosov2010multiscale}
\begin{equation}
\mu = \frac{\sqrt{3}\,k_{\mathrm{B}}T_s}{4\lambda\, l_0}\left[\frac{x_0}{2(1-x_0)^3} - \frac{1}{4(1-x_0)^2} + \frac{1}{4}\right] + \frac{\sqrt{3}\,(m+1)\,k_p}{4\,l_0^{m+1}}, \qquad x_0 = \frac{l_0}{l_{\max}},
\end{equation}
and the WLC and power-law terms are scaled by a common factor so that $\mu$ equals the target $\mu_0$. The springs carry no dissipative or random force of their own. Bending acts on the dihedral angle $\theta_j$ between adjacent triangles, $U_{\mathrm{bend}} = \sum_j k_b\left[1-\cos(\theta_j-\theta_0)\right]$ with $\theta_0 = 0$, which corresponds to a continuum bending modulus $k_c = \sqrt{3}\,k_b/2$, so that $k_b = 2k_c/\sqrt{3}$~\cite{fedosov2010multiscale}. Area and volume are constrained by
\begin{equation}
U_{\mathrm{area}} = \frac{k_a (A-A_0)^2}{2A_0} + \sum_{t} \frac{k_d (A_t-A_{t,0})^2}{2A_{t,0}}, \qquad U_{\mathrm{vol}} = \frac{k_v (V-V_0)^2}{2V_0},
\end{equation}
where $A_{t,0}$ is the area of triangle $t$ in the initial mesh. Table~\ref{tab:membrane} lists the values.

\textit{Vacuoles.} Each vacuole is a closed triangulated vesicle with the same energy terms. The single vacuoles of the filter series use $154$ vertices with $A_0 = \pi D^2$ and $V_0 = \pi D^3/6$; across that series the median as-run diameter is within $0.004~\mu$m of nominal at every size from $1.0$ to $2.4~\mu$m. Simulations with several vacuoles per cell use $34$-vertex meshes: $A_0 = 1.13$ and $V_0 = 0.11$ at $D = 0.6~\mu$m, a reduced volume of $0.97$, and $A_0 = 3.80$ and $V_0 = 0.697$ at $D = 1.1~\mu$m. The coarser mesh does not enclose the full target volume, so the eight $D = 1.1~\mu$m vacuoles start about $15\%$ below it; volume ratios for that series are therefore quoted against the first frame of the run rather than against the target sphere (section~\ref{sec:filter}).

The resting cell also limits how large a vacuole it can hold. Placed in the rim, a vacuole of $D = 2.45~\mu$m is still fully contained while $D = 2.50~\mu$m is not (fraction of vertices inside $0.987$, falling to $0.766$ at $D = 2.70~\mu$m). At the centre of the dimple even $D = 2.40~\mu$m is only half contained ($0.519$), so a large vacuole can be hosted only in the rim. The filter series therefore spans the full range of sizes a resting cell can carry.

\textit{Interactions.} Membranes and vacuoles interact with the solvent and with each other only through the pair forces of table~\ref{tab:pairs}. There is no adhesion between cells, between vacuoles, or between a vacuole and its host membrane. The membrane--vacuole barrier keeps every vacuole vertex inside the host, which we check in every frame (section~\ref{sec:checks}); its range leaves a gap of $0.55$--$0.75~\mu$m between membrane and vacuole vertices in free flow, narrowing to $0.12~\mu$m where the largest vacuole is squeezed through the pore. The membrane is impermeable to solvent at rest and exchanges it symmetrically during a squeeze (section~\ref{sec:checks}). Because cytosol and plasma are the same fluid, the higher viscosity of the cytosol in vivo is not represented.

\begin{table}[htbp]
\small
\centering
\renewcommand\arraystretch{1.25}
\caption{\textbf{Membrane and vacuole parameters.} Physical values use $E_0 = 2.197\times10^{-20}$~J and $L_0 = 1~\mu$m (section~\ref{sec:units}).}
\label{tab:membrane}
\begin{tabular}{p{6.0cm}p{5.0cm}p{4.0cm}}
\toprule
Parameter & Reduced & Physical \\
\midrule
\multicolumn{3}{l}{\textit{Cell membrane}}\\
Vertices $N_v$ & 500 & -- \\
Area $A_0$; volume $V_0$ & 132.87; 92.46 & $132.9~\mu$m$^2$; 92.5~fL \\
$k_{\mathrm{B}}T_s$; $l_{\max}/l_0$; $m$ & 0.093; 2.2; 2 & -- \\
Shear modulus $\mu_0$ & 215 & $4.73~\mu$N/m \\
Bending constant $k_b$ & 12.6 & $k_c = 2.4\times10^{-19}$~J \\
$k_a$; $k_d$; $k_v$ & 4900; 100; 5000 & -- \\
\multicolumn{3}{l}{\textit{Vacuole membrane}}\\
Vertices & 154 (one vacuole per cell); 34 (several) & -- \\
Area $A_0$; volume $V_0$ & $\pi D^2$; $\pi D^3/6$ (154 vertices); 1.13; 0.11 at $D = 0.6~\mu$m and 3.80; 0.697 at $D = 1.1~\mu$m (34 vertices) & -- \\
Shear modulus $\mu_0$ & 120 & $2.6~\mu$N/m \\
Bending constant $k_b$ & 6.025 & $k_c = 1.1\times10^{-19}$~J \\
$k_a$; $k_d$; $k_v$ & 4900; 100; 5000, and $5\times10^4$ or $5\times10^5$ in the filter controls & -- \\
\bottomrule
\end{tabular}
\end{table}

\section{Driving the flow and the neutral buoyancy of vacuoles}\label{sec:drive}

In the filter, the tubes and the vessel simulations, a uniform body force $f$ along the flow acts on every solvent particle and every membrane vertex. In a fluid of density $\rho$ this is a pressure gradient $\rho f$ in reduced units, or $0.02197\,\rho f$~Pa/$\mu$m. A vacuole displaces fluid that would receive a force $\rho V_0 f$, so each of its $N$ vertices receives
\begin{equation}
f_v = \frac{\rho V_0 f}{N}
\end{equation}
and the vacuole is neutrally buoyant with respect to the drive. The reference densities used to set $f_v$ are $4.29$ in the encounter tube, $4.59$ in the cascade, $4.325$ in the filter and $4.73$ in the vessel simulations. The resulting values are $f_v = 0.006418$ in the encounter tube, $0.003781$ in the survey tube at $f = 0.28$, $0.007651$ in the vessels and $0.265955$ for the eight $D = 1.1~\mu$m vacuoles of the filter ($3 \times 4.325 \times 0.697/34$). Applying the full $f$ to every vacuole vertex instead would put $Nf$ on the vacuole, $68$ times the neutral force at $D = 1.0~\mu$m, and drag it forward through the cytosol. In a pilot run done that way a $D = 1.0~\mu$m vacuole rode $5.0~\mu$m ahead of the cell centroid, whereas with the neutral force it sits $3.6~\mu$m behind it (medians over the run). The transit time of the cell itself was similar in the two. In the plane-shear cell the walls drive the flow and no body force acts.

\section{Fusion}\label{sec:fusion}

The rules below are those implemented in the driver used for the production runs, which was validated on forced merges, on a natural cascade and on a third-generation product, in every case with all vacuoles contained in their host and no bond overstretched. A run proceeds in segments of $140$, $440$ and $1520$ steps for $t_f = 1.2$, $3.6$ and $11.9$~ms, each a little under half of $t_f$. Between segments the driver reads the vacuole trajectories, decides which pairs fuse, rebuilds the topology and restarts the run.

\textit{Contact and persistence.} Two vacuoles are in contact while the smallest distance between a vertex of one and a vertex of the other is below $\delta_f$, evaluated every $20$ steps. A pair fuses once it has been in contact without interruption for $t_f$; one frame out of contact resets its clock. The driver converted $t_f$ to steps with a nominal time scale of $3.284$~ms per $1000$ steps and rounded up to the sampling interval, giving $320$, $920$ and $3060$ steps. On the measured time scale of that simulation, $3.89$~ms per $1000$ steps (section~\ref{sec:units}), these are $1.2$, $3.6$ and $11.9$~ms. We quote $t_f$, like every other time, on the measured scale. A contact that spans the end of a segment continues only if the pair is in contact in the last frame of that segment and in the first frame of the next. A pair that has reached $t_f$ but is out of contact in the last frame does not fuse and is counted as missed. Across the ten runs four fusions were missed in this way, three of them in one run at $\delta_f = 0.5~\mu$m, $t_f = 1.2$~ms and one at $\delta_f = 0.5~\mu$m, $t_f = 3.6$~ms; none was missed at $t_f = 11.9$~ms. Because the barrier keeps vertices of different vacuoles about $0.43~\mu$m apart, $\delta_f$ was swept over $0.5$, $0.7$ and $0.9~\mu$m.

\textit{Spin-up of the drive.} The body force starts from rest, so the early fusions happen in a flow that is still accelerating. Averaging the axial velocity over every frame of \texttt{state.lammpstrj}, which carries velocities every $140$, $440$ or $1520$ steps at $t_f = 1.2$, $3.6$ and $11.9$~ms, the solvent reaches half of its terminal mean speed at $11.5$--$11.8$~ms and $95\%$ of it at $112$--$122$~ms, and the cell at $8.6$--$9.4$ and $47$--$62$~ms. Those ranges are over the six runs that carry the full $233$~ms; the terminal speed has to be read from a record that reaches the plateau, and taking it from the last third of a truncated run puts the $95\%$ crossing far too early.

\textit{Eligibility.} Both vacuoles must lie inside the same cell, with at least $99\%$ of their vertices inside its membrane (section~\ref{sec:checks}).

\textit{Product.} The product has the volume of its parents, $V_0 = V_a + V_b$, and their membrane area, $A_0 = A_a + A_b$. Area-conserving products are not spherical: fusing $k$ base vacuoles gives a reduced volume of $0.97/\sqrt{k}$. As a sensitivity test, a second rule gives the product the area of a sphere of volume $V_0/0.97$, keeping it round. Compared on the validation merges, the two rules fused the same pair at the same step and differed only in the product (at $k = 2$, area $2.26~\mu$m$^2$, radius $0.424~\mu$m and reduced volume $0.69$ against $1.80~\mu$m$^2$, $0.378~\mu$m and $0.97$, with the same volume). Whole cascades were run under both rules only in the repeated-passage campaign, three runs per rule in each of its three arms (Fig.~\mref{fig:passage}D). There the choice mattered for the cells at rest, which stopped at three vacuoles with summed-area products, each still holding qualified pairs whose products could not be placed, but went on to one or two with near-spherical ones. Driven cells reached one or two under either rule, and five of the six driven summed-area runs then failed on a vacuole-mesh bond. The cascade of table~\ref{tab:cascade} was run under the summed-area rule only, so its floor of four or five vacuoles belongs to that rule. The product mesh has $N = \mathrm{round}(34\,A_0/1.13)$ vertices placed quasi-uniformly on a sphere of area $A_0$ and triangulated by their convex hull, and starts at the volume-weighted centroid of its parents with their mass-weighted mean velocity. If a product vertex would lie outside the host, or closer to a membrane vertex than a clearance of half the membrane--vacuole cutoff, the product is moved towards the host centroid in steps of $0.1~\mu$m, up to $1~\mu$m. If it still does not fit, the fusion is deferred to the next segment boundary and retried. Solvent particles inside the product are removed and the same number is inserted at random positions inside the host, at least $0.4~\mu$m from every other particle and with the velocities of the removed particles, so the cytosol is conserved. Vacuoles never divide.

\textit{The runs.} Table~\ref{tab:cascade} lists all ten. Deferral, not the supply of contacts, is what ends a cascade: the five-parent fusion waited $28.9$--$100.4$~ms after qualifying in the six runs that executed it, and in two runs it never executed, the driver retrying $26$ and $421$ times to the end of the record. Four runs stopped early when a worm-like-chain bond of a \emph{vacuole} mesh exceeded its maximum extension; in two of them this happened $0.4$ and $0.5$~ms after the five-parent fusion, in a third $2.0$~ms after four fusions executed together, and in the fourth $35$~ms after the last fusion. The host membrane was intact in all four. Because the worm-like chain has no rupture criterion, such a failure is a breakdown of the model rather than a prediction about membranes.

\begin{table}[htbp]
\small
\centering
\renewcommand\arraystretch{1.2}
\caption{\textbf{The ten fusion cascade runs.} Times use $\tau = 3.89$~ms per $10^3$ steps. ``Due'' is the time at which a pair had been in contact for $t_f$; ``executed'' the time at which its fusion was carried out; the maximum wait is the largest gap between the two in that run. The composition column gives the number of original vacuoles in each body at the end of the run.}
\label{tab:cascade}
\resizebox{\textwidth}{!}{\begin{tabular}{lccccccccl}
\toprule
Run & $\delta_f$ ($\mu$m) & $t_f$ (ms) & Complete & End (ms) & Fusions & Last due & Last executed & Max wait & Composition at the end \\
\midrule
$0.5$ / $1$ & 0.5 & 1.2 & no & 67.9 & 6 & 29.3 & 67.5 & 38.3 & 5, 2, 2, 1 \\
$0.5$ / $3$ & 0.5 & 3.6 & yes & 232.8 & 6 & 18.0 & 73.6 & 55.6 & 5, 2, 2, 1 \\
$0.5$ / $10$ & 0.5 & 11.9 & yes & 230.6 & 5 & 31.6 & 41.4 & 9.8 & 3, 2, 2, 2, 1 \\
$0.7$ / $1$ & 0.7 & 1.2 & no & 145.3 & 6 & 10.1 & 110.6 & 100.4 & 5, 2, 2, 1 \\
$0.7$ / $3$ & 0.7 & 3.6 & yes & 232.8 & 6 & 13.8 & 78.7 & 64.9 & 5, 2, 2, 1 \\
$0.7$ / $10$ & 0.7 & 11.9 & no & 19.8 & 4 & 13.6 & 17.7 & 5.8 & 2, 2, 2, 2, 1, 1 \\
$0.7$ / $10$, seed 2 & 0.7 & 11.9 & no & 77.4 & 6 & 48.0 & 76.9 & 28.9 & 5, 2, 2, 1 \\
$0.9$ / $1$ & 0.9 & 1.2 & yes & 233.6 & 5 & 2.9 & 3.3 & 0.4 & 3, 2, 2, 2, 1 \\
$0.9$ / $3$ & 0.9 & 3.6 & yes & 232.8 & 6 & 13.8 & 71.9 & 58.0 & 5, 2, 2, 1 \\
$0.9$ / $10$ & 0.9 & 11.9 & yes & 230.6 & 5 & 29.6 & 35.5 & 5.8 & 3, 2, 2, 2, 1 \\
\bottomrule
\end{tabular}}
\\[4pt]
\parbox{0.97\textwidth}{\footnotesize The four incomplete runs stopped on an over-extended bond of a vacuole mesh: $0.5$ / $1$ at $0.4$~ms after its five-parent fusion, $0.7$ / $10$ seed~2 at $0.5$~ms after its own, $0.7$ / $10$ at $2.0$~ms after four fusions executed together, and $0.7$ / $1$ at $35$~ms after its last fusion, when the solver reported missing bond atoms. In $0.5$ / $10$ and $0.9$ / $1$ the five-parent fusion became due at $82.8$ and $4.9$~ms and was retried $26$ and $421$ times without ever being placed. In $0.9$ / $10$ it never became due at all: that pair was in contact in only $1.8\%$ of frames, for at most $3.3$~ms at a time. All ten runs share one initial placement of the ten vacuoles, in which $2$, $5$ and $8$ of the $45$ pairs have their surfaces within $0.5$, $0.7$ and $0.9~\mu$m, so the early fusions are common to the set and the runs are not independent samples. The first fusion occurred at $2.2$, $1.6$ and $1.6$~ms in the three runs at $t_f = 1.2$~ms, at $5.1$~ms at $t_f = 3.6$~ms and at $17.7$~ms at $t_f = 11.9$~ms. Measured from their meshes, the volume-equivalent diameters of the products are $0.76$, $0.86$ and $1.04~\mu$m at $k = 2$, $3$ and $5$, against $D_0k^{1/3} = 0.756$, $0.865$ and $1.026~\mu$m, and $0.58~\mu$m for an unfused vacuole.}
\end{table}

\section{Contact analysis}\label{sec:contact}

Vacuole positions were written every $20$ steps in the single-cell simulations and every $100$ steps in the vessel ones. For each vacuole we computed its centroid, and for each pair of vacuoles in the same cell the centroid distance $d$ and the smallest vertex--vertex distance $s$, both under the minimum-image convention. A pair is in contact while $d < d_c$, or, in the surface version, while $s < s_c$. For each threshold we report four things. The occupancy is the fraction of the observed frames, taken over all pairs, in which the pair is in contact. The entry rate is the number of transitions into contact per pair per second. Durations are given as the median and 90th percentile over contacts that both begin and end within the record. Last come the number of pairs ever in contact and the number in contact in more than $90\%$ of the frames in which they are observed. Contacts already under way when the record starts, or still under way when it ends, are censored and counted separately. The thresholds swept were $d_c \in \{1.0, 1.1, 1.2, 1.3, 1.4, 1.6, 1.8, 2.0\}~\mu$m and $s_c \in \{0.3, 0.4, 0.5, 0.6, 0.8, 1.0\}~\mu$m.

A sweep is needed rather than a single criterion. Vertices of different vacuoles repel within the $0.5~\mu$m cutoff of their barrier (table~\ref{tab:pairs}), so the centroids of two vacuoles of $D = 0.6~\mu$m do not come closer than $0.82~\mu$m once a run is under way, measured over every frame of all six encounter runs. The shared initial placement starts them closer, at $0.74~\mu$m. Pairs in sustained contact sit at a centroid separation whose median is $1.00$--$1.08~\mu$m and whose 95th percentile is $1.12$--$1.17~\mu$m in every simulation. We quote $d_c = 1.2~\mu$m throughout, which is above that band, and give the full sweep in table~\ref{tab:dcsweep}. Both the occupancy and the entry rate depend on $d_c$: the entry rate falls sharply between $1.1$ and $1.2~\mu$m as the threshold stops cutting through the distribution of bonded separations, and the occupancy continues to rise with $d_c$ over the whole range. The comparisons in the main text are therefore made between environments at one threshold, and we do not treat the flatness of the occupancy between $1.2$ and $1.4~\mu$m as evidence for anything.

\textit{The entry rate is not an absolute measurement.} Vacuoles repel one another through the barrier of table~\ref{tab:pairs} rather than colliding, so an entry into contact is a crossing of the surface that $d_c$ defines and not an event that could be counted independently of it. A pair resting against that barrier fluctuates across any threshold set within its jitter, and that jitter straddles $d_c \approx 1.0$--$1.1~\mu$m. Table~\ref{tab:dcsweep_enc} gives the consequence in the two encounter flows: the entry rate falls from $13.8$ and $11.4$ per pair per second at $1.0~\mu$m to $0.29$ and $2.35$ at $1.2~\mu$m, a factor of $48$ in the tube. The ordering of the two flows reverses over that step, the tube having the higher rate at $1.0~\mu$m and the lower at $1.2~\mu$m. Above $1.6~\mu$m the rate rises again as the surface reaches pairs that are merely near one another. We therefore use entry rates only to compare two flows analysed at the same $d_c$, never as a rate in their own right. The conclusions drawn in the main text rest on the number of pairs held in contact for more than $90\%$ of a record and on the number of contacts that both begin and end within it.

\textit{Per-seed counts, and the two record lengths.} Table~\ref{tab:encseeds} gives the per-seed counts quoted in the main text, and with them the one asymmetry in the comparison. The two encounter flows were run for the same $200\,000$ steps, but their time scales differ, so the plane-shear records are $1296$--$1302$~ms against $745$--$747$~ms in the tube. Counts that accumulate over a record---pairs ever in contact, and contacts that both begin and end within it---are therefore not matched between the flows as they stand. The fraction of frames a pair spends in contact, on which the $>90\%$ criterion rests, is the easier to satisfy over the shorter record. The lower block of the table re-scores the plane-shear runs over the tube's window. Every comparison drawn in the main text survives it: $21$, $31$ and $26$ pairs ever in contact against $12$, $9$ and $10$, and $32$, $66$ and $50$ completed contacts against $2$, $1$ and $1$. The persistence comparison is the one that narrows, from $8$, $9$ and $8$ pairs against none, none and two, to $8$, $9$ and $8$ against $3$, $2$ and $3$. The main text gives both figures, and its Discussion the more conservative of them.

\textit{Start-up of the vessel simulations.} The bifurcation, the curved vessel and the dilation are equilibrated with the plain soft pair between vacuoles, which is replaced by the barrier of table~\ref{tab:pairs} at the last stage, after which the clock is reset and the production record begins. The set of bonded pairs therefore assembles while the cell interiors relax into their production interaction. Of the $272$, $818$ and $236$ bonded pairs of the bifurcation, the curved vessel and the dilation, only $2$, $6$ and $0$ are in contact at the first frame in which the pair is observed. Over the set as a whole, first contact comes at a median of $8.04$, $3.66$ and $3.54$~ms, with 95th percentiles of $56.2$, $20.2$ and $12.6$~ms. Thus $68$, $88$ and $91\%$ of them have formed within $10$~ms, and within $50$~ms $94$ and $98\%$ of them in the first two vessels and all but one of the $236$ in the dilation. We report this as a start-up transient and make no claim that the flow brought those vacuoles together.

\textit{The vessel occupancies have not converged.} The three vessel records differ in length, $1206$, $785$ and $758$~ms, and in none of them has the occupancy converged. Between the first and the last quarter of the $758$~ms they share, it rises from $0.081$ to $0.105$ in the bifurcation, from $0.086$ to $0.168$ in the curved vessel and from $0.085$ to $0.191$ in the dilation. Restricted to that common window the three become $0.095$, $0.126$ and $0.137$, so the ordering survives, but each is an average over a rising curve and none of them is a plateau. The two tube simulations retain only the aggregate per-pair record and cannot be re-windowed this way.

\begin{table}[htbp]
\small
\centering
\renewcommand\arraystretch{1.2}
\caption{\textbf{Dependence of the vacuole--vacuole contact statistics on the threshold $d_c$.} The threshold is on the centroid separation, and the simulations are those of Fig.~\mref{fig:survey}. Each cell gives the contact occupancy and, in parentheses, the entry rate per pair per second.}
\label{tab:dcsweep}
\begin{tabular}{lccccc}
\toprule
 & Tube & Tube & & Curved & \\
$d_c$ ($\mu$m) & $118~\mathrm{s}^{-1}$ & $1168~\mathrm{s}^{-1}$ & Bifurcation & vessel & Dilation \\
\midrule
1.0 & 0.065 (5.73) & 0.012 (2.54) & 0.020 (2.17) & 0.034 (5.51) & 0.033 (5.82) \\
1.1 & 0.117 (5.33) & 0.095 (15.66) & 0.092 (2.08) & 0.117 (3.00) & 0.121 (4.50) \\
1.2 & 0.131 (0.28) & 0.131 (0.28) & 0.100 (0.12) & 0.128 (0.41) & 0.137 (0.72) \\
1.3 & 0.131 (0.24) & 0.131 (0.28) & 0.101 (0.11) & 0.129 (0.31) & 0.139 (0.52) \\
1.4 & 0.132 (0.24) & 0.132 (0.24) & 0.101 (0.10) & 0.129 (0.30) & 0.141 (0.50) \\
1.6 & 0.133 (0.04) & 0.132 (0.12) & 0.103 (0.09) & 0.132 (0.32) & 0.145 (0.61) \\
1.8 & 0.150 (2.09) & 0.134 (0.12) & 0.107 (0.25) & 0.142 (0.83) & 0.159 (1.30) \\
2.0 & 0.196 (0.32) & 0.137 (0.36) & 0.116 (0.40) & 0.169 (1.88) & 0.186 (1.93) \\
\bottomrule
\end{tabular}
\\[4pt]
\parbox{0.97\textwidth}{\footnotesize Entry rates are converted with the $\tau$ of each simulation (table~\ref{tab:tau}). The two tubes are single runs of one cell holding ten vacuoles, that is $45$ pairs of which only six and seven ever make contact, so their columns carry no seed spread. The three vessels are single runs of many cells, and the pair-to-pair spread within each is quoted where it is used. The encounter tube and plane-shear simulations of Fig.~\mref{fig:contacts}, three seeds each, are swept over the same thresholds in table~\ref{tab:dcsweep_enc}.}
\end{table}

\begin{table}[htbp]
\small
\centering
\renewcommand\arraystretch{1.2}
\caption{\textbf{Threshold dependence of the contact statistics in the two encounter flows.} The flows are those of Fig.~\mref{fig:contacts}. Occupancy is the fraction of observed frames, over all pairs, in which the pair is in contact, and the entry rate is the number of transitions into contact per pair per second; both are the mean $\pm$ SD over the three seeds of each flow.}
\label{tab:dcsweep_enc}
\begin{tabular}{lcccc}
\toprule
& \multicolumn{2}{c}{Occupancy} & \multicolumn{2}{c}{Entries per pair per second} \\
\cmidrule(lr){2-3}\cmidrule(lr){4-5}
$d_c$ ($\mu$m) & Tube & Plane shear & Tube & Plane shear \\
\midrule
$1.0$ & 0.059 $\pm$ 0.020 & 0.066 $\pm$ 0.001 & 13.80 $\pm$ 4.10 & 11.37 $\pm$ 1.06 \\
$1.1$ & 0.184 $\pm$ 0.017 & 0.265 $\pm$ 0.029 & 6.51 $\pm$ 0.94 & 11.34 $\pm$ 2.68 \\
$1.2$ & 0.197 $\pm$ 0.019 & 0.320 $\pm$ 0.042 & 0.29 $\pm$ 0.06 & 2.35 $\pm$ 0.48 \\
$1.3$ & 0.198 $\pm$ 0.019 & 0.331 $\pm$ 0.044 & 0.26 $\pm$ 0.09 & 1.63 $\pm$ 0.31 \\
$1.4$ & 0.198 $\pm$ 0.019 & 0.339 $\pm$ 0.046 & 0.26 $\pm$ 0.09 & 1.56 $\pm$ 0.32 \\
$1.6$ & 0.202 $\pm$ 0.021 & 0.359 $\pm$ 0.053 & 0.28 $\pm$ 0.17 & 1.95 $\pm$ 0.47 \\
$1.8$ & 0.232 $\pm$ 0.023 & 0.420 $\pm$ 0.071 & 1.93 $\pm$ 1.56 & 4.29 $\pm$ 1.11 \\
$2.0$ & 0.244 $\pm$ 0.016 & 0.505 $\pm$ 0.085 & 0.94 $\pm$ 0.39 & 4.29 $\pm$ 0.62 \\
\bottomrule
\end{tabular}
\\[4pt]
\parbox{0.97\textwidth}{\footnotesize The three seeds of a flow share their initial placement of the ten vacuoles and differ only in thermal noise, so the spread is over the noise and not over the configuration. The entry rate is not a measurement of a collision frequency and is given here to show how strongly it depends on the threshold: between $1.0$ and $1.2~\mu$m it falls by a factor of $48$ in the tube and by a factor of $5$ in plane shear, and which of the two flows has the higher rate reverses. The occupancy does not reverse anywhere in the range.}
\end{table}

\begin{table}[htbp]
\small
\centering
\renewcommand\arraystretch{1.2}
\caption{\textbf{Per-seed contact statistics of the two encounter flows at $d_c = 1.2~\mu$m.} The flows are those of Fig.~\mref{fig:contacts}. One cell carrying ten vacuoles of $D = 0.6~\mu$m, that is $45$ intracellular pairs; the three seeds of a flow share their initial placement of the vacuoles and differ only in thermal noise. A completed contact is one that both begins and ends within the record; contacts already under way when it starts, or still under way when it ends, are censored and are not counted here. In the tube the median is taken over one or two completed contacts and is reported for completeness rather than as a distribution. The lower block re-scores the same three plane-shear records over the tube's window, $746.9$~ms.}
\label{tab:encseeds}
\begin{tabular}{lrrrrrrr}
\toprule
 & Record & \multicolumn{4}{c}{Pairs in contact (of $45$)} & \multicolumn{2}{c}{Completed contacts} \\
\cmidrule(lr){3-6}\cmidrule(lr){7-8}
Flow, seed & (ms) & at $t=0$ & ever & by $10$~ms & $>90\%$ & number & median (ms) \\
\midrule
Tube, seed 1        & $746.6$  & $3$ & $12$ & $9$  & $8$ & $2$   & $32.5$ \\
Tube, seed 2        & $745.5$  & $3$ & $9$  & $9$  & $9$ & $1$   & $24.2$ \\
Tube, seed 3        & $746.9$  & $3$ & $10$ & $9$  & $8$ & $1$   & $33.7$ \\
\midrule
Plane shear, seed 1 & $1296.0$ & $6$ & $38$ & $11$ & $0$ & $88$  & $87.8$ \\
Plane shear, seed 2 & $1301.6$ & $6$ & $40$ & $11$ & $0$ & $148$ & $87.2$ \\
Plane shear, seed 3 & $1299.9$ & $6$ & $44$ & $11$ & $2$ & $118$ & $106.0$ \\
\midrule
\multicolumn{8}{l}{\textit{the same plane-shear records, truncated to the tube's window}} \\
Plane shear, seed 1 & $746.8$  & $6$ & $21$ & $11$ & $3$ & $32$  & $107.9$ \\
Plane shear, seed 2 & $746.9$  & $6$ & $31$ & $11$ & $2$ & $66$  & $100.1$ \\
Plane shear, seed 3 & $746.8$  & $6$ & $26$ & $11$ & $3$ & $50$  & $94.2$ \\
\bottomrule
\end{tabular}
\end{table}

\section{Containment, temperature and run quality}\label{sec:checks}

\textit{Containment.} In every frame, each membrane and each vacuole was unwrapped across the periodic boundaries along its bonds. For every vacuole vertex $\mathbf{p}$ we computed the generalized winding number with respect to the membrane triangulation of each cell, $w(\mathbf{p}) = (4\pi)^{-1}\sum_t \Omega_t(\mathbf{p})$, where $\Omega_t$ is the signed solid angle of triangle $t$ seen from $\mathbf{p}$; $w \approx 1$ inside a closed membrane and $w \approx 0$ outside. A vertex is inside when $w > 0.5$, and the host of a vacuole is the cell that contains its centroid. A run was used only if every vacuole kept at least $99\%$ of its vertices inside its host in every frame. Where a bond exceeded its maximum extension, the run was not used beyond that step.

\textit{Cells that leave the lumen.} The wall of a vessel is a layer of frozen particles, and a driven membrane can be pushed through it. No cell leaves the lumen of any of the three vessels. In the curved vessel every membrane vertex was tested against the tube itself, since the distance to the nearest wall particle cannot see a cell that has crossed the wall completely. At every $10\,000$-step anchor all of them lie inside the wall, the outermost $0.50~\mu$m inside the centres of its particles. A first build of this vessel had placed seven of its $240$ cells across the wall, and four of them drifted out of the lumen during its run, while no cell that started inside left. The seven were removed and the vessel run again with the other $233$, and that run is the one reported. None left the dilation: at every $10\,000$-step anchor no membrane vertex lies beyond the wall, and the closest stays $0.34$--$0.79~\mu$m from the nearest wall particle. No cell left the bifurcation, where the cells stay within $2.27~\mu$m of the mid-plane of the $10~\mu$m channel. At a drive of $f = 2$, tried first, the body force pushed membrane vertices through the wall outright, which is why all the vessels are run at $f = 0.5$.

\textit{Position of the cell in a tube.} In the encounter tube the cell was placed $0.08$--$0.09~\mu$m off the axis and settled at a median offset of $0.26$--$0.30~\mu$m after step $20\,000$, with a maximum of $0.48$--$0.49~\mu$m. In the survey tubes the median offset is $0.41~\mu$m (maximum $0.63$) at the low drive and $0.07~\mu$m (maximum $0.17$) at the high one, so the cell rides closer to the axis the harder it is driven.

\textit{Temperature.} The drive does work on the solvent that the thermostat must remove, so we check how far the solvent is from the bath temperature of $0.1884$. We measure it from the solvent velocities directly, subtracting a local \emph{linear} velocity field---for each sampled particle, a least-squares fit of $\mathbf{v} = \mathbf{b} + \mathsf{A}\,(\mathbf{r}-\mathbf{r}_i)$ over its nearest neighbours---rather than the mean of a fixed bin. The distinction matters in the vessels: a bin large enough to hold a usable number of particles also spans a rotating or strongly sheared flow, and the velocity spread inside it is then counted as thermal motion. The solver's own binned estimate reads $1.20$ of the bath value for the bifurcation and $4.14$ for the curved vessel at step $100\,000$, where the local fit to the same solvent at the same steps gives $1.022 \pm 0.007$ and $1.78$. Even the local fit misleads when it is taken over all the solvent, because two of the vessels have solvent on the far side of their frozen walls. In the curved vessel $1.6\%$ of the solvent is outside the tube at the start and, since the wall leaks, $12\%$ at the end (table~\ref{tab:tau}); in the dilation $1.6\%$ lies outside the vessel throughout. Little resists the drive there, so by the end of the records this solvent flows $13$ and $6.5$ times faster than the solvent inside, and the estimator reads it at $11$ and $36$ times the bath value. Averaged in with the rest, it raises the curved vessel to $1.78$ at step $100\,000$ and $2.27$ at step $150\,000$, and the dilation to $1.51$--$1.65$ from step $90\,000$ on. Table~\ref{tab:temp} gives the solvent inside the vessels, where the cells are, and there the drive barely warms it: at every $10\,000$ steps the curved vessel's lumen lies between $1.01$ and $1.02$, and the dilation between $1.00$ and $1.04$.

\begin{table}[htbp]
\small
\centering
\renewcommand\arraystretch{1.2}
\caption{\textbf{Solvent temperature relative to the bath value of $0.1884$, from a local linear fit to the solvent velocities.}}
\label{tab:temp}
\begin{tabular}{lcp{9.6cm}}
\toprule
Simulation & $T/T_{\mathrm{bath}}$ & Note \\
\midrule
Encounter tube & 0.98--1.03 & over the three seeds \\
Plane shear & 1.00--1.01 & over the three seeds \\
Fusion cascade & 1.01--1.05 & over the ten runs \\
Microsphere filter & 1.11--1.12 & binned estimate, $0.2089$--$0.2112$ over $22$ runs; its bins are the finest of any simulation, $2 \times 1.9 \times 1.9~\mu$m \\
Bifurcation & 1.02 & plateau from step $12\,000$, and the same in the single-channel spans of the duct \\
Curved vessel & 1.01--1.02 & solvent inside the lumen, every $10\,000$ steps; over all the solvent, $2.27$ at step $150\,000$ \\
Dilation & 1.00--1.04 & solvent inside the vessel, every $10\,000$ steps; over all the solvent, $1.56$ at step $150\,000$ \\
\bottomrule
\end{tabular}
\end{table}

\textit{Solvent exchange across the membrane, and the density of the cytosol.} On the time scale of a simulation the membrane is impermeable to solvent at rest and weakly but symmetrically permeable during a squeeze. Without a driving force a cell expels $70$ of its $382$ interior beads in the first $1000$ steps and $96$ in all, then stops: over the last $8000$ steps not one bead crosses in either direction and the interior count is constant. The same $70$-bead loss occurs at every driving force, so it is the initial packing relaxing---beads placed inside the membrane's excluded-volume shell are pushed out---and not a leak. At the strongest drive a two-way traffic of about six beads per $1000$ steps in each direction appears while the cell is in the filter. It exchanges roughly a third of the cytosol over a run but drains none of it, the interior count rising while the cell volume falls by $2\%$.

The solvent is a stiff DPD fluid and layers next to every repulsive surface. Within $0.4~\mu$m of a membrane vertex its density is below $1$ on both sides; at $0.4$--$1.0~\mu$m it overshoots; and only beyond $1~\mu$m from any surface does it settle at $4.65$--$4.70$, the ambient value. A cell interior about $2~\mu$m across is thinner than this layering length, so the cytosol never reaches a plateau. Averaged over the geometric interior its density is only $3.1$--$4.1$, because most of that volume lies inside the depleted shell. In the part a vacuole can occupy, more than $0.6~\mu$m from any membrane or vacuole vertex, it is $7.5$--$8.3$ at rest and $8.2$--$8.9$ during a squeeze. A grid of $0.2$ rather than $0.3~\mu$m changes every value by at most $4\%$.

\section{Geometries}\label{sec:geometry}

Every box is periodic in all three directions, and flow is along $x$.

\textit{Microsphere filter.} The box measures $20\times9.5\times9.5~\mu$m$^3$. Four beads, each a filled sphere of $3232$ frozen particles lying within $2.008~\mu$m of its centre, sit at $x = 10.01~\mu$m on a square lattice of pitch $4.75~\mu$m in $y$ and $z$; with their periodic images they tile a layer between $x = 8.01$ and $12.01~\mu$m. Adjacent beads are $0.73~\mu$m apart between particle centres. The pore at the centre of four beads has a throat of $2.72~\mu$m between bead particle centres and is the widest straight channel through the layer; at the mid-plane of the beads, $39.6$ of the $90.25~\mu$m$^2$ cross-section ($44\%$) is open geometrically. Each bead packs its particles to a number density of $95~\mu$m$^{-3}$ against $4.3$ in the fluid, so the repulsion excludes solvent out to $r = 2.69~\mu$m from a bead centre, the edge being a step rather than a ramp ($10\%$ to $90\%$ of bulk within $0.025~\mu$m) followed by a first solvent layer at $4.9$ times bulk. Measured over the passes, a membrane vertex comes within $2.483~\mu$m of a bead centre, giving an effective throat of $1.75~\mu$m, and the route between two adjacent beads is closed, since the membrane standoff exceeds half the $4.75~\mu$m spacing. The cell starts as a biconcave disc in the $x$--$z$ plane, centred on a pore at $(y, z) = (4.75, 4.73)~\mu$m, with its membrane between $x = 0.11$ and $7.90~\mu$m. A single vacuole starts at $(1.4, 4.75, 4.75)~\mu$m, where the smallest membrane--vacuole distance is $0.75~\mu$m at $D = 1.0~\mu$m and $0.28~\mu$m at $D = 2.0~\mu$m. The body force is $f = 3$, which at $\rho = 4.325$ is $0.29$~Pa/$\mu$m, about $1.1$~Pa across the layer. The membrane--vacuole barrier is ramped from $a = 150$ to $600$ over the first $2000$ steps, after which the run proceeds for $150\,000$ steps.

\begin{figure}[htbp]
\centering
\includegraphics[width=0.95\textwidth]{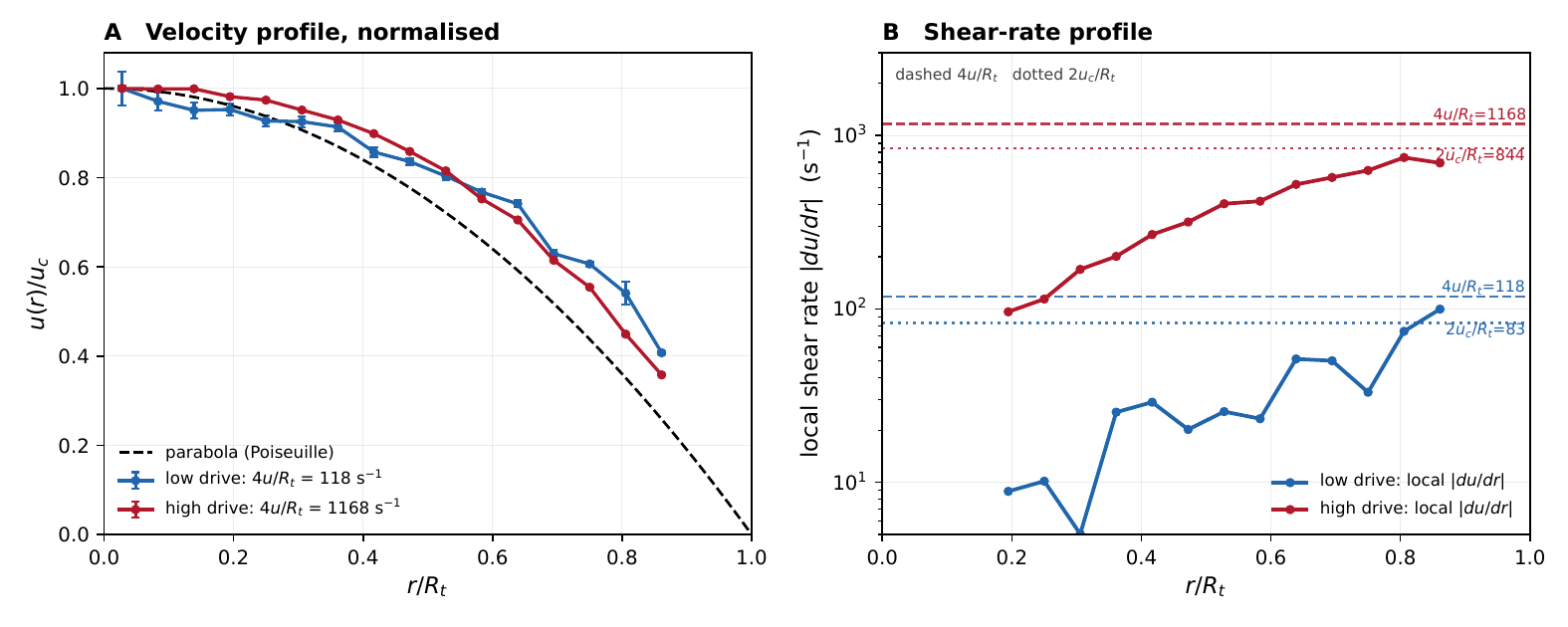}
\caption{\textbf{Radial velocity profile of the two survey tube drives.} The drives, $f = 0.28$ and $f = 2.5$, are those of the microvascular comparison of Fig.~\mref{fig:survey}. These are separate simulations from the encounter tube, which is driven at $f = 0.45$, and their shear rates differ accordingly. Profiles are averaged over the last five dumps (steps $110\,000$--$150\,000$) in $18$ radial bins, and are measured from the trajectories. The area-weighted wall shear rate $4u/R_t$ is $117.6$ and $1168.2~\mathrm{s}^{-1}$ at the two drives; the centreline convention $2u_c/R_t$ gives $82.9$ and $843.8~\mathrm{s}^{-1}$; and the bluntness $u/u_c$ is $0.710$ and $0.692$, against $0.50$ for a parabolic profile.}
\label{fig:tubeshear}
\end{figure}

\textit{Encounter tube and survey tubes.} Frozen wall particles, the innermost at a radius of $5.41~\mu$m, bound a cylindrical lumen of diameter $10.8~\mu$m along $x$ in a box of $30.4\times16\times16~\mu$m$^3$. One cell carries ten vacuoles of $D = 0.6~\mu$m. Solvent is also seeded outside the lumen, in the corners of the box, and must be excluded when the lumen density or velocity is measured. The encounter tube is driven at $f = 0.45$. Its profile is blunted by wall slip and by the cell, so the mean velocity is $0.70$ of the centreline rather than the $0.50$ of a parabola. The two conventional wall shear rates therefore differ: $4u/R_t = 194\pm1~\mathrm{s}^{-1}$ from the area-weighted mean velocity against $2u_c/R_t = 138\pm1~\mathrm{s}^{-1}$ from the centreline. The velocity gradient measured directly over the outermost sampled shell is $111\pm1~\mathrm{s}^{-1}$. The two survey tubes use the same lumen at $f = 0.28$ and $f = 2.5$ (fig.~\ref{fig:tubeshear}).

\textit{Plane shear.} Two slabs of particles, at $z = 0$--$5$ and $11$--$16~\mu$m in a $20\times13\times16~\mu$m$^3$ box, move at $\pm v_w$ along $x$ with the forces on them set to zero, and shear a $6~\mu$m layer of solvent holding one cell with ten vacuoles. The slabs meet across the periodic boundary in $z$; forces between them have no effect because their velocity is prescribed. The walls move at $v_w = \pm3.0$, that is $\pm463~\mu$m/s. The profile across the gap is linear, giving a measured shear rate of $140.4\pm0.9~\mathrm{s}^{-1}$ against the nominal $2v_w/H$; the difference is wall slip, the solvent at the wall lagging it by $0.28$ of $3.0$. Matched on $2u_c/R_t$, the tube and plane-shear arms agree to within $2\%$.

\textit{Bifurcation.} A channel $10~\mu$m high, between frozen plates at $z = 0$--$1.5$ and $11.5$--$13~\mu$m, runs along a $141\times47\times13~\mu$m$^3$ box. Upstream and downstream the lumen is a single channel $20~\mu$m wide, between side walls at $y = 0.5$--$13.5$ and $33.5$--$46.5~\mu$m. An obstacle, a hexagon $55~\mu$m long centred at $x = 70.5~\mu$m and spanning the full channel height, splits it into two branches that rejoin downstream. Its sides taper straight to a point at each end, and over $x = 60.5$--$80.5~\mu$m it has its full width, $y = 13.5$--$33.5~\mu$m. There the side walls have drawn back to $y = 0.5$--$3.5$ and $43.5$--$46.5~\mu$m, so each branch is $10~\mu$m wide and the two together carry the width of the inlet. It holds $80$ cells carrying $762$ vacuoles.

\textit{Curved vessel.} A lumen of $29.8~\mu$m follows an $\Omega$-shaped centreline in a $121\times102\times42~\mu$m$^3$ box: a quarter bend of centreline radius $36.0~\mu$m turns the flow from the direction of the drive to the perpendicular, a semicircle of radius $24.0~\mu$m carries it over the apex, and a second quarter bend of $36.0~\mu$m turns it back. Both are measured from the frozen wall: the lumen diameter is twice the mean inner radius over $15^\circ$ sectors about the local centre, $14.91~\mu$m (range $13.75$--$16.30$, all $24$ sectors occupied), and the centreline is the set of midpoints between the two sides of the wall in the plane $z = 21.0~\mu$m, which lie within $0.05~\mu$m of the three arcs. It holds $233$ cells carrying $2301$ vacuoles.

\textit{Aneurysm-like dilation.} A lumen of $25.0~\mu$m opens into a sac of $50.0~\mu$m in a $161\times60\times52~\mu$m$^3$ box, a bulge-to-neck ratio of $2.0$. Both diameters come from least-squares fits to the $116\,260$ frozen wall particles. The tube shell lies at $r = 12.47$--$12.53~\mu$m about its axis. The $27\,162$ wall particles above $y = 29~\mu$m lie $25.00~\mu$m from the centre of a sphere that sits $2.5~\mu$m above the tube's upper wall and meets it between $x = 55.7$ and $105.5~\mu$m, where the sections of Fig.~\mref{fig:survey} are cut ($x = 55$ and $106~\mu$m). These are wall-to-wall diameters between wall particle centres; as at the filter, the opening a membrane can enter is narrower. The start-up ramps the timestep over $50\,000$ steps, because a shorter start-up tore a membrane, and the $150\,000$-step record was run in two jobs, the second continuing from the configuration at step $90\,000$. It holds $190$ cells in the lumen, $89$ of them built carrying the $856$ vacuoles, and $100$ vacuole-free cells pre-filling the sac. At the first frame one vacuole lies inside the membranes of two overlapping cells and is assigned to the one that carries no other vacuole, so the analyses count $90$ host cells. The body force acts on all fluid and cells, including those in the sac.

\section{The slit runs}\label{sec:slit}

\textit{Geometry and drive.} A periodic box of $30\times12.5\times12.5~\mu$m$^3$ is closed at $x = 12.0~\mu$m by a wall of five planes of frozen particles on a simple cubic lattice of spacing $0.219~\mu$m, $0.88~\mu$m thick between its outer planes and pierced by one rectangular aperture of $2.85\times5.48~\mu$m between particle centres. The solvent stands off the wall particles by $0.67~\mu$m. The section it fills within the aperture, counted from its particles, is equivalent to $1.74\times3.14~\mu$m, or $5.5~\mu$m$^2$, and its flow is constricted over $2.5~\mu$m of length; the slit of Li et al.\ is $1.2\times4.0~\mu$m, $4.8~\mu$m$^2$, and $1.9~\mu$m long~\cite{li2021How}. The cell starts edge-on to the aperture, its disc normal along the short side, centred $6.0~\mu$m upstream of the wall with its front $2.1~\mu$m from it, and the vacuole sits $2.6~\mu$m behind or ahead of the cell centre. Each cell was relaxed with its membrane permeable to the solvent while its volume target was raised at fixed area, $A_0 = 132.87~\mu$m$^2$, to $92.46$, $108.91$ or $119.70~\mu$m$^3$, and then sealed; the measured surface-to-volume ratios are $1.41$, $1.20$ and $1.09~\mu$m$^{-1}$. The body force acts on membrane vertices and cytosol only. Driving the plasma as well makes the wall hold a pressure drop $\rho f L_x$, and the fluid cavitates behind a plugged aperture as that approaches the fluid's own pressure, which caps the drive near $f = 0.45$. At that drive even the cell at $1.41~\mu$m$^{-1}$ was still held after $600$~ms. With only the cell driven, the cytosol leaks through the membrane mesh at the default membrane--solvent repulsion, $31\%$ of it in $70\,000$ steps at $f = 4$, so that repulsion is raised to $a = 100$, at which a cell held for $150\,000$ steps kept all of its cytosol. The vacuole's vertices together carry $f N_{\mathrm{def}}$, where $N_{\mathrm{def}}$ is the number of solvent particles the vacuole displaces, counted in a separate $16~\mu$m box as the deficit within $4$--$6~\mu$m of its centroid. It is $15.7$ and $33.4$ for $D = 0.8$ and $1.4~\mu$m, against $\rho V_0 = 1.2$ and $6.2$; the difference is the layer the solvent leaves around the vacuole's vertices. Inside a driven cell the choice matters little. In a wall-free box, with the plasma carrying the opposite force, the offset of a $1.4~\mu$m vacuole behind the membrane's centroid fell from $2.8$ to $0.6$--$1.4~\mu$m as the drive started, less with a larger share. Thereafter the vacuole drifted relative to the membrane by less than $0.2\%$ of the cell's speed for every share from none to $N_{\mathrm{def}}$. A run at $f = 8$ with the nominal share crossed the slit in the same $49.9$~ms. The total force on the cell is $150$--$174$~pN at $f = 8$ and $75$--$87$~pN at $f = 4$. The membrane--vacuole barrier has the $0.7~\mu$m cutoff. After $2000$ undriven steps each run proceeds for $100\,000$ steps at $f = 8$ or $150\,000$ at $f = 4$, $370$ or $554$~ms at the $3.695$~ms per $1000$ steps of the solvent density $4.325$. With three seeds each, the trailing vacuole was run at every ratio with both diameters, $36$ runs, and the leading vacuole at $1.41~\mu$m$^{-1}$ with both diameters and at $1.09~\mu$m$^{-1}$ with $D = 1.4~\mu$m, $18$ runs.

\textit{Measurements.} Membrane and vacuole vertices were written every $500$ steps, $1.85$~ms, with unwrapped coordinates. With $x_0$ and $x_1$ the upstream and downstream particle planes of the wall, $0.88~\mu$m apart, a cell reaches the slit when its front passes $x_0 - 0.8~\mu$m and leaves it when its rear passes $x_1 + 0.8~\mu$m. A vacuole reaches and leaves the slit by the same rule applied to its front and rearmost vertices. Times are counted from the start of the drive, with $\tau = 3.695$~ms per $10^3$ steps. A cell that has not left by the last frame is held, and its time at the slit, like that of a vacuole that reached the slit without leaving it, is a lower bound. The membrane past the slit is the fraction of membrane vertices downstream of $x_1$ in the last frame. All $54$ runs reached their last step, none had a membrane vertex inside the wall solid, and all are listed in table~\ref{tab:slitruns}.

\textit{What the table adds.} Every held cell had reached the slit $5.5$~ms after the drive started, so each time at the slit of a held cell is censored at $548.7$~ms. The two cells held at $f = 4$ and $1.41~\mu$m$^{-1}$ with a trailing vacuole ended with $4.8$ and $2.2\%$ of their membrane past the wall and their vacuole short of the slit; the twelve held at $1.20$ and $1.09~\mu$m$^{-1}$ ended with none. Of the two cells held with a leading $0.8~\mu$m vacuole, one had pushed the vacuole through the wall with $22\%$ of its membrane after it: the rearmost vertex of the vacuole ended $0.63~\mu$m past $x_1$, short of the $0.8~\mu$m at which it would have left, which is why its time at the slit is censored. In the other, the vacuole ended at the mouth of the aperture, its front vertex $0.06~\mu$m short of $x_0$. With the vacuole leading at $f = 8$, a cell at $1.41~\mu$m$^{-1}$ spent $48$--$52$~ms at the slit with a $0.8~\mu$m vacuole and $57$--$63$~ms with a $1.4~\mu$m one, against $50$--$52$~ms with the latter trailing; at $1.09~\mu$m$^{-1}$ a leading $1.4~\mu$m vacuole gave $74$--$78$~ms, against $74$--$76$~ms trailing.

\textit{What the runs can and cannot say.} The model varies the surface reserve alone, so it shows that the slit discriminates on the cell, not how far ageing moves a cell towards the margin. The plasma is not driven, because the DPD fluid, pushed against a plugged aperture, cavitates behind the wall before the pressure drop can squeeze a cell through. A held cell is therefore pressed against the wall by the force on its own contents, and it withdraws its tongue from the aperture where a cell held by a pressure drop would keep it there. What the model supports is the comparison between surface reserves and vacuole positions at a common drive, not the force at which a real cell is held.

\begin{table}[htbp]
\small
\centering
\renewcommand\arraystretch{1.15}
\caption{\textbf{The slit runs, three seeds per condition.} Where the seeds differ their values are listed in seed order; where they agree, one value is given. A time preceded by $>$ is a lower bound, the run having ended first, and a dash means that the vacuole never reached the slit.}
\label{tab:slitruns}
\resizebox{\textwidth}{!}{\begin{tabular}{cccccccc}
\toprule
$f$ & Vacuole & $S/V$ ($\mu$m$^{-1}$) & $D$ ($\mu$m) & Passed & Cell at slit (ms) & Vacuole at slit (ms) & Membrane past \\
\midrule
8 & trailing & 1.41 & 0.8 & 3/3 & 49.9 / 48.0 / 49.9 & 5.5 & 1 \\
8 & trailing & 1.41 & 1.4 & 3/3 & 49.9 / 51.7 / 51.7 & 7.4 / 7.4 / 3.7 & 1 \\
8 & trailing & 1.20 & 0.8 & 3/3 & 55.4 & 5.5 / 5.5 / 3.7 & 1 \\
8 & trailing & 1.20 & 1.4 & 3/3 & 55.4 & 5.5 / 3.7 / 5.5 & 1 \\
8 & trailing & 1.09 & 0.8 & 3/3 & 75.7 / 73.9 / 77.6 & 3.7 & 1 \\
8 & trailing & 1.09 & 1.4 & 3/3 & 73.9 / 75.7 / 73.9 & 5.5 / 5.5 / 3.7 & 1 \\
8 & leading & 1.41 & 0.8 & 3/3 & 48.0 / 49.9 / 51.7 & 20.3 / 20.3 / 22.2 & 1 \\
8 & leading & 1.41 & 1.4 & 3/3 & 62.8 / 57.3 / 57.3 & 38.8 / 33.3 / 31.4 & 1 \\
8 & leading & 1.09 & 1.4 & 3/3 & 73.9 / 75.7 / 77.6 & 40.6 / 40.6 / 44.3 & 1 \\
\midrule
4 & trailing & 1.41 & 0.8 & 2/3 & 194.0 / 173.7 / $>$548.7 & 5.5 / 7.4 / -- & 1 / 1 / 0.048 \\
4 & trailing & 1.41 & 1.4 & 2/3 & 354.7 / 247.6 / $>$548.7 & 12.9 / 14.8 / -- & 1 / 1 / 0.022 \\
4 & trailing & 1.20 & 0.8 & 0/3 & $>$548.7 & -- & 0 \\
4 & trailing & 1.20 & 1.4 & 0/3 & $>$548.7 & -- & 0 \\
4 & trailing & 1.09 & 0.8 & 0/3 & $>$548.7 & -- & 0 \\
4 & trailing & 1.09 & 1.4 & 0/3 & $>$548.7 & -- & 0 \\
4 & leading & 1.41 & 0.8 & 1/3 & $>$548.7 / 382.4 / $>$548.7 & $>$546.9 / 315.9 / $>$546.9 & 0.222 / 1 / 0 \\
4 & leading & 1.41 & 1.4 & 0/3 & $>$548.7 & $>$546.9 & 0 / 0 / 0.014 \\
4 & leading & 1.09 & 1.4 & 0/3 & $>$548.7 & $>$546.9 & 0 \\
\bottomrule
\end{tabular}}
\\[4pt]
\parbox{0.97\textwidth}{\footnotesize $f$ is the body force on membrane and cytosol, and $S/V$ the measured surface-to-volume ratio. The runs last $100\,000$ steps after the drive starts at $f = 8$, $370$~ms, and $150\,000$ steps at $f = 4$, $554$~ms. A cell is at the slit from the frame at which its front comes within $0.8~\mu$m of the upstream plane of the wall to the frame at which its rear is $0.8~\mu$m past the downstream plane. A vacuole is timed likewise, with its front and rearmost vertices. ``Membrane past'' is the fraction of membrane vertices downstream of the wall in the last frame.}
\end{table}

\section{The filter runs}\label{sec:filter}

\textit{Measurements.} Transit runs from the frame at which the cell front reaches the bead layer to the frame at which its rear leaves it, with frames every $1000$ steps. The advance is the net displacement of the membrane centroid over the run, and a cell that advanced less than $15~\mu$m is scored as held. The gap is the smallest distance between a membrane vertex and a vacuole vertex, taken as its median on the approach, and the envelope is $D + 2\times\mathrm{gap}$, the size that must pass the throat. $V_{\min}$ and $A_{\max}$ are the smallest volume and the largest area a vacuole reaches over the run, relative to its value in the first frame.

\textit{One vacuole} (table~\ref{tab:filter1}). At the default volume constraint, $k_v = 5000$, every diameter passed, with the transit rising from $225.4\pm7.4$~ms at $D = 1.0~\mu$m to $259.9\pm2.1$~ms at $D = 2.4~\mu$m against $208.2\pm2.1$~ms for a cell with no vacuole. Passage is accompanied by a large deflation, $V_{\min} = 0.695$--$0.849$. That deflation is not physical. The constraint $U_{\mathrm{vol}} = k_v(V-V_0)^2/2V_0$ with $k_v = 5000$ opposes a $28\%$ compression with only about $31$~Pa, whereas the osmotic pressure of a cytosol-like interior is of order $2.9\times10^5$~Pa. Even at $31$~Pa, a water permeability in the measured range for lipid membranes, $P_f = 30$--$150~\mu$m/s, that is $L_p \approx 2\times10^{-13}$--$1\times10^{-12}$~m\,s$^{-1}$\,Pa$^{-1}$, would need hours to move that volume, and tens of hours at the $\sim\!1$~Pa the flow actually applies. Raising $k_v$ to $5\times10^5$ holds the volume to $1.000$, and the cell still passed at $D = 1.0$ and $1.4~\mu$m, but now the vacuole membrane stretches in area by $5$--$15\%$, up to several times the few percent at which bilayers rupture~\cite{rawicz2000effect,olbrich2000water}. At $D \geq 1.8~\mu$m with the volume held, all twelve runs failed the containment criterion between steps $65\,000$ and $71\,000$, at the pore: the membrane--vacuole barrier gives way, and no vacuole was still hosted at the end. We report those runs as a limit of the model and draw no passage conclusion from them. An intermediate constraint, $k_v = 5\times10^4$ at $D = 2.2~\mu$m, behaves the same way (three runs, all invalid from steps $67\,000$--$68\,000$, $V_{\min} = 0.996$--$0.997$). One run at $D = 2.4~\mu$m and $k_v = 5000$ exited at step $57\,000$ and was replaced by a fourth seed.

\textit{Displaced-fluid drive and the leading rim.} Every run of table~\ref{tab:filter1} placed the vacuole in the trailing rim, at $(1.4, 4.75, 4.75)~\mu$m, and drove it with the nominal share of the force (section~\ref{sec:drive}). In a control set of twelve runs, three seeds each, a vacuole of $D = 1.4$ or $2.2~\mu$m is driven instead with the fluid it displaces, $N_{\mathrm{def}} = 33.4$ and $78.1$ (the second scaled from the calibration of section~\ref{sec:slit} by the volume of the vacuole together with its $0.52~\mu$m solvent layer), and placed either in the trailing rim or in the leading rim at $(6.6, 4.75, 4.75)~\mu$m. In the trailing rim this drive changed no outcome and the transit by $3$--$5\%$: a cell with a $1.4~\mu$m vacuole crossed in $222$--$229$~ms against $229$--$233$~ms with the nominal share, and one with a $2.2~\mu$m vacuole in $236$--$240$ against $248$--$262$~ms. In the leading rim the same vacuoles reached the layer first. A $1.4~\mu$m vacuole there held its cell at the layer for the whole run in two of three seeds, and the third cell crossed only after $510$~ms; a $2.2~\mu$m one delayed every crossing, to $384$--$414$~ms, and left the layer one frame before the rear of the cell. We have not resolved why the smaller body is the more obstructive.

\begin{table}[htbp]
\small
\centering
\renewcommand\arraystretch{1.2}
\caption{\textbf{The filter runs with one vacuole or none.} Transit is the mean $\pm$ SD over the valid runs of that set; $V_{\min}$, $A_{\max}$, the gap, the envelope and the advance are ranges or means over all runs of the set.}
\label{tab:filter1}
\resizebox{\textwidth}{!}{\begin{tabular}{lccccccc}
\toprule
Set & Valid & Transit (ms) & $V_{\min}$ & $A_{\max}$ & Gap ($\mu$m) & Envelope ($\mu$m) & Advance ($\mu$m) \\
\midrule
No vacuole & 3/3 & $208.2 \pm 2.1$ & -- & -- & -- & -- & 37.6--38.5 \\
$k_v = 5000$, $D = 1.0$ & 3/3 & $225.4 \pm 7.4$ & 0.837--0.849 & 1.024--1.026 & 0.654 & 2.30 & 35.1--36.3 \\
$k_v = 5000$, $D = 1.4$ & 3/3 & $231.6 \pm 2.1$ & 0.769--0.779 & 1.032--1.034 & 0.644 & 2.69 & 28.0--34.6 \\
$k_v = 5000$, $D = 1.8$ & 3/3 & $242.7 \pm 4.3$ & 0.715--0.723 & 1.032--1.034 & 0.637 & 3.07 & 27.6--29.0 \\
$k_v = 5000$, $D = 2.0$ & 3/3 & $246.4 \pm 4.3$ & 0.709--0.748 & 1.036--1.044 & 0.633 & 3.27 & 25.8--26.8 \\
$k_v = 5000$, $D = 2.2$ & 3/3 & $252.5 \pm 8.5$ & 0.708--0.723 & 1.038--1.050 & 0.631 & 3.46 & 25.5--25.6 \\
$k_v = 5000$, $D = 2.4$ & 3/3 & $259.9 \pm 2.1$ & 0.695--0.733 & 1.036--1.059 & 0.632 & 3.67 & 25.3--25.6 \\
$k_v = 5\times10^5$, $D = 1.0$ & 3/3 & $222.9 \pm 2.1$ & 1.000 & 1.048--1.053 & 0.652 & 2.32 & 33.7--35.3 \\
$k_v = 5\times10^5$, $D = 1.4$ & 3/3 & $235.3 \pm 2.1$ & 1.000 & 1.123--1.153 & 0.640 & 2.70 & 28.3--30.2 \\
$k_v = 5\times10^5$, $D = 1.8$ & 0/3 & -- & 1.000 & 1.142--1.147 & 0.633 & 3.10 & 23.0--23.4 \\
$k_v = 5\times10^5$, $D = 2.0$ & 0/3 & -- & 1.000 & 1.093--1.103 & 0.630 & 3.30 & 22.9--23.3 \\
$k_v = 5\times10^5$, $D = 2.2$ & 0/3 & -- & 1.000 & 1.082--1.093 & 0.629 & 3.50 & 22.8--23.2 \\
$k_v = 5\times10^5$, $D = 2.4$ & 0/3 & -- & 1.000 & 1.072--1.085 & 0.629 & 3.70 & 22.9--23.3 \\
$k_v = 5\times10^4$, $D = 2.2$ & 0/3 & -- & 0.996--0.997 & 1.075--1.079 & 0.630 & 3.49 & 22.9--23.3 \\
\bottomrule
\end{tabular}}
\\[4pt]
\parbox{0.97\textwidth}{\footnotesize Diameters are in $\mu$m. ``Valid'' counts the runs in which the vacuole stayed hosted and at least $99\%$ inside the membrane in every frame. The invalid sets failed between steps $65\,000$ and $71\,000$ in every run, at the pore, and no transit is quoted for them. The $k_v = 5\times10^5$ sets are the corresponding $k_v = 5000$ simulations rerun from step $0$ with the same data and seeds and only $k_v$ changed. The envelope reaches $3.67~\mu$m at the largest size, about twice the effective throat of $1.75~\mu$m, so the vacuole passes only by deforming.}
\end{table}

\textit{Eight vacuoles} (table~\ref{tab:filter8}). Eight vacuoles of $D = 1.1~\mu$m have the volume of one of $D = 2.2~\mu$m. Of the $17$ runs, $13$ kept every vacuole inside the host to the end, and all $13$ held their cell at the bead layer, advancing $4.36 \pm 0.37~\mu$m (range $3.96$--$5.08$) against $25.53 \pm 0.10~\mu$m for the three runs carrying a single $2.2~\mu$m vacuole at the same barrier cutoff, all of which passed. The four runs that lost a vacuole all passed, which is why the loss is a disqualification rather than a result. The outcome does not depend on the vacuole--vacuole pair (barrier or plain soft), on the membrane--vacuole cutoff ($0.6$ or $0.7~\mu$m) or on the volume constraint ($5000$ or $5\times10^5$). One structural caveat belongs with this comparison: the repulsive shells the model puts around eight vacuoles exclude $43$--$55~\mu$m$^3$ of the $92.5~\mu$m$^3$ cell, against $22~\mu$m$^3$ for a single $2.2~\mu$m vacuole, so part of the effect is excluded volume rather than the number of bodies as such.

\begin{table}[htbp]
\small
\centering
\renewcommand\arraystretch{1.15}
\caption{\textbf{The filter runs with eight vacuoles of $D = 1.1~\mu$m.} ``Hosted'' is the number still inside the cell at the end; the first bad step is where the containment criterion first failed. $V_{\min}$ and $V_{\mathrm{end}}$ are relative to the first frame of the run, in which the $34$-vertex meshes sit about $15\%$ below their target volume.}
\label{tab:filter8}
\begin{tabular}{llcccccc}
\toprule
Setting & Run & Hosted & First bad step & Advance ($\mu$m) & Outcome & $V_{\min}$ & $V_{\mathrm{end}}$ \\
\midrule
Cutoff 0.6 & s1 & 8/8 & -- & 3.96 & held & 1.000 & 1.079 \\
Cutoff 0.6 & s2 & 8/8 & -- & 3.98 & held & 1.000 & 1.095 \\
Cutoff 0.6 & s3 & 6/8 & $64\,000$ & 24.96 & passed & 0.911 & 1.092 \\
Cutoff 0.6 & s4 & 8/8 & -- & 4.07 & held & 1.000 & 1.085 \\
Cutoff 0.6 & s5 & 8/8 & -- & 4.06 & held & 1.000 & 1.085 \\
Cutoff 0.6, soft pair & s1 & 7/8 & $88\,000$ & 24.01 & passed & 0.932 & 1.096 \\
Cutoff 0.6, soft pair & s2 & 8/8 & -- & 4.05 & held & 1.000 & 1.090 \\
Cutoff 0.6, soft pair & s3 & 5/8 & $70\,000$ & 24.65 & passed & 0.951 & 1.092 \\
Cutoff 0.7 & s1 & 8/8 & -- & 4.48 & held & 1.000 & 1.086 \\
Cutoff 0.7 & s2 & 8/8 & -- & 4.76 & held & 1.000 & 1.086 \\
Cutoff 0.7 & s3 & 8/8 & -- & 4.88 & held & 1.000 & 1.085 \\
Cutoff 0.7, soft pair & s1 & 8/8 & -- & 4.14 & held & 1.000 & 1.100 \\
Cutoff 0.7, soft pair & s2 & 8/8 & -- & 4.53 & held & 1.000 & 1.101 \\
Cutoff 0.7, soft pair & s3 & 8/8 & -- & 4.41 & held & 1.000 & 1.098 \\
Cutoff 0.7, $k_v = 5\times10^5$ & s1 & 6/8 & $72\,000$ & 24.40 & passed & 1.000 & 1.188 \\
Cutoff 0.7, $k_v = 5\times10^5$ & s2 & 8/8 & -- & 5.08 & held & 1.000 & 1.188 \\
Cutoff 0.7, $k_v = 5\times10^5$ & s3 & 8/8 & -- & 4.29 & held & 1.000 & 1.187 \\
\bottomrule
\end{tabular}
\\[4pt]
\parbox{0.97\textwidth}{\footnotesize At $k_v = 5000$ the eight vacuoles end at $1.08$--$1.10$ of their initial volume, about $93\%$ of the target sphere; at $k_v = 5\times10^5$ they end at $1.187$--$1.188$, that is at the target. The largest area reached is $1.094$--$1.106$ of the initial value at $k_v = 5000$ and $1.136$--$1.189$ at $k_v = 5\times10^5$. ``Soft pair'' replaces the vacuole--vacuole barrier by the plain soft DPD pair ($a = 30$, $\gamma = 15$, $\sigma = 2.3774$, $r_c = 1$).}
\end{table}

\section{Repeated passage}\label{sec:passage}

\textit{Set-up.} The cell of the fusion cascade, its ten vacuoles of $D = 0.6~\mu$m placed as there, was turned edge-on to the aperture of the slit set-up (section~\ref{sec:slit}), centred $6.0~\mu$m upstream of the wall, and relaxed for $8000$ steps with its membrane permeable to the solvent and $5000$ sealed, with the membrane--vacuole barrier of $0.6~\mu$m cutoff. The relaxed cell has $11$ of its $45$ pairs within $0.5~\mu$m. It is driven as at the slit, at $f = 8$ on membrane and cytosol, each vacuole carrying $f N_{\mathrm{def}}$ with $N_{\mathrm{def}} = 9.9$. The box is periodic, so a driven cell returns to the wall once per lap. A second set-up pierces the same wall with an aperture of $9.87\times11.18~\mu$m between particle centres, wider than the cell, and in the third arm the cell is not driven. The runs without fusion proceed for $300\,000$ steps after $2000$ undriven ones, three seeds per arm. The runs with fusion proceed at half the time step, $0.0005$, in five undriven segments of $800$ steps and up to $625$ driven ones of $960$, which is the same $1108$~ms of drive, with vacuole vertices written every $40$ steps. The rule is that of the cascade at $\delta_f = 0.7~\mu$m and $t_f = 3.6$~ms on this set-up's time scale. Each arm was run with three seeds under each of two area rules: the cascade's, in which the product takes the summed area of its parents, and one in which it takes the area of a sphere of volume $V_0/0.97$, $0.97$ being the reduced volume of an unfused vacuole, so that it is a near-sphere with less membrane than its parents together (section~\ref{sec:fusion}). A product is built as a sphere of its target area. Over the segment after the fusion its volume target is brought geometrically from the volume of that sphere to the summed volume in $16$ stages. Its drive is $f N_{\mathrm{def}}$ scaled by the volume of the product grown by $0.52~\mu$m, the solvent layer that sets $N_{\mathrm{def}}$ for an unfused vacuole.

\textit{Runs without fusion.} Vacuole vertices were written every $20$ steps. For each of the $45$ pairs, the surface separation $s$ is the smallest distance between a vertex of one vacuole and a vertex of the other, under the minimum-image convention. A pair is in contact while $s < \delta_f = 0.7~\mu$m, and a contact is a maximal run of such frames. The statistics are taken over the $1108.6$~ms that follow the start of the drive, $2000$ steps into the run, and over the same window for the cell at rest. A contact already under way when the window opens has no onset and is not counted as new; a new contact qualifies once it has lasted $t_f = 3.6$~ms, including one still under way at the end. The separation of pairs in contact is summarized by its fifth percentile over every frame of every pair in contact, $d_5$, and the closest approach of the record by $d_{\min}$. A pass is counted each time the unwrapped membrane centroid crosses the mid-plane of the wall in a new lap; a centroid that falls back across that plane counts again only after it has crossed it forward again. Table~\ref{tab:passcontact} lists the nine runs.

\textit{Runs with fusion.} Each run is a chain of segments whose step counters restart at zero. Times are counted from the start of the drive, so the fusions made during the five undriven segments, $7.4$~ms in all, fall at negative times, and a fusion is executed at the end of the segment in which it qualifies. Because the rebuild of the topology resets the image flags at every segment, passes are counted from the membrane centroid in the last frame of each segment, chained across segments by taking the nearest lap. A pending pair is one whose fusion had qualified but was still deferred at the end of the record, because its product could not be placed inside the host. A run that failed was restarted from the start of the failed segment with a new thermostat seed, at most three times over the run, and a fourth failure ended it; the record of a run is the segments it completed. Table~\ref{tab:passfusion} lists the $18$ runs. They are a second set: a first, at $\Delta t = 0.001$ and with the volume target of each new product set to its final value at once rather than in stages, was replaced after all six of its driven runs ended early, four of them on an over-extended bond of a vacuole mesh.

\textit{What the fusion runs add.} With near-spherical products the last fusion had come by $69$~ms in seven of the nine runs, before a squeezed cell reached the slit for the second time, and the latest, at $291$~ms, was in a cell at rest. With summed areas each cell at rest still held one to three qualified pairs whose products could not be placed. Five of the six driven runs failed on a vacuole-mesh bond $113$--$311$~ms into the drive and $6$--$153$~ms after their last fusion, and the sixth, restarted twice, kept two vacuoles to the end.

\begin{table}[htbp]
\small
\centering
\renewcommand\arraystretch{1.15}
\caption{\textbf{Repeated passage, runs without fusion.} Contacts are counted over the $1108.6$~ms after the drive starts, a pair being in contact while the surfaces of its two vacuoles are within $\delta_f = 0.7~\mu$m.}
\label{tab:passcontact}
\resizebox{\textwidth}{!}{\begin{tabular}{lccccccccc}
\toprule
Arm & Seed & Passes & In contact & New contacts & Lasting $t_f$ & Per $100$~ms & Pairs & $d_5$ ($\mu$m) & $d_{\min}$ ($\mu$m) \\
\midrule
Rest & 1 & 0 & 13.40 & 101 & 11 & 0.99 & 9 & 0.455 & 0.418 \\
Rest & 2 & 0 & 15.49 & 48 & 14 & 1.26 & 8 & 0.453 & 0.419 \\
Rest & 3 & 0 & 13.03 & 11 & 6 & 0.54 & 5 & 0.454 & 0.419 \\
\midrule
Squeezed & 1 & 16 & 11.46 & 230 & 189 & 17.05 & 43 & 0.441 & 0.404 \\
Squeezed & 2 & 15 & 9.44 & 269 & 197 & 17.77 & 33 & 0.442 & 0.403 \\
Squeezed & 3 & 16 & 8.38 & 180 & 147 & 13.26 & 24 & 0.442 & 0.364 \\
\midrule
Wide & 1 & 67 & 17.69 & 72 & 32 & 2.89 & 17 & 0.441 & 0.406 \\
Wide & 2 & 68 & 23.13 & 123 & 55 & 4.96 & 33 & 0.446 & 0.404 \\
Wide & 3 & 63 & 22.10 & 310 & 93 & 8.39 & 34 & 0.447 & 0.405 \\
\bottomrule
\end{tabular}}
\\[4pt]
\parbox{0.97\textwidth}{\footnotesize ``In contact'' is the mean number of the $45$ pairs in contact; ``new contacts'' are those that began within the window, and ``lasting $t_f$'' those of them that lasted $t_f = 3.6$~ms, also given per $100$~ms; ``pairs'' is the number of distinct pairs with at least one such contact. $d_5$ is the fifth percentile of the surface separation over all frames of all pairs in contact and $d_{\min}$ the smallest separation of the record; both are close to the separation of about $0.43~\mu$m at which the vacuole--vacuole barrier holds the vertices of different vacuoles (section~\ref{sec:fusion}). The wide arm crosses an opening of $9.87\times11.18~\mu$m between particle centres, wider than the cell.}
\end{table}

\begin{table}[htbp]
\small
\centering
\renewcommand\arraystretch{1.1}
\caption{\textbf{Repeated passage, runs with fusion at $\delta_f = 0.7~\mu$m and $t_f = 3.6$~ms.} Times are from the start of the drive, or from the same step for the cell at rest.}
\label{tab:passfusion}
\resizebox{\textwidth}{!}{\begin{tabular}{llcccccccc}
\toprule
Arm & Product area & Seed & At drive start & At end & Last fusion (ms) & Pending & Passes & Restarts & End (ms) \\
\midrule
Rest & summed area & 1 & 6 & 3 & 505.5 & 3 & 0 & 0 & 1108.5 \\
Rest & summed area & 2 & 6 & 3 & 567.6 & 1 & 0 & 1 & 1108.5 \\
Rest & summed area & 3 & 6 & 3 & 133.0 & 2 & 0 & 0 & 1108.5 \\
\cmidrule(lr){2-10}
Rest & near-sphere & 1 & 5 & 1 & 69.2 & 0 & 0 & 0 & 1108.5 \\
Rest & near-sphere & 2 & 5 & 2 & 46.1 & 0 & 0 & 0 & 1108.5 \\
Rest & near-sphere & 3 & 5 & 2 & 290.9 & 0 & 0 & 0 & 1108.5 \\
\midrule
Squeezed & summed area & 1 & 6 & 1 & 138.3 & 0 & 2 & 3 & 178.9$^{*}$ \\
Squeezed & summed area & 2 & 6 & 1 & 196.9 & 0 & 3 & 3 & 203.0$^{*}$ \\
Squeezed & summed area & 3 & 6 & 2 & 23.1 & 0 & 3 & 3 & 176.1$^{*}$ \\
\cmidrule(lr){2-10}
Squeezed & near-sphere & 1 & 5 & 1 & 69.2 & 0 & 15 & 0 & 1108.5 \\
Squeezed & near-sphere & 2 & 5 & 1 & 37.2 & 0 & 15 & 0 & 1108.5 \\
Squeezed & near-sphere & 3 & 5 & 2 & 12.4 & 0 & 15 & 0 & 1108.5 \\
\midrule
Wide & summed area & 1 & 5 & 1 & 47.9 & 0 & 6 & 3 & 112.8$^{*}$ \\
Wide & summed area & 2 & 5 & 1 & 198.6 & 0 & 16 & 3 & 310.8$^{*}$ \\
Wide & summed area & 3 & 6 & 2 & 23.1 & 0 & 61 & 2 & 1108.5 \\
\cmidrule(lr){2-10}
Wide & near-sphere & 1 & 5 & 2 & 12.4 & 0 & 57 & 0 & 1108.5 \\
Wide & near-sphere & 2 & 5 & 2 & 12.4 & 0 & 59 & 0 & 1108.5 \\
Wide & near-sphere & 3 & 6 & 1 & 145.4 & 0 & 56 & 0 & 1108.5 \\
\bottomrule
\end{tabular}}
\\[4pt]
\parbox{0.97\textwidth}{\footnotesize ``Summed area'' gives each product the area of its two parents together, the rule of the cascade; ``near-sphere'' gives it the area of a sphere of volume $V_0/0.97$. The vacuole counts are those when the drive started, $7.4$~ms into the run, and at the end of the record. ``Pending'' counts the pairs whose fusion had qualified and was still deferred at the end because the product could not be placed. In the three runs at rest with summed areas their fusions had been deferred $1134$, $615$ and $1160$ times in all, beginning at $92.2$, $19.5$ and $24.8$~ms. ``Restarts'' counts the restarts after a failed segment. $^{*}$The run stopped when a worm-like-chain bond of a vacuole mesh exceeded its maximum extension for the fourth time; the time given is that of the failure, in the segment after the last one completed.}
\end{table}

\section{Reynolds and capillary numbers}\label{sec:reynolds}

For each geometry we report $\mathrm{Re} = \rho_f u L/\eta_p$, with $\rho_f = 1000~\mathrm{kg/m^3}$, $\eta_p = 1.2$~mPa\,s, the measured mean velocity $u$ converted with the $\tau$ of that geometry, and the channel size $L$; and $\mathrm{Ca} = \eta_p \dot\gamma R_0/\mu_0$, with the wall shear rate $\dot\gamma$ and the cell radius $R_0 = \sqrt{A_0/4\pi} = 3.25~\mu$m. In the encounter tube $\mathrm{Re} = 2.4\times10^{-3}$ and $\mathrm{Ca} = 0.114$ at $2u_c/R_t$; in plane shear $\mathrm{Re} = 2.3\times10^{-3}$ and $\mathrm{Ca} = 0.116$, so the two arms are matched in capillary number as closely as in shear rate. In the vessels $\mathrm{Re} = 1.6\times10^{-3}$ (bifurcation), $1.3\times10^{-2}$ (curved vessel, over the solvent inside the lumen; table~\ref{tab:tau}) and $1.9\times10^{-2}$ (dilation, measured over the parent lumen rather than the sac). These are the Reynolds numbers of plasma at the mapped size and speed, and all are deep in the Stokes regime; those of the simulated fluid are not (below). In a tube we give the wall shear rate both as $4u/R_t$, from the mean velocity $u$, and as $2u_c/R_t$, from the centreline velocity $u_c$, with $R_t$ the tube radius; the two agree for Poiseuille flow and differ when the cell blunts the profile.

\textit{The Reynolds number of the simulated fluid.} The mapping of section~\ref{sec:units} fixes length, energy and time, and with them the unit of mass, $E_0\tau^2/L_0^2$. A solvent of $\rho$ particles of unit mass per $\mu$m$^3$ then has the density $\rho E_0\tau^2/L_0^5$, between $1.3$ and $4.0\times10^6~\mathrm{kg/m^3}$ in our geometries, $1300$ to $4000$ times that of plasma. Its own Reynolds number, $\rho u L/\eta^{\mathrm{DPD}}$ in reduced units, exceeds the values above by the same factor: it is about $3$ in the encounter tube, $7$ in plane shear, $6$ in the bifurcation, $29$ in the curved vessel and $40$ in the dilation. Along a straight tube or between the plates of plane shear the undisturbed flow of a Newtonian fluid is the same at any Reynolds number below transition; where the lumen splits, turns or widens it is not. We have not measured how much the flow changes in the bifurcation or the curved vessel. In the dilation the change decides the flow in the sac.

\textit{The flow in the dilation's sac.} To separate what the geometry imposes from what the model's inertia adds, we solved the steady, incompressible Navier--Stokes equations without cells in the dilation of section~\ref{sec:geometry}, taken as the union of the tube, of radius $12.5~\mu$m, and the sphere, of radius $25~\mu$m and centred $15~\mu$m above the tube's axis. The domain is periodic along $x$ over $161~\mu$m and driven, like the DPD simulation, by a uniform body force; its walls are no-slip and meshed as a staircase of $0.5~\mu$m cubes, $1.03\times10^6$ of them in the fluid. We used \texttt{simpleFoam} from OpenFOAM (ESI v2512), laminar, with linear-upwind convection, and iterated each case until the residuals fell below $10^{-7}$ for pressure and $10^{-8}$ for velocity, which took $1181$ to $1335$ iterations. The kinematic viscosity is that of the parent lumen's solvent at step $150\,000$, $\nu = \eta^{\mathrm{DPD}}/\rho = 11.08/3.70 = 2.99$ in reduced units. We applied three forces. The first makes the lumen's mean speed equal to that of the DPD solvent over steps $50\,000$--$150\,000$, $5.06$, so that $\mathrm{Re} = uD/\nu = 42$ as in the DPD ($40$ with the speed at step $120\,000$ and the density of table~\ref{tab:tau}). The second drives the lumen $15\%$ faster, and the third brings it down to $0.023$, close to the $0.019$ of plasma in the dilation. Each field was averaged on the $2~\mu$m bins of the DPD field and scaled to the same lumen mean, so that only the shape of the flow is compared (table~\ref{tab:foam}).

At the DPD fluid's Reynolds number the cell-free sac holds a vortex like the DPD's. It turns the same way; its fluid moves upstream in $59\%$ of the sac against $72\%$ in the DPD, at up to $0.09$ of the lumen's mean speed against $0.11$; and its axial velocity correlates with the DPD's at $r = 0.93$. Near plasma's Reynolds number there is no reverse flow anywhere in the sac, and its fluid moves downstream at $13\%$ of the lumen's mean speed, where the DPD's barely moves. The vertical velocity then anticorrelates with the DPD's ($r = -0.73$): fluid enters the sac on its upstream side and leaves on its downstream side instead of turning over. The vortex is therefore an effect of the model's inertia rather than of the geometry, and so may be the retention of cells in the sac (main text). The comparison leaves out the cells that fill the DPD sac, and the DPD's vortex is the stronger and reaches lower: on the sac's vertical axis its flow reverses $6.2~\mu$m above the tube's top wall, against $9.9~\mu$m without cells.

\begin{table}[htbp]
\centering
\renewcommand\arraystretch{1.2}
\caption{\textbf{Flow in the dilation's sac: the DPD solvent against cell-free Navier--Stokes flow through the same geometry.}}
\label{tab:foam}
\begin{tabular}{lcccccc}
\toprule
Field & Re & $V_{\mathrm{rev}}$ & $u_{\mathrm{rev}}$ & $\bar{u}_{\mathrm{sac}}$ & $h_{\mathrm{rev}}$ ($\mu$m) & $r(v_x)$ / $r(v_y)$ \\
\midrule
DPD solvent, steps $50\,000$--$150\,000$ & $42$ & $0.72$ & $0.113$ & $-0.007$ & $6.2$ & -- \\
Navier--Stokes & $42$ & $0.59$ & $0.091$ & $0.044$ & $9.9$ & $0.93$ / $0.52$ \\
Navier--Stokes & $49$ & $0.61$ & $0.098$ & $0.038$ & $9.2$ & $0.94$ / $0.66$ \\
Navier--Stokes & $0.023$ & $0$ & -- & $0.132$ & -- & $0.80$ / $-0.73$ \\
\bottomrule
\end{tabular}
\\[4pt]
\parbox{0.97\textwidth}{\footnotesize $\mathrm{Re} = uD/\nu$, with the parent lumen's mean speed $u$ over the same bins in every field, $D = 25~\mu$m and $\nu = 2.99$; velocities are fractions of $u$. The sac is the fluid inside the sphere and outside the tube. $V_{\mathrm{rev}}$ is the fraction of its volume in which the fluid moves upstream, and $u_{\mathrm{rev}}$ the fastest such motion. $\bar{u}_{\mathrm{sac}}$ is its mean axial speed at $x = 55$--$106~\mu$m, the region of table~\ref{tab:tau}, here averaged over the record rather than taken at one step. $h_{\mathrm{rev}}$ is the height above the tube's top wall at which the flow reverses, going up the vertical line through the sphere's centre, and $r$ is the correlation with the DPD field over the sac; both are taken in the $4~\mu$m slab through the tube's axis. A dash under $u_{\mathrm{rev}}$ and $h_{\mathrm{rev}}$ means that there is no reverse flow.}
\end{table}

\section{The population model}\label{sec:pop}

Table~\ref{tab:popmodel} sets the best point of the scan described below against the published record.

\begin{table}[htbp]
\small
\centering
\renewcommand\arraystretch{1.15}
\caption{\textbf{The population model against the published record.}
Best point of a $32\,400$-combination scan scored against every observable except the four that
describe the sub-visible pool: $J_0 = 0.60$/day, $\tau = 35$~d, $\phi = 0$, birth-size median $90$~nm with
log-standard-deviation $0.65$, $f_{ab} = 0.01$, $m_J = 25$, $m_d = 1$, and splenic clearance rising
$5.6$-fold over the cell's life. Only the $144$~h endpoint of the six-day curve was a scored
target. Density fractions are
listed lightest to densest, that is youngest to oldest. The healthy prevalence of the largest
cohort is given as mean $\pm$ SE. The two sub-visible entries are
predictions; the pool is compared with Reinhart and Chien's estimate corrected for section
thickness, as set out in the main text.}
\label{tab:popmodel}
\begin{tabular}{p{5.4cm}p{3.3cm}p{3.3cm}p{2.5cm}}
\toprule
Observable & Model & Observed & Source \\
\midrule
Post-splenectomy acquisition, $0/48/96/144$~h (\%) & $0$ / $1.38$ / $2.66$ / $3.88$ & $0$ / $1.31$ / $2.91$ / $3.71$ & \cite{sills1988physiologic} \\
\quad \emph{only $144$~h fitted; raw counts $0.4$ / $1.7$ / $3.3$ / $4.1\%$,} & & & \\
\quad \emph{corrected to the cells free at $t=0$} & & & \\
In vitro acquisition by fraction, $144$~h (\%) & $8.95$ / $3.51$ / $1.75$ & $10.3$ / $3.3$ / $2.4$ & \cite{sills1988formation} \\
In vitro top-to-bottom ratio & $5.13$ & $4.29$ & \cite{sills1988formation} \\
In vivo top-to-bottom ratio, spleen present & $9.02$ & $13.5$ & \cite{sills1988formation} \\
Healthy prevalence & $0.48\%$ & $0.52\pm0.04\%$ ($n = 84$) & \cite{buchanan1987formation} \\
\quad \emph{scored against $0.5$--$2.6\%$, five series} & & & \cite{holroyde1970acquisition,sills1988formation,buchanan1987pocked,colin1991spontaneous} \\
Asplenic prevalence & $41.1\%$ & $20$--$70\%$ & \cite{holroyde1970acquisition,sills1988formation,buchanan1987pocked,buchanan1987formation} \\
\quad \emph{scored against $25$--$55\%$, series means} & & & \cite{holroyde1970acquisition,sills1988formation} \\
Time after splenectomy to $20\%$ / to plateau & $37$ / $120$~d & $20$--$40$ / $60$--$100$~d & \cite{buchanan1987formation} \\
\quad \emph{level fitted, approach not} & & & \\
Largest body & $1.57~\mu$m & $1.5~\mu$m & \cite{holroyde1970acquisition} \\
Most loaded cell & $26$ bodies & one to over $20$ & \cite{holroyde1970acquisition} \\
Vacuole membrane of a minority cell at $120$~d & $24\%$ of $140~\mu$m$^2$ & must be $<100\%$ & physical \\
\midrule
Sub-visible pool, healthy (\emph{predicted}) & $5.6$ per cell & $7.7$--$9.6$ (corrected) & \cite{reinhart1988red} \\
Sub-visible rise after splenectomy (\emph{predicted}) & $3.2\times$ & $4\times$ & \cite{reinhart1988red} \\
\bottomrule
\end{tabular}
\end{table}

\textit{Numerical implementation.} Ages are represented on $121$ points, one per day from $0$ to $120$~d, and integrated with trapezoidal weights; diameters on $400$ points spaced logarithmically from $0.04$ to $2.0~\mu$m. The birth-size law is lognormal in diameter, truncated to that grid and normalized on it. The splenic gate is the logistic
\begin{equation}
g(D) = \frac{1}{1 + \exp[-(D - D_g)/w]}, \qquad D_g = 0.35~\mu\mathrm{m}, \quad w = 0.05~\mu\mathrm{m}.
\end{equation}
With the floor $\phi$ of the source, the inner integral of the main text has a second term,
\begin{equation}
\int_0^a J(s)\,e^{-R(a-s)}\,\mathrm{d}s = J_0\left[\frac{e^{-a/\tau} - e^{-Ra}}{R - 1/\tau} + \phi\,\frac{1 - e^{-Ra}}{R}\right],
\end{equation}
in which the second fraction takes its limit $a$ where $R = 0$, as it is wherever the spleen is absent, and the first its limit $a\,e^{-Ra}$ where $R = 1/\tau$. As in the main text, $R$ is evaluated at the cell's current age. A body is visible above the DIC limit of $0.30~\mu$m, and the sub-visible pool is the expected number of bodies between the foot of the grid and that limit. The membrane of a cell's bodies is $\pi D^2$ summed over their expected numbers and is compared with $140~\mu$m$^2$ for the cell. The extremes are taken in a population of $10^4$ cells without a spleen. The largest body is the largest diameter $D$ for which the expected number of cells carrying a body of at least $D$ reaches one; the most loaded cell is the largest number $m$ for which the expected number of cells carrying at least $m$ visible bodies reaches one. The six-day acquisition is the fraction of the cells free of visible bodies at the start that carry one six days later. Each cell keeps the nucleation rate of its age over the six days, so it acquires a visible body with probability $1 - \exp[-J(a)\,p_{\mathrm{vis}}\,t]$, $p_{\mathrm{vis}}$ being the fraction of births above the DIC limit. The probability is averaged over the whole population with the spleen present at the start for the acquisition after splenectomy, and over a density fraction for the acquisition in vitro.

\textit{The scan.} Table~\ref{tab:popgrid} gives the nine scanned parameters and their levels, $32\,400$ combinations in all, every one of which was evaluated without failure. The decay time $\tau = 10^6$~d is the control of a source that never switches off; $m_J = m_d = 1$ makes the minority identical to the other cells; and $\tau_R = \infty$ is a clearance that does not depend on age. The rise of clearance over the life of the cell is $e^{120/\tau_R}$.

\begin{table}[htbp]
\small
\centering
\renewcommand\arraystretch{1.15}
\caption{\textbf{The parameters of the population scan.}}
\label{tab:popgrid}
\begin{tabular}{llc}
\toprule
Parameter & Levels & Number \\
\midrule
Peak nucleation rate $J_0$ (per day) & $0.3$, $0.6$, $1.2$, $2.5$, $5.0$ & $5$ \\
Decay time of the source $\tau$ (d) & $5$, $10$, $20$, $35$, $10^6$ & $5$ \\
Floor of the source $\phi$ & $0$, $0.02$, $0.10$ & $3$ \\
Median birth diameter $d_{50}$ (nm) & $90$, $110$ & $2$ \\
Log-standard-deviation of birth diameters $\sigma$ & $0.50$, $0.65$, $0.80$ & $3$ \\
Fraction of fast-nucleating cells $f_{ab}$ & $0.001$, $0.003$, $0.01$ & $3$ \\
Rate multiplier of that fraction $m_J$ & $1$, $25$, $50$ & $3$ \\
Size multiplier of that fraction $m_d$ & $1.0$, $1.6$ & $2$ \\
Age scale of splenic clearance $\tau_R$ (d) & $\infty$, $200$, $110$, $70$ & $4$ \\
\quad \emph{rise of clearance over $120$~d, $e^{120/\tau_R}$} & $1$, $1.82$, $2.98$, $5.55$ & \\
\bottomrule
\end{tabular}
\\[4pt]
\parbox{0.97\textwidth}{\footnotesize Fixed, not scanned: $R_{\mathrm{pass}} = 2$ per day, $D_g = 0.35~\mu$m, $w = 0.05~\mu$m, the DIC limit of $0.30~\mu$m (varied in table~\ref{tab:popsens}), the lifespan $T_L = 120$~d, the reticulocyte lifetime $T_r = 1.8$~d, the cell area of $140~\mu$m$^2$ and the $10^4$ cells over which the extremes are taken.}
\end{table}

\textit{Scoring.} Each score is a sum of terms, one per target (table~\ref{tab:popscore}). A point target $t$ contributes $[\ln(x/t)]^2$ for a model value $x$; a range contributes nothing inside it and $[\ln(\mathrm{lo}/x)]^2$ or $[\ln(x/\mathrm{hi})]^2$ outside; and a model value of zero, for which the logarithm is undefined, contributes $100$. The asplenic age profile is scored as a direction only: it contributes nothing when the top-to-bottom ratio without a spleen is at most $1$, and $[\ln x]^2$ when it exceeds $1$. The all-targets score sums all thirteen terms and the first-hand score the first nine. The in vitro ratio enters the first at $4.3$ and the second at $4.29$, the ratio of the published fractions, a difference that changes no point's score by more than $0.0101$. The asplenic prevalence is scored against $25$--$55\%$, the span of the series means from $25.3\%$~\cite{sills1988formation} to $54.3\%$~\cite{holroyde1970acquisition}; table~\ref{tab:popmodel} quotes instead the reference range for individuals, $20$--$70\%$~\cite{buchanan1987pocked}. Each of the three further rankings changes one term. ``First-hand, faster rate'' scores the six-day acquisition at the $10.47\%$ that Colin and Schrier obtained in vitro on an explicitly unmodified copy of the protocol~\cite{colin1991spontaneous}, in place of the $3.71\%$ measured in vivo in six splenectomized patients, which the same laboratory's two in vitro series flank at $3.70$ and $4.61\%$. ``First-hand, cohort healthy'' scores the healthy prevalence against $0.44$--$0.60\%$, the mean $\pm\,2$ standard errors of $84$ subjects~\cite{buchanan1987formation}. ``All targets, pool corrected'' scores the sub-visible pool against the range $7.7$--$9.6$ per cell in place of the point $20$.

\begin{table}[htbp]
\small
\centering
\renewcommand\arraystretch{1.15}
\caption{\textbf{The targets of the population model.} All thirteen enter the all-targets score; the first nine make up the first-hand score.}
\label{tab:popscore}
\resizebox{\textwidth}{!}{\begin{tabular}{rllcc}
\toprule
 & Observable & Model quantity & Target & Source \\
\midrule
1 & Healthy prevalence (\%) & prevalence of visible bodies, spleen present & $0.5$--$2.6$ & \cite{holroyde1970acquisition,sills1988formation,buchanan1987pocked,colin1991spontaneous} \\
2 & Asplenic prevalence (\%) & the same, spleen absent & $25$--$55$ & \cite{holroyde1970acquisition,sills1988formation} \\
3 & Six-day acquisition (\%) & cells free at splenectomy carrying a visible body $144$~h later & $3.71$ & \cite{sills1988physiologic} \\
4 & Largest body ($\mu$m) & largest diameter carried by one cell in $10^4$, spleen absent & $1.5$ & \cite{holroyde1970acquisition} \\
5 & Most loaded cell & most visible bodies carried by one cell in $10^4$, spleen absent & $20$--$60$ & \cite{holroyde1970acquisition} \\
6 & In vivo top-to-bottom ratio & prevalence, light over dense fraction, spleen present & $13.5$ & \cite{sills1988formation} \\
7 & In vitro top-to-bottom ratio & six-day acquisition, light over dense fraction & $4.3$ / $4.29$ & \cite{sills1988formation} \\
8 & In vitro acquisition, light fraction (\%) & six-day acquisition in the light fraction & $10.3$ & \cite{sills1988formation} \\
9 & Asplenic age profile & prevalence, light over dense fraction, spleen absent & $\leq 1$ & \cite{reinhart1988red,dehaan1988vacuoles} \\
\midrule
10 & Sub-visible pool, healthy (per cell) & bodies below the DIC limit, spleen present & $20$ & \cite{reinhart1988red} \\
11 & Rise of the pool after splenectomy & pool without a spleen over pool with one & $4$ & \cite{reinhart1988red} \\
12 & Age-flatness of the pool, healthy & pool at $120$~d over pool at $20$~d, spleen present & $1$ & \cite{reinhart1988red} \\
13 & Age-flatness of the pool, asplenic & the same, spleen absent & $1$ & \cite{reinhart1988red} \\
\bottomrule
\end{tabular}}
\\[4pt]
\parbox{0.97\textwidth}{\footnotesize A range is written lo--hi. Target 7 is $4.3$ in the all-targets score and $4.29$ in the first-hand score. The most loaded cell is scored against $20$--$60$ for the ``one to over $20$'' of Holroyde and Gardner~\cite{holroyde1970acquisition}. Targets 10--13 describe the pool below the optical limit and come from one paper; the first-hand score sets them aside.}
\end{table}

\textit{Best points.} Table~\ref{tab:popbest} gives the best point of each ranking, with its parameters and observables, and table~\ref{tab:poprise} how the hundred best points of each ranking divide by the rise of clearance, the best score attainable at each rise, and how each ranking's best point scores under the other four. Three features bear on the main text. First, every best point has $m_d = 1$: the minority differs from the other cells in how fast it nucleates bodies rather than in their size. The preference is weak, since $29$ of the first-hand ranking's hundred best points have $m_d = 1.6$, and it does not survive a change of the DIC limit (below). Second, the first-hand rankings place their best point at $\tau_R = 70$~d, the largest rise of this grid, and their best attainable score falls monotonically with the rise, from $0.609$ at constant clearance to $0.220$ at $5.6$-fold under the first-hand score. This grid therefore does not locate an optimum rise; what it establishes is the lower bound used in the main text, and the extended scan below shows how the score continues beyond it. The all-targets rankings run the other way, the best all-targets score rising from $1.634$ at constant clearance to $2.880$ at $5.6$-fold. Third, the best point of the cohort-healthy ranking differs from the first-hand best point in $\tau$, $\phi$, $d_{50}$ and $f_{ab}$, but scores $0.221$ under the first-hand score against $0.220$, which is the sense in which the main text calls it within $1\%$ of the point we report. The best all-targets point lies at the edge of the scan in $J_0$, $\tau$ and $m_J$, and a minority cell without a spleen then carries vacuole membrane equal to $84\%$ of its own area by $120$~d.

\begin{table}[htbp]
\small
\centering
\renewcommand\arraystretch{1.1}
\caption{\textbf{The best point of each ranking: its parameters (top) and its observables (bottom).}}
\label{tab:popbest}
\begin{tabular}{rlccccc}
\toprule
 & & All & All, corr. & FH & FH, rate & FH, cohort \\
\midrule
 & Peak nucleation rate $J_0$ (per day) & 5.00 & 2.50 & 0.60 & 1.20 & 0.60 \\
 & Decay time of the source $\tau$ (d) & 5 & 5 & 35 & 35 & 10 \\
 & Floor of the source $\phi$ & 0.02 & 0.02 & 0 & 0 & 0.1 \\
 & Median birth diameter $d_{50}$ (nm) & 110 & 90 & 90 & 90 & 110 \\
 & Log-standard-deviation of birth diameters $\sigma$ & 0.50 & 0.65 & 0.65 & 0.65 & 0.65 \\
 & Fraction of fast-nucleating cells $f_{ab}$ & 0.01 & 0.01 & 0.01 & 0.01 & 0.001 \\
 & Rate multiplier of that fraction $m_J$ & 50 & 25 & 25 & 25 & 25 \\
 & Size multiplier of that fraction $m_d$ & 1.0 & 1.0 & 1.0 & 1.0 & 1.0 \\
 & Age scale of splenic clearance $\tau_R$ (d) & $\infty$ & $\infty$ & 70 & 70 & 70 \\
 & Rise of clearance over $120$~d, $e^{120/\tau_R}$ & 1.00 & 1.00 & 5.55 & 5.55 & 5.55 \\
 & Score under its own ranking & 1.634 & 1.425 & 0.220 & 0.588 & 0.204 \\
\midrule
1 & Healthy prevalence (\%) & 0.95 & 0.62 & 0.48 & 0.94 & 0.44 \\
2 & Asplenic prevalence (\%) & 49.6 & 41.7 & 41.1 & 63.3 & 43.7 \\
3 & Six-day acquisition (\%) & 4.07 & 3.20 & 3.88 & 7.21 & 4.15 \\
4 & Largest body ($\mu$m) & 1.09 & 1.57 & 1.57 & 1.69 & 1.68 \\
5 & Most loaded cell & 53 & 24 & 26 & 47 & 23 \\
6 & In vivo top-to-bottom ratio & 7.33 & 8.29 & 9.02 & 8.87 & 8.93 \\
7 & In vitro top-to-bottom ratio & 6.85 & 7.65 & 5.13 & 4.96 & 3.91 \\
8 & In vitro acquisition, light fraction (\%) & 14.65 & 12.05 & 8.95 & 16.42 & 10.48 \\
9 & Asplenic age profile & 0.69 & 0.68 & 0.43 & 0.48 & 0.49 \\
10 & Sub-visible pool, healthy (per cell) & 19.5 & 8.9 & 5.6 & 11.2 & 2.4 \\
11 & Rise of the pool after splenectomy & 2.2 & 2.0 & 3.2 & 3.2 & 3.6 \\
12 & Age-flatness of the pool, healthy & 0.55 & 0.63 & 0.13 & 0.13 & 0.21 \\
13 & Age-flatness of the pool, asplenic & 1.39 & 1.39 & 2.22 & 2.22 & 2.07 \\
 & Membrane of a minority cell, asplenic, $120$~d (\%) & 84 & 22 & 24 & 48 & 22 \\
\bottomrule
\end{tabular}
\\[4pt]
\parbox{0.97\textwidth}{\footnotesize The rankings are all targets (All), all targets with the pool corrected (All, corr.), first-hand (FH), first-hand with the faster rate (FH, rate) and first-hand with the cohort's healthy prevalence (FH, cohort). The score is each ranking's own. The observables are numbered as in table~\ref{tab:popscore}; All, corr.\ scores observable 10, FH, rate observable 3 and FH, cohort observable 1 against the targets given in the text. The membrane is that of the bodies of a minority cell at $120$~d without a spleen, as a percentage of $140~\mu$m$^2$. The FH column is the point of table~\ref{tab:popmodel} and Figs.~\mref{fig:popmodel} and~\mref{fig:forced}.}
\end{table}

\begin{table}[htbp]
\small
\centering
\renewcommand\arraystretch{1.15}
\caption{\textbf{How the rankings divide by the rise of splenic clearance over the life of the cell.}}
\label{tab:poprise}
\resizebox{\textwidth}{!}{\begin{tabular}{lccccccccccccc}
\toprule
 & \multicolumn{4}{c}{Top hundred, by rise} & \multicolumn{4}{c}{Best score, by rise} & \multicolumn{5}{c}{Best point scored under} \\
\cmidrule(lr){2-5}\cmidrule(lr){6-9}\cmidrule(lr){10-14}
Ranking & $1$ & $1.82$ & $2.98$ & $5.55$ & $1$ & $1.82$ & $2.98$ & $5.55$ & All & All, corr. & FH & FH, rate & FH, cohort \\
\midrule
All & 74 & 23 & 3 & 0 & 1.634 & 2.146 & 2.372 & 2.880 & 1.634 & 2.134 & 0.826 & 1.710 & 1.038 \\
All, corr. & 72 & 24 & 4 & 0 & 1.425 & 1.644 & 1.922 & 2.808 & 2.073 & 1.425 & 0.621 & 2.002 & 0.623 \\
FH & 0 & 3 & 7 & 90 & 0.609 & 0.347 & 0.263 & 0.220 & 6.731 & 5.204 & 0.220 & 1.202 & 0.218 \\
FH, rate & 0 & 0 & 4 & 96 & 1.334 & 1.009 & 0.780 & 0.588 & 6.111 & 5.793 & 0.890 & 0.588 & 1.088 \\
FH, cohort & 0 & 1 & 7 & 92 & 0.623 & 0.347 & 0.263 & 0.204 & 7.669 & 4.533 & 0.221 & 1.067 & 0.204 \\
\bottomrule
\end{tabular}}
\\[4pt]
\parbox{0.97\textwidth}{\footnotesize The rise is $e^{120/\tau_R}$, $1$ at constant clearance. The first block is Fig.~\mref{fig:forced}D. The second gives, at each rise, the lowest score of the ranking among the $8100$ points with that rise. The third scores each ranking's best point under all five rankings, abbreviated as in table~\ref{tab:popbest}, so that its diagonal repeats the scores given there.}
\end{table}

\textit{The scan extended, and the DIC limit moved.} Two choices behind the scan were tested by repeating it (table~\ref{tab:popsens}). The grid of $\tau_R$ stops at $70$~d, where the first-hand rankings place their best points, so it was extended to $\tau_R = 50$, $40$ and $30$~d, rises of $11.0$, $20.1$ and $54.6$-fold. The DIC limit of $0.30~\mu$m is the permissive end of what has been published: it is the diameter above which Reinhart and Chien classed a vacuole as large, while the smallest vacuole they could detect by light microscopy measured $390$~nm~\cite{reinhart1988red}. The extended scan was therefore repeated with the limit at $0.40~\mu$m, the strict end, and at $0.20~\mu$m, beyond the permissive one. Each repetition covers $56\,700$ points, and the one at $0.30~\mu$m reproduces the $32\,400$ points of the published grid exactly.

At $0.30~\mu$m the extension turns the first-hand score over. Its best attainable value falls from $0.220$ at $5.6$-fold to $0.075$ at $20$-fold and rises again to $0.107$ at $55$-fold. The best point there differs from the one we report only in $\tau_R = 40$~d and $m_J = 50$. It scores $19.1$ under all targets, against $6.7$, and a minority cell without a spleen then carries vacuole membrane equal to $48\%$ of its own area by $120$~d, against $24\%$. Under the cohort variant the best scores at $20$- and $55$-fold are nearly equal, $0.056$ and $0.058$, and under the faster-rate variant the score is still falling at $55$-fold. The all-target rankings keep rising, to $11.6$ at $55$-fold. The first-hand record therefore sets a lower bound on the rise and favours rises of ten-fold and more, without excluding the largest scanned.

Moving the DIC limit leaves three conclusions of the main text in place. At neither limit do the hundred best points of any ranking include a source that never switches off, and no more than three of any hundred have a single population. The split between the rankings persists: the all-target rankings put at least $95$ of their hundred best points at constant or $1.8$-fold clearance, and the first-hand rankings none at constant clearance. What moves is the size of the rise that matches the measured pair (the last row of each block of table~\ref{tab:popsens}). Constant clearance separates the two ratios at most $1.57$-fold at $0.20~\mu$m and $2.27$-fold at $0.40~\mu$m, against the measured $3.14$; at $0.40~\mu$m a $1.8$-fold rise reaches $3.11$, and at $0.20~\mu$m a three-fold rise reaches only $2.73$. The three-fold bound of the main text is therefore about two-fold at the strict limit and between three- and $5.6$-fold at $0.20~\mu$m.

Two features of the best points do not survive. The first is $m_d = 1$. At $0.20~\mu$m the first-hand best point gives the minority bodies $1.6$ times larger, as do $41$ to $53$ of the hundred best points under the first-hand score and its cohort variant, and at $0.40~\mu$m so does the best first-hand point once its membrane is capped. Whether the minority's bodies are larger as well as more frequent is therefore left open. The second is the physical admissibility of the first-hand best point at $0.40~\mu$m, with which a minority cell without a spleen carries vacuole membrane equal to $387\%$ of its own area, more than the cell has. Among the points that keep it within the cell's area, the best first-hand score is $0.377$ at an $11$-fold rise, little below the $0.406$ and $0.393$ at three- and $5.6$-fold, with $m_d = 1.6$ and the largest body at the $2.0~\mu$m top of the size grid. At $0.20~\mu$m, finally, every first-hand best point, on either grid, has the lowest nucleation rate scanned, $J_0 = 0.3$ per day, so the grid chosen at $0.30~\mu$m does not bracket them.

\begin{table}[htbp]
\small
\centering
\renewcommand\arraystretch{1.1}
\caption{\textbf{The scan extended to larger rises of clearance and repeated at two further DIC limits.}}
\label{tab:popsens}
\resizebox{\textwidth}{!}{\begin{tabular}{llccccccccc}
\toprule
 & & \multicolumn{7}{c}{Best score, by rise of clearance over $120$~d} & \multicolumn{2}{c}{Best point} \\
\cmidrule(lr){3-9}\cmidrule(lr){10-11}
DIC limit & Ranking & $1$ & $1.82$ & $2.98$ & $5.55$ & $11.0$ & $20.1$ & $54.6$ & $m_d$ & Membrane (\%) \\
\midrule
$0.20~\mu$m & All & 3.543 & 3.566 & 4.592 & 5.446 & 7.156 & 9.654 & 15.470 & 1.0 & 23 \\
 & All, corr. & 2.762 & 2.796 & 3.373 & 3.971 & 5.456 & 7.630 & 13.119 & 1.0 & 17 \\
 & FH & 1.090 & 0.685 & 0.473 & 0.319 & 0.212 & 0.080 & 0.025 & 1.6 & 18 \\
 & FH, rate & 1.983 & 1.510 & 1.289 & 0.963 & 0.669 & 0.530 & 0.388 & 1.0 & 13 \\
 & FH, cohort & 2.042 & 1.471 & 1.184 & 0.919 & 0.743 & 0.678 & 0.413 & 1.0 & 8 \\
 & \emph{Largest separation} & 1.57 & 2.16 & 2.73 & 3.54 & 4.49 & 5.55 & 7.92 & & \\
\midrule
$0.30~\mu$m & All & 1.634 & 2.146 & 2.372 & 2.880 & 4.438 & 6.718 & 11.559 & 1.0 & 84 \\
 & All, corr. & 1.425 & 1.644 & 1.922 & 2.808 & 4.431 & 6.837 & 11.630 & 1.0 & 22 \\
 & FH & 0.609 & 0.347 & 0.263 & 0.220 & 0.103 & 0.075 & 0.107 & 1.0 & 48 \\
 & FH, rate & 1.334 & 1.009 & 0.780 & 0.588 & 0.444 & 0.386 & 0.364 & 1.0 & 36 \\
 & FH, cohort & 0.623 & 0.347 & 0.263 & 0.204 & 0.075 & 0.056 & 0.058 & 1.0 & 48 \\
 & \emph{Largest separation} & 1.82 & 2.49 & 3.12 & 3.96 & 5.32 & 6.73 & 9.20 & & \\
\midrule
$0.40~\mu$m & All & 1.495 & 1.926 & 2.129 & 2.813 & 4.457 & 6.777 & 12.065 & 1.0 & 124 \\
 & All, corr. & 1.324 & 1.670 & 2.051 & 2.991 & 4.625 & 6.903 & 12.033 & 1.6 & 78 \\
 & FH & 0.594 & 0.435 & 0.406 & 0.304 & 0.274 & 0.317 & 0.461 & 1.0 & 387 \\
 & FH, rate & 1.235 & 0.997 & 0.788 & 0.612 & 0.513 & 0.487 & 0.528 & 1.0 & 83 \\
 & FH, cohort & 0.582 & 0.389 & 0.340 & 0.260 & 0.198 & 0.216 & 0.331 & 1.0 & 387 \\
 & \emph{Largest separation} & 2.27 & 3.11 & 3.88 & 4.92 & 6.13 & 7.71 & 10.65 & & \\
\bottomrule
\end{tabular}}
\\[4pt]
\parbox{0.97\textwidth}{\footnotesize The first four rises are those of the published grid (table~\ref{tab:popgrid}); the last three add $\tau_R = 50$, $40$ and $30$~d, $8100$ points each, so that each DIC limit is a scan of $56\,700$ points. A best score is the lowest score of the ranking among the points at that rise, as in table~\ref{tab:poprise}, whose second block the first four columns at $0.30~\mu$m repeat. The best point is the ranking's best over all seven rises; its membrane is that of the bodies of a minority cell at $120$~d without a spleen, as a percentage of $140~\mu$m$^2$. Scores are taken over all points whatever their membrane; capped at the cell's area, the all-targets best point at $0.40~\mu$m stays at constant clearance (score $1.538$, membrane $62\%$). The largest separation is the largest ratio of the in vivo to the in vitro top-to-bottom ratio reached by any point at that rise; the measured pair, $13.5$ and $4.3$, is $3.14$-fold apart. Rankings are abbreviated as in table~\ref{tab:popbest}.}
\end{table}

\textit{The coalescence model behind the volume budget.} The nucleation rate of the volume budget, and the effect of cell-to-cell spread quoted with it in the main text, come from a second, stochastic model in which vacuoles are born small and grow only by fusing, which is the hypothesis the budget tests. Each cell is simulated event by event over $120$~d. Vesicles are nucleated at the cell's own rate $J$, drawn once per cell from a gamma distribution of mean $J_m$ and a given coefficient of variation and capped at $50\,J_m$, with volumes drawn from the size law of the budget, lognormal with $130\pm72$~nm and truncated to $40$--$330$~nm~\cite{reinhart1988red}. Any two bodies of diameters $d_i$ and $d_j$ fuse at the rate $\alpha K_{ij}/V$, where $K_{ij} = (2k_{\mathrm{B}}T/3\eta)(1/d_i + 1/d_j)(d_i + d_j)$ is the Brownian collision kernel at $310$~K in a cytosol of $\eta = 6$~mPa\,s and $V = 90~\mu$m$^3$, and the product has the summed volume. The efficiency $\alpha$ is the same for every pair and is set for each $J_m$ so that a cell at the mean rate holds about $20$ sub-visible bodies, the published estimate~\cite{reinhart1988red}: $\alpha = J_m V/(400\,K_{11})$ with $K_{11}$ the kernel of two $130$~nm bodies, which is $6.8\times10^{-7}$ at $J_m = 0.5$ per day. With a spleen every body is removed at the rate $R_{\mathrm{pass}}\,g(D)$ of the population model, without its age dependence, and a cell is pocked when it holds a body above $0.30~\mu$m. Volumes are kept on a grid of four bins per octave from $40$~nm to $2.56~\mu$m, and a volume that falls between two bins is assigned to one of them at random so that its expectation is exact.

Table~\ref{tab:coalescence} gives the twelve conditions run, each with $2000$ cells. Prevalences are averages over a uniform age distribution, taken from the state of every cell at $20$, $40$, \ldots, $120$~d. For the acquisition, $2000$ cells were grown to each of those ages with the spleen and then followed for six days without it, each at its own rate, and the statistic is that of the measurement, $(p_t - p_0)/(100 - p_0)$. At $J_m = 0.5$ per day and a coefficient of variation of $1.5$ the modelled acquisition is $1.41$, $2.53$ and $3.76\%$ at $48$, $96$ and $144$~h against the measured $1.31$, $2.91$ and $3.71\%$, with both prevalences in range; this is the rate of the budget. At that rate the largest body in any cell without a spleen is $0.76~\mu$m, and widening the distribution raises it to $1.08~\mu$m only as the asplenic prevalence falls to $12.0\%$. At $1.0$ per day the largest body reaches $1.21$--$1.81~\mu$m, and at a coefficient of variation of $2$ both prevalences remain in range and the most loaded cell without a spleen holds $28$ visible bodies at $120$~d. The six-day acquisition of that condition, $8.18\%$, is $2.2$ times the in vivo measurement and below the $10.47\%$ of the second laboratory~\cite{colin1991spontaneous}. The fraction of cells whose rate clears the budget of a $1.5~\mu$m body over a whole life follows from the gamma distribution alone: $1.4\times10^{-4}$ at $15.9$ times the mean and a coefficient of variation of $1.5$, and $3.0\times10^{-2}$ at $6.3$ times the mean and a coefficient of variation of $2$. The budget of the main text is therefore a statement about a cell at the mean rate. At the slower of the two published rates a single broad distribution of rates does not let coalescence reach the observed sizes, and at the faster one it does. This model has a constant source and no density fractions, so it cannot be scored against the targets of the scan above; the model of the next paragraph can.

\textit{Coalescence given the source and the minority of the population model.} To ask whether those targets tell bodies born large from bodies assembled by fusion, the coalescence model was given everything the population model has except its size law. Each cell nucleates at $J(a) = J_c\,(e^{-a/\tau} + \phi)$, with $J_c = J_0$ in the majority and $m_J J_0$ in a fraction $f_{ab}$ of cells. Bodies are born on the size law of the budget and grow only by fusion, at the rate $\alpha K_{ij}/V$ above. The spleen removes a body at $R_{\mathrm{pass}}\,g(D)\,e^{a/\tau_R}$, and the density fractions are the same tilts in age. The efficiency $\alpha$ is free and takes the place of the birth-size median and its spread: eleven values from $10^{-7}$ to $10^{-2}$, on the population model's grid of $J_0$, $\tau$, $\phi$, $f_{ab}$, $m_J$ and clearance rise, $19\,932$ combinations in all, each assembled from $3000$ simulated cells of the majority and $300$ of the minority. Twenty-four minorities whose lifetime volume exceeds the largest bin were not simulated. Every combination was scored with the first-hand score, unchanged, and the ten best were run again with four times the cells and new random numbers. The rule of comparison was fixed before the scan: coalescence reaches the born-large model if its best confirmed point scores no worse than the hundredth point of the born-large scan, $0.373$. It fails by the standard applied to that model's own controls if it scores worse than the better of them, the $1.674$ of the best point without a minority.

It reaches it. The best confirmed point scores $0.349$ ($J_0 = 5.0$ per day, $\tau = 5$~d, $\phi = 0.02$, $\alpha = 3\times10^{-6}$, $f_{ab} = 0.01$, $m_J = 50$, a three-fold rise in clearance) and a second $0.368$ ($J_0 = 2.5$ per day, $\tau = 20$~d, $\alpha = 10^{-6}$, a $5.6$-fold rise), against $0.220$ for the best born-large point. The other eight lie between $0.376$ and $0.474$. At the best point the prevalence is $0.82\%$ with a spleen and $77.4\%$ without, the acquisition $0.99$, $1.86$ and $2.68\%$ at $48$, $96$ and $144$~h, the in vivo top-to-bottom ratio $11.8$, the in vitro top fraction $8.6\%$ and ratio $5.4$, the largest body $1.71~\mu$m, and the most loaded cell holds $44$ visible bodies. Its weak terms are the asplenic prevalence, above the observed range where the born-large fit gives $41.1\%$, and the six-day acquisition, below the measurement; the second point has the acquisition, $1.44$, $2.71$ and $3.85\%$, and overshoots the prevalence again, at $75.1\%$. The controls fail as they do for the born-large model: the best point without a minority scores $1.90$, with a largest body of $0.60~\mu$m, and the best with a constant source $5.56$. Without an age-dependent clearance the best score of the scan is $0.82$, against $0.33$ at a three-fold rise and $0.35$ at $5.6$-fold. Scores below $0.5$ occur only for $\alpha$ between $10^{-6}$ and $10^{-5}$, the order of the value set from the sub-visible pool in the constant-source model above. The fast cells are asked for more vacuole volume than in the born-large fit, though for no more membrane: at the two points that reach the born-large range a minority cell holds $3.4$ and $5.2~\mu$m$^3$ of vacuole at $120$~d, in bodies whose surface is $14$ and $24\%$ of the host's membrane area, against $1.45~\mu$m$^3$ and $24\%$. No combination whose minority holds less than the volume of a $1.5~\mu$m body scores better than $0.456$. The first-hand record therefore requires a source concentrated in young cells, a fast minority and a clearance that rises with age whichever way a vacuole reaches its size, and does not by itself tell large bodies born large from large bodies assembled by fusion in the fast cells.

\begin{table}[htbp]
\small
\centering
\renewcommand\arraystretch{1.15}
\caption{\textbf{The coalescence model: mean nucleation rate and cell-to-cell spread.} Vacuoles are born at $130\pm72$~nm and grow only by fusion; the rate of each cell is drawn from a gamma distribution of mean $J_m$ and coefficient of variation CV. Acquisition is the fraction of cells free of visible bodies at splenectomy that carry one $48$, $96$ and $144$~h later; prevalence is the fraction of cells carrying a body above $0.30~\mu$m, averaged over cell age; the largest body is the largest in any of the $2000$ cells without a spleen at any of six ages. The last row gives the measurements: acquisition in six splenectomized patients~\cite{sills1988physiologic}, the ranges of prevalence quoted in the main text, and the largest body reported in splenectomized patients~\cite{holroyde1970acquisition}.}
\label{tab:coalescence}
\begin{tabular}{rrrrrrrr}
\toprule
 & & \multicolumn{3}{c}{Acquisition (\%)} & \multicolumn{2}{c}{Prevalence (\%)} & Largest \\
\cmidrule(lr){3-5}\cmidrule(lr){6-7}
$J_m$ (per day) & CV & $48$~h & $96$~h & $144$~h & spleen & no spleen & body ($\mu$m) \\
\midrule
$0.3$ & $1.5$ & $0.80$ & $1.79$ & $2.65$ & $0.63$ & $20.3$ & $0.60$ \\
$0.3$ & $2.0$ & $0.76$ & $1.45$ & $2.32$ & $0.73$ & $17.4$ & $0.64$ \\
$0.3$ & $3.0$ & $0.83$ & $1.56$ & $2.06$ & $0.73$ & $13.0$ & $0.85$ \\
$0.3$ & $4.0$ & $0.71$ & $1.33$ & $2.09$ & $0.60$ & $9.4$ & $0.76$ \\
\midrule
$0.5$ & $1.5$ & $1.41$ & $2.53$ & $3.76$ & $0.87$ & $29.4$ & $0.76$ \\
$0.5$ & $2.0$ & $1.37$ & $2.72$ & $3.87$ & $0.97$ & $24.5$ & $0.85$ \\
$0.5$ & $3.0$ & $1.16$ & $2.30$ & $3.63$ & $1.10$ & $16.4$ & $1.02$ \\
$0.5$ & $4.0$ & $1.30$ & $2.21$ & $3.00$ & $1.15$ & $12.0$ & $1.08$ \\
\midrule
$1.0$ & $1.5$ & $2.65$ & $5.39$ & $8.24$ & $2.50$ & $44.3$ & $1.21$ \\
$1.0$ & $2.0$ & $2.84$ & $5.57$ & $8.18$ & $2.36$ & $35.7$ & $1.61$ \\
$1.0$ & $3.0$ & $2.49$ & $4.45$ & $6.25$ & $2.81$ & $23.1$ & $1.81$ \\
$1.0$ & $4.0$ & $1.75$ & $3.40$ & $4.75$ & $2.48$ & $16.2$ & $1.81$ \\
\midrule
\multicolumn{2}{l}{Measured} & $1.31$ & $2.91$ & $3.71$ & $0.5$--$2.6$ & $20$--$70$ & $1.5$ \\
\bottomrule
\end{tabular}
\end{table}

\section{Measured values compared with the population model}\label{sec:fig5data}

Every point, bar and reference line in Fig.~\mref{fig:popmodel} that comes from a measurement is listed in tables~\ref{tab:measA} to~\ref{tab:measD} with its source, its reported dispersion and the number of subjects behind it, so that the figure can be checked or re-plotted without the source papers to hand. Model curves are not tabulated; they follow from the equations of section~\ref{sec:pop} at the parameters given in table~\ref{tab:popmodel}, panel D carrying the same model through the months after splenectomy.

\begin{table}[htbp]
\small
\centering
\renewcommand\arraystretch{1.2}
\caption{\textbf{Healthy adults with an intact spleen.} Plotted in Fig.~\mref{fig:popmodel}A. From Table~1 of Sills et al.~\cite{sills1988formation}: ten healthy adults; at least $500$ consecutive cells scored per subject by Nomarski interference contrast; values are mean $\pm$ SD. The reticulocyte column is from the same table and is what makes the horizontal axis an age axis. \emph{The ages in the last column are not measured.} They are the mean age of each layer under the model's reconstruction of the gradient, and are given only so that the plotted abscissae can be reproduced. The three layers are the uppermost and lowermost two to three layers and several middle layers of a continuous gradient, not three equal thirds, which is why they do not average to the unseparated value; only ratios between fractions are compared with the model.}
\label{tab:measA}
\begin{tabular}{lcccc}
\toprule
Density fraction & Cells with a vacuole (\%) & Reticulocytes (\%) & Mean cell age (d) \\
\midrule
lightest (youngest) & $5.4 \pm 2.3$ & $10.5$ & $16.2$ \\
middle              & $1.3 \pm 1.3$ & $1.2$  & $64.4$ \\
densest (oldest)    & $0.4 \pm 0.5$ & $0.2$  & $88.0$ \\
unseparated         & $0.5 \pm 0.5$ & $1.5$  & -- \\
\bottomrule
\end{tabular}
\end{table}

\begin{table}[htbp]
\small
\centering
\renewcommand\arraystretch{1.2}
\caption{\textbf{In vivo acquisition with the spleen just removed.} Plotted in Fig.~\mref{fig:popmodel}B. From Table~1, group V, of Sills et al.~\cite{sills1988physiologic}: six patients splenectomized for trauma, bled within the day before surgery and at $48$, $96$ and $144$~h after; samples fixed immediately, without incubation; $500$ consecutive cells by interference contrast; mean $\pm$ SD; $P = 0.005$ across the series. The quantity plotted is the acquisition among cells that carried no vacuole at time zero, $(v(t)-v(0))/(100-v(0))$, which is the statistic the model computes. \emph{The dispersions in the last column are the reported SDs of the counts carried through the same factor} $1/(100-v(0)) = 1/99.6$; they are not errors propagated through the difference, and they should be read as the spread of the raw counts. \emph{Only the $144$~h value was a scored target}; the $48$ and $96$~h points are free.}
\label{tab:measB}
\begin{tabular}{lcc}
\toprule
Time after splenectomy & Pocked cells as counted (\%) & Plotted acquisition (\%) \\
\midrule
$0$ (within $24$~h before surgery) & $0.4 \pm 0.2$ & $0 \pm 0.20$ \\
$48$~h  & $1.7 \pm 0.7$ & $1.31 \pm 0.70$ \\
$96$~h  & $3.3 \pm 1.6$ & $2.91 \pm 1.61$ \\
$144$~h & $4.1 \pm 1.7$ & $3.71 \pm 1.71$ \\
\bottomrule
\end{tabular}
\end{table}

\begin{table}[htbp]
\small
\centering
\renewcommand\arraystretch{1.2}
\caption{\textbf{The observed size ranges, drawn as horizontal bars.} Plotted in Fig.~\mref{fig:popmodel}C. From Holroyde and Gardner~\cite{holroyde1970acquisition}: at least $500$ cells scanned per subject; mean $\pm$ SD. The bars mark the reported diameter ranges only; their heights in the figure carry no information and are chosen to separate them. Two vertical reference lines are drawn in the same panel and neither is a measurement of the plotted distribution: $0.30~\mu$m is the diameter above which Reinhart and Chien~\cite{reinhart1988red} classed a vacuole as large (their Table~II) and is the visibility threshold the model uses, and $0.35~\mu$m is the model's splenic gate, which is fixed rather than fitted. \emph{The $0.30~\mu$m threshold is the permissive end of what has been published.} The same authors measured their own smallest light-microscope-detectable vacuole at $390$~nm, and in one subject examined both ways found vacuoles in $13.4\%$ of cell sections by electron microscopy against $0.7\%$ of cells by light microscopy. A stricter threshold would lower the prevalence the model predicts.}
\label{tab:measC}
\resizebox{\textwidth}{!}{
\begin{tabular}{lccc}
\toprule
Group & Vacuole diameter ($\mu$m) & Vacuoles per cell & Cells with a vacuole (\%) \\
\midrule
healthy adults, spleen intact & $0.2$--$0.8$ & usually one; none above five & $2.6 \pm 1.1$ \\
splenectomized, $n = 20$      & $0.2$--$1.5$ & one to over twenty           & $54.3 \pm 13.1$ \\
\bottomrule
\end{tabular}}
\end{table}

\begin{table}[htbp]
\small
\centering
\renewcommand\arraystretch{1.2}
\caption{\textbf{The rise after splenectomy, drawn as a bar and a box.} Plotted in Fig.~\mref{fig:popmodel}D. From the Results and Fig.~1 of Buchanan et al.~\cite{buchanan1987formation}: six children splenectomized for haematological disease---autoimmune haemolytic anaemia, immune thrombocytopenic purpura (two), sickle cell anaemia, lupus with thrombocytopenia and hereditary spherocytosis---in all but one of whom red cell survival was judged normal or near normal. One drop of blood fixed in $3\%$ buffered glutaraldehyde; $500$ consecutive cells scored by Nomarski interference contrast for one or more vacuoles; duplicate counts agree within $\pm20\%$. The two rows are the authors' own summary of the six per-patient curves of their Fig.~1. No per-patient values were digitized, so the figure draws exactly what is stated: a lower bound of $20\%$ across $20$--$40$ days, and a box of $40$--$60\%$ across $60$--$100$ days. \emph{Only the plateau level was a scored target}, as the asplenic prevalence of table~\ref{tab:popmodel}; the approach to it was not fitted. Two further series are not plotted, because only second-hand summaries of them are available to us: thirty-one adults who reached a plateau at $9$--$12$ weeks, quoted in the same paper, and a continuous rise between $10$ and $130$ days reported by de~Haan et al.~\cite{dehaan1988vacuoles}.}
\label{tab:measD}
\begin{tabular}{lcc}
\toprule
As stated by the authors & Days after splenectomy & Pocked cells (\%) \\
\midrule
``rose progressively to over a value of 20\%''        & $20$--$40$  & over $20$ \\
``the maximal value of 40--60\% was usually reached'' & $60$--$100$ & $40$--$60$ \\
\bottomrule
\end{tabular}
\end{table}
\section{Unstained dots in reticulocytes and in cells at rest}\label{sec:dots}

Two exploratory observations on human blood bear on the account of the main text. Both count compact dots, brighter than the cytoplasm in bright-field images, that do not take up the reticulocyte stain. Cells that the experimenter had labelled pocked carry them (below), but the same cells have not been imaged by DIC, so we do not call the dots vacuoles. All images are bright-field colour images taken on an Echo Revolution microscope with a $60\times/1.2$ water-immersion objective; a pixel is $0.0593~\mu$m, calibrated from stage moves.

\textit{Do reticulocytes carry the dots?} Blood from a donor with sickle cell disease (HbSS) was stained with new methylene blue and imaged unfixed in $17$ fields. In an earlier blind reading of the colour images, two readers had classed $590$ of its cells from the stain as reticulocytes (R) or not (N): all $86$ cells the experimenter had staged as reticulocytes, the $104$ further cells of strongest stain colour, and $400$ drawn at random from the rest. The same two readers then scored dots on the same cells in the blue channel, where the stain is not seen, each alone and without the stain calls: $0$ to $4$ dots or uncountable, clear or equivocal, on tiles $11.9~\mu$m wide in random order, $20$ of them shown twice. The analysis was fixed before the scores were returned. A cell is dotted if both readers scored at least one dot and is left out if either called it uncountable; R and N are the cells both readers called R ($67$) or N ($343$). $P(\mathrm{dot}\,|\,R)$ and $P(\mathrm{dot}\,|\,N)$ carry Horvitz--Thompson weights ($8.51$ for the random stratum, $1$ otherwise), and the interval of their ratio comes from a stratified bootstrap of $10\,000$ resamples.

The population model of the main text sets the expectation. With the spleen absent, over red cell lifespans of $10$ to $25$ days, reticulocytes taken as the youngest $1.1$ to $3.0\%$ of cells (the share the stain reading found in this sample) and a visibility limit of $0.3$ to $0.5~\mu$m, its best point gives a ratio of $0.029$ ($0.012$--$0.19$ over those ranges). A body made after release has had little time to appear in a cell a day or two old. The ratio hardly depends on how fast the source decays, $0.029$ to $0.074$ at the best points for decay times of $5$ to $35$ days. If the bodies are present at release, the ratio is $1$.

The readers agreed moderately: Cohen's $\kappa$ for $0$, $\ge 1$ or uncountable was $0.58$ between them, and $0.77$ and $0.90$ within each on the repeated tiles. Reticulocytes were often uncountable, $45$ of $67$ against $63$ of $343$ mature cells, mostly as crumpled or folded cells, so $22$ R and $280$ N cells were scored. Two of the $22$ carried a dot by both readers, both called equivocal, against $18$ of the $280$: a ratio of $1.5$ ($95\%$ CI $0$--$4.4$; $1.8$ within fields by Mantel--Haenszel), against the $0.04$ dotted reticulocytes the model expects, which makes two or more a $0.001$ event. With equivocal dots counted as none, no reticulocyte carries a dot, and the reading cannot separate $0.03$ from $1$ (table~\ref{tab:dots}). Every variant that counts equivocal dots gives $1.4$ to $2.9$; none restricted to clear dots has the power to decide. A second reading by one reader, made after the first results were known, gave $5$ of $23$ against $27$ of $280$ (ratio $2.3$, $0.6$--$4.8$) and agreed with that reader's first at $\kappa = 0.53$. The dots are not reader noise: $13$ of $38$ scored cells that the experimenter had labelled pocked carried one by both readers, against $4$ of $244$ she had left unclassified. The reading leans towards dots already present in reticulocytes, which the population model, whose source starts at release, does not include; but it rests on two equivocal calls, and a larger sample of reticulocytes, imaged by DIC, is needed to settle it.

\begin{table}[htbp]
\small
\centering
\renewcommand\arraystretch{1.2}
\caption{\textbf{Dots in reticulocytes and in mature cells of one HbSS sample.} R and N are the cells that both readers called reticulocytes or not from the stain; a cell is scored unless either reader called its dots uncountable. Fractions carry Horvitz--Thompson weights; intervals are $95\%$ from a stratified bootstrap of $10\,000$ resamples, or a one-sided $95\%$ bound where no reticulocyte carries a dot. The population model gives a ratio of $0.029$ ($0.012$--$0.19$) for bodies made after release and $1$ for bodies present at release (section~\ref{sec:dots}).}
\label{tab:dots}
\begin{tabular}{lccccc}
\toprule
Analysis & R dotted/scored & N dotted/scored & $P(\mathrm{dot}\,|\,R)$ & $P(\mathrm{dot}\,|\,N)$ & Ratio ($95\%$ CI) \\
\midrule
Dot by both readers (main) & $2/22$ & $18/280$ & $0.091$ & $0.062$ & $1.5$ ($0$--$4.4$) \\
Dot by either reader & $7/22$ & $30/280$ & $0.32$ & $0.11$ & $2.9$ ($1.1$--$5.6$) \\
Equivocal dots counted as none & $0/22$ & $7/280$ & $0$ & $0.026$ & $0$ ($\le 4.9$) \\
Clear calls only & $0/2$ & $7/126$ & $0$ & $0.059$ & $0$ ($\le 13$) \\
Stain class from reader 1 alone & $3/26$ & $18/294$ & $0.090$ & $0.059$ & $1.5$ ($0$--$4.6$) \\
Stain class from reader 2 alone & $2/24$ & $19/286$ & $0.083$ & $0.062$ & $1.4$ ($0$--$4.2$) \\
\bottomrule
\end{tabular}
\end{table}

\begin{figure}[htbp]
\centering
\includegraphics[width=\textwidth]{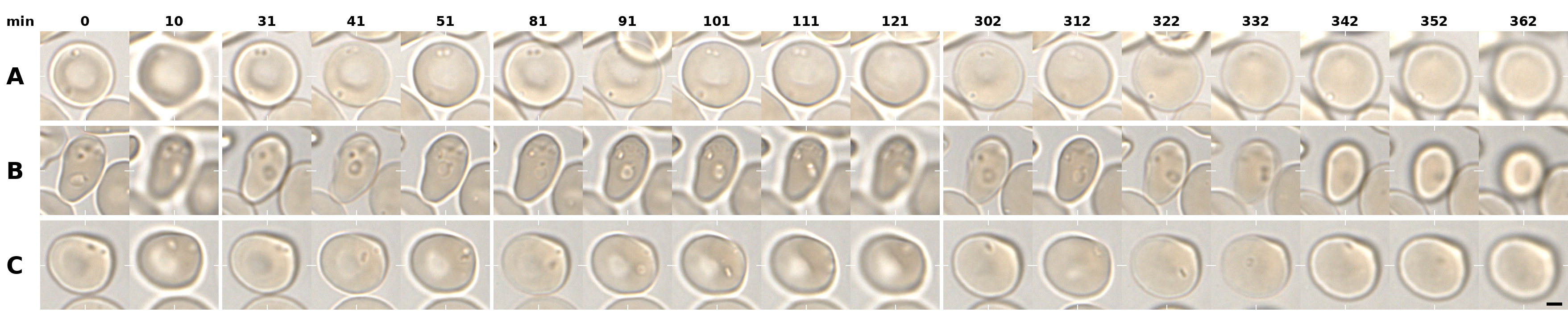}
\caption{\textbf{Three cells followed for six hours at rest.} Rows as the readers saw them: each row follows one cell, columns are minutes after the first frame, and white gaps separate the four imaging runs. Tiles are $11.9~\mu$m wide and fixed on the glass, so a moving cell moves within its row; white ticks mark the tile centre. \textbf{(A)} A pair of dots near the top and a third near the lower rim, seen in the focused frames from $31$ to $312$~min; at $0$~min the pair reads as one dot. \textbf{(B)} A dot at upper left and a ring-shaped body below it that resolves into two dots at $101$--$111$~min; the cell rounds up from $342$~min. Both readers called no merge in (A) and (B). \textbf{(C)} Two dots from $0$ to $101$~min and one from $302$~min; the frames in which the second was lost, $111$--$121$~min, are blurred, and both readers called the merge uncertain. Scale bar, $2~\mu$m.}
\label{fig:dotstrips}
\end{figure}

\textit{Do dots merge in a cell at rest?} Unstained blood was imaged at $10$ fixed stage positions in four runs, $17$ frames between $0$ and $362$~min. Cells were segmented, registered against the glass and linked from frame to frame automatically, and $816$ were followed through at least $10$ frames. An automatic screen for compact dots then selected the $69$ cells that showed two or more dots, each found again in the next or previous frame, in at least one frame. The screen finds candidates rather than counting: in cells with dots visible by eye it found them in only $2$ to $6$ of $15$ to $17$ frames, and it does not look within $1.2~\mu$m of the cell edge, so it supplies no denominator for a rate over all cells. The two readers judged each selected cell on its time strip (fig.~\ref{fig:dotstrips}), alone and without the screen's marks: whether one intact, in-focus cell could be followed; the most dots in any one frame; and whether two dots seen apart became one that stayed single in at least two frames, with the cell in focus and unfolded in the frames compared (a merge), or uncertain when focus, folding or an overlap prevented the call.

Each reader found $51$ of the $69$ cells usable, $47$ of them in common ($\kappa = 0.70$), and both saw two or more dots in one frame in $29$ of those ($\kappa = 0.61$ for the count). Neither reader called a merge in any cell. In $15$ of the $29$ at least one reader called the merge uncertain ($\kappa = 0.48$ for the merge call over all $69$), in each case a pair that read as one dot as the focus softened or the cell changed shape (fig.~\ref{fig:dotstrips}C). Pairs that alternated between one and two dots with the focus were called no merge by both. With no merge in $29$ cells, the one-sided $95\%$ bound on the chance that a cell carrying two or more dots shows a visible merge within six hours at rest is $0.10$. The record cannot be longer: from about $300$~min some cells crenate, round up or become ghosts.

\section{Limitations of the model}\label{sec:limits}

\textit{The nucleation rate.} The nucleation rate on which the volume budget and the fitted source both rest is one laboratory's six-day curve, and a second laboratory running the same protocol measured adult acquisition two to three times faster. For a cell at the mean rate the budget's case against coalescence survives that (main text, Results). Its agreement with the healthy size range does not, and at the higher rate a lifetime's fused output sits at the top of the observed range rather than in the middle of it. Where cells differ in rate the case is weaker. At the higher rate a coalescence model with a broad distribution of rates reaches the observed sizes (table~\ref{tab:coalescence}). At either rate the fast minority of the population model makes about the volume of the largest bodies, $1.45~\mu$m$^3$ in a lifetime against the $1.77~\mu$m$^3$ of a $1.5~\mu$m body. In those cells the budget therefore does not decide between bodies born large and bodies assembled by fusion. Nor does the scored record: given the same source, minority and clearance, a coalescence model reaches a first-hand score of $0.349$ against the $0.220$ of the born-large fit (section~\ref{sec:pop}). Which rate is right is unresolved, and the two series differ in shape as well as level, a lag to $96$~h then a jump in one and a steady ramp from $48$~h in the other (section~\ref{sec:pop}).

\textit{The population model.} The population model is a mean-field description of a Poisson birth-and-clearance process, with no spatial structure inside the cell and no representation of how a body is made; it cannot say why a source should decay at all, only that it must. Its weakest match is the in vivo top-to-bottom ratio, $9.0$ against a measured $13.5$, and its fraction \emph{levels} run $2.7$- to $4.0$-fold low even though the unseparated prevalence is right. The splenic passage rate and the gate diameter are fixed rather than fitted, and the age dependence of clearance is a single exponential chosen for want of any measurement of its shape. The scan is a grid, so the reported best point is the best of $32\,400$ rather than an optimum, and no confidence region is quoted. The re-scoring shows directly how flat the surface is in some directions: a point with a source decay time of $10$ rather than $35$ days, a birth-size median of $110$ rather than $90$~nm and a minority fraction ten times smaller scores within one per cent of the one we report. Parameters should be read as the region the record admits, not as estimates (section~\ref{sec:pop}).

\textit{Fusion and resolution.} The fusion criterion is phenomenological: two surfaces held within $\delta_f$ for $t_f$ are merged, with no model of bilayer fusion, and $d_c = 1.2~\mu$m scores a gap of about $0.6~\mu$m, far above molecular apposition. The vacuoles we simulate, $0.6~\mu$m and larger, sit at the resolution limit of the DPD fluid (particle spacing $\sim\!0.6~\mu$m, cutoff $1~\mu$m), so the sub-$300$~nm vacuoles from which the coalescence hypothesis starts are not resolved at all. The same coarseness makes the cytosol layer against every repulsive surface over about $1~\mu$m, and makes constrictions effectively narrower than their geometry (sections~\ref{sec:checks} and~\ref{sec:geometry}).

\textit{Shared placements and single realizations.} The cascade runs share one initial placement, so their first fusions are common to all of them, and four runs stopped on a vacuole-mesh bond failure (section~\ref{sec:fusion}). The repeated-passage runs share another, packed so that half their vacuoles fuse before the drive starts. Under the summed-area rule five of their six driven runs failed on a vacuole-mesh bond before the end of the record, four of them after reaching a single vacuole (section~\ref{sec:passage}). There, as everywhere, the vacuole--vacuole barrier rather than the flow sets how closely two vacuoles are pressed. The vascular simulations were run once each, and they begin immediately after the vacuole--vacuole barrier is switched on. In none of the three has the contact occupancy converged, so the comparison between geometries is one between rising curves sampled over a window they share (section~\ref{sec:contact}).

\textit{Units, drive and duration.} Under our unit mapping the DPD solvent is two to four thousand times denser than plasma, so the vessel flows run at Reynolds numbers of $6$ to $40$ rather than plasma's $10^{-3}$ to $10^{-2}$. In the dilation that inertia makes the vortex in the sac (section~\ref{sec:reynolds}). The time scale of each geometry carries an uncertainty of about $10\%$, since measuring the solvent density that sets it only away from the cell shortens it by $8$--$13\%$ (section~\ref{sec:units}). At the slit only the cell is driven, and a held cell withdraws its tongue from the aperture where one held by a pressure drop would not (section~\ref{sec:slit}); and the cytosol has the viscosity of plasma. Each run covers at most $1.3$~s of flow, against the weeks over which vacuoles accumulate in patients, and the model contains no source of new vacuoles.

\clearpage
\section*{Captions for Movies S1 and S2}
\noindent Movies S1 and S2 are the ancillary files \texttt{Movie\_S1.mp4} and \texttt{Movie\_S2.mp4} of this article.

\medskip
\noindent\textbf{Movie S1. A fusion cascade in tube flow.} The cell of Fig.~\mref{fig:fusion}A, its ten vacuoles of $D_0 = 0.6~\mu$m, at $\delta_f = 0.7~\mu$m and $t_f = 3.6$~ms, seen along the tube axis by a camera that moves with the cell; one frame per segment of the run, every $1.71$~ms, over its first $103$~ms, shown at $8$ frames per second. Colour gives the number of original vacuoles a body contains, as in Fig.~\mref{fig:fusion}A: purple $1$, green $2$, orange $3$, red $5$. Five fusions are executed by $10.3$~ms and the sixth at $78.7$~ms, after which the cell holds four vacuoles for the rest of its run.

\medskip
\noindent\textbf{Movie S2. Repeated passage through the slit.} The cell of Fig.~\mref{fig:passage}A, its ten vacuoles of $D = 0.6~\mu$m each in its own colour, driven at $f = 8$ through the slit of Fig.~\mref{fig:arrest}A in a run without fusion, seen obliquely with the near half of the wall cut away by a camera that moves with the cell, so that the wall comes past once per lap of the periodic box. One frame every $1.85$~ms, from $7.4$~ms before the drive starts to $1108$~ms after it, shown at $24$ frames per second; the cell crosses the aperture $16$ times, about every $70$~ms.

\end{document}